\documentclass[aps,twoside,reprint,amsmath,amssymb,prb]{revtex4-1}
\usepackage{amsfonts}%
\usepackage{amsmath}%
\usepackage{subfigure}
\usepackage{amssymb}%
\usepackage{graphicx}
\providecommand{\U}[1]{\protect\rule{.1in}{.1in}}

\newcommand\diverg{\nabla\cdot}

\newcommand{\mn}[1]{\left\langle #1 \right\rangle}

\begin{document}


\title{Optical and DC conductivity of the two-dimensional Hubbard model in the pseudogap regime and
across the antiferromagnetic quantum critical point, including vertex corrections}

\author{Dominic Bergeron}
\email{dominic.bergeron@usherbrooke.ca}
\author{Vasyl Hankevych}
\author{Bumsoo Kyung}
\author{A.-M. S. Tremblay}
\email{tremblay@physique.usherbrooke.ca}
\affiliation{D\'epartement de physique, Regroupement Qu\'eb\'ecois sur les Mat\'eriaux de Pointe (RQMP) and Canadian Insitute for Advanced Research (CIFAR),  Universit\'e de Sherbrooke, Qu\'ebec, Canada}

\pacs{PACS number}

\begin{abstract}
The conductivity of the two-dimensional Hubbard model is particularly relevant for high-temperature superconductors. Vertex corrections are expected to be important because of strongly momentum dependent self-energies. To attack this problem, one must also take into account the Mermin-Wagner theorem, the Pauli principle and crucial sum rules in order to reach non-perturbative regimes. Here, we use the Two-Particle Self-Consistent approach that satisfies these constraints. This approach is reliable from weak to intermediate coupling. A functional derivative approach ensures that vertex corrections are included in a way that satisfies the f sum-rule. The two types of vertex corrections that we find are the antiferromagnetic analogs of the Maki-Thompson and Aslamasov-Larkin contributions of superconducting fluctuations to the conductivity but, contrary to the latter, they include non-perturbative effects. The resulting analytical expressions must be evaluated numerically. The calculations are impossible unless a number of advanced numerical algorithms are used. These algorithms make extensive use of fast Fourier transforms, cubic splines and asymptotic forms. A maximum entropy approach is specially developed for analytical continuation of our results. These algorithms are explained in detail in appendices. The numerical results are for nearest neighbor hoppings. In the pseudogap regime induced by two-dimensional antiferromagnetic fluctuations, the effect of vertex corrections is dramatic. Without vertex corrections the resistivity increases as we enter the pseudogap regime. Adding vertex corrections leads to a drop in resistivity, as observed in some high temperature superconductors. At high temperature, the resistivity saturates at the Ioffe-Regel limit. At the quantum critical point and beyond, the resistivity displays both linear and quadratic temperature dependence and there is a correlation between the linear term and the superconducting transition temperature. A hump is observed in the mid-infrared range of the optical conductivity in the presence of antiferromagnetic fluctuations.

\end{abstract}
\maketitle

\section{Introduction}

The calculation of transport quantities in strongly correlated electron systems is particularly challenging, but is a necessary step to make contact with a wide class of experiments. Even for the simplest model, namely the single-band Hubbard model, this is a formidable task. Taking up the challenge is all the more important for the two-dimensional case, that acts as the minimal model for the high temperature cuprate superconductors,\cite{Anderson:1987} layered organic superconductors,~\cite{Powell:2006} and a number of other materials.

Even in cases where one has a good handle on the single-particle Green's function, the difficulty of calculating transport in the 2D Hubbard model stems from the fact that one cannot neglect the effect of vertex corrections when strong momentum-dependent correlations are present. Those vertex corrections are the analog of the self-energy, but for the two-particle response functions. When vertex corrections are not included, conservation laws can be violated and results inaccurate. In the case of small finite systems tractable by exact diagonalization or quantum Monte Carlo calculations (QMC), the correlation function is directly evaluated and vertex corrections are not an issue. However, those results are more relevant for finite frequency conductivity and strong coupling, where correlations are mainly local.\cite{Nakano:1999,Tohyama:2005,Dagotto:1992,Riera:1994,Li:1994,Scalapino92_1}

Consider for example the electrical conductivity for the two-dimensional Hubbard model. For the infinite system, optical and DC conductivity calculations have been performed without vertex corrections using Dynamical Mean-Field Theory (DMFT)~\cite{comanac:2008} and Cellular-DMFT (CDMFT)~\cite{Chakraborty:2008}. Those calculations have also been done with the composite operator method~\cite{Mancini:2004} but vertex corrections cannot all be taken into account. For the $t-J$ model, the strong coupling limit of the Hubbard model, a number of approaches have been used, in particular the extended dynamical cluster approximation \cite{Maier:2003, Hettler:2000}, but vertex corrections \cite{Haule:2007,Haule:2007_2} have been neglected. However, recent optical conductivity calculations for the Hubbard model with the dynamical cluster approximation (DCA) took vertex corrections into account\cite{Lin:2009_1,Okamoto:2010}. The effects were found to be important only at high frequency. Despite this recent advance, the calculation of vertex corrections with DCA or CDMFT, considered the best available ones at strong coupling,~\cite{Kotliar:2001,Maier:2005,KotliarRMP:2006} is still an open problem.

At weak coupling, the Boltzmann equation offers a tractable approach that satisfies conservation laws when one includes scattering-in terms. For linear response, its variational formulation has been used for example to compute the effect of spin fluctuations within the self-consistent renormalized approach~\cite{Moriya:1990} and also to investigate the resistivity near the quantum critical point in the clean~\cite{Hlubina:1995} and disordered cases.\cite{Rosch:1999,RoschRMP:2007} The drawback of this approach is that it assumes the existence of quasiparticles and that this assumption is not valid in two dimensions, especially near the pseudogap regime and quantum critical point. Green's function approaches that do not assume quasiparticles are preferable. Hence, some resistivity calculations without vertex corrections were done with the \textit{T}-matrix approximation \cite{Wermbter:1993} and with the fluctuation-exchange (FLEX) approximation\cite{Dahm:1994}. Other FLEX calculations take into account some vertex corrections due to spin and charge fluctuations \cite{Kontani:2007,Kontani:1999}. In these works, only the antiferromagnetic Maki-Thompson (MK) diagrams were included. The antiferromagnetic analogs of the Aslamasov-Larkin (AL) diagrams were neglected, claiming that the latter are negligible. We will see that in the DC case this is true only in the presence of particle-hole symmetry. A review on those calculations is given in Ref. \onlinecite{Kontani:2008}. In Ref.~\onlinecite{Yanase:2002} that takes into account superconducting fluctuations, AL diagrams are taken into account for superconducting fluctuations but they are neglected for the spin fluctuations. A recent calculation using field-theory methods of the conductivity at the quantum critical point \cite{Hartnoll:2011} includes all vertex corrections but only at $T=0$. The renormalized classical regime where a pseudogap appears has also been considered but focussing only on the hot spots \cite{Sadovskii:2002} or neglecting the AL contribution \cite{LinMillis:2011}. There are also analytical results for the conductivity with vertex corrections using Fermi liquid theory \cite{Maebashi:1997,Maebashi:1998}.

Despite all these results, the electrical resistivity at weak to intermediate coupling has not been reliably computed for all dopings and temperatures because particle-hole symmetry, where the AL term vanishes, does not generally hold and because there are regimes where Fermi liquid theory is no-longer valid. Fermi-liquid theory breaks down in the pseudogap regime, near the antiferromagnetic quantum critical point and in the Ioffe-Regel limit. Hence, in this paper, we extend the Two-Particle Self-Consistent (TPSC) approach\cite{Vilk:1994,Vilk:1997,Allen:2003,TremblayMancini:2011} to include the effect of vertex corrections in the calculation of the resistivity and optical conductivity of the one-band, square lattice, nearest-neighbor two-dimensional Hubbard model for weak to intermediate coupling. This regime corresponds to values of the interaction strength $U$ below the critical value for the Mott transition. We present numerical results as examples and discuss possible links with experiments on cuprates. In particular, we consider the origin of the mid-infrared hump in the electron-doped materials, the Ioffe-Regel limit, insulating behavior in the pseudogap regime and the link between linear resistivity, quantum critical behavior and superconductivity.

The TPSC approach has the following strengths that make it a good choice for the present purposes. In two dimensions, the Mermin-Wagner theorem\cite{Mermin:1966,Hohenberg:1966} prevents the occurrence of antiferromagnetic long-range order at finite temperature. Not many theories can handle that constraint. Because long-range order is prohibited, there is a wide range of temperatures where there are huge antiferromagnetic [or spin-density wave (SDW)] fluctuations in the paramagnetic state. It is in this regime that a fluctuation-induced pseudogap can appear.\cite{Vilk:1995,Vilk:1997,Moukouri:2000,Kyung:2004} The standard way to treat fluctuations in many-body theory, the Random Phase Approximation (RPA), leads instead to long-range order and misses this effect. The RPA also violates the Pauli principle in an important way.\cite{Vilk:1994} The FLEX\cite{Bickers:1989,Bickers_dwave:1989} approximation and the self-consistent renormalized theory of Moriya-Lonzarich \cite{Moriya:2003,Lonzarich:1985,Moriya:1985} satisfy the Mermin-Wagner theorem but they do not satisfy the Pauli principle and consistency between one and two-particle quantities. Strengths and weaknesses of these approaches are discussed further in Refs.~\onlinecite{Vilk:1997,Allen:2003}. Weak coupling renormalization group approaches become uncontrolled when the antiferromagnetic fluctuations begin to diverge \cite{Dzyaloshinskii:1987,Schulz:1987,Lederer:1987,Honerkamp:2001}. Other approaches include the effective spin-Hamiltonian approach \cite{Logan:1996}. The TPSC approach does not assume a Migdal theorem for spin fluctuations and Kanamori-Brueckner renormalization of the bare interaction is included without adjustable parameter. The conditions for the appearance of a pseudogap induced by antiferromagnetic fluctuations have been verified experimentally in electron-doped cuprates.\cite{Motoyama:2007} In addition to the above theoretical considerations, the TPSC approach has been extensively benchmarked against Quantum Monte Carlo calculations in regimes where the latter is available.\cite{Vilk:1994,Vilk:1997,Veilleux:1995,Moukouri:2000,Kyung:2003a,LTP:2006,TremblayMancini:2011} The agreement between the one-particle spectral function of the approach and QMC calculations in the pseudogap regime is remarkable\cite{Moukouri:2000}. The TPSC approach however fails when temperature is too far below that where the pseudogap appears.

In the TPSC method, the calculation proceeds in two steps. At the first step, spin-spin and charge-charge correlation functions are obtained with irreducible vertices that are determined self-consistently. That is the origin of the name of the approach. At the second level of approximation, a non-trivial self-energy that is consistent with the spin and charge fluctuations and that can explain fluctuation-induced momentum-dependent pseudogaps is then calculated. The charge-charge correlation function at the first level of approximation satisfies the \textit{f}-sum rule with the Green's function at the same level, but it misses lifetime effects necessary to obtain non-trivial conductivity. What is needed is a calculation of the current-current correlation function that includes Green's functions dressed at the second level of approximation with the corresponding irreducible vertices. What has been missing up to now is an expression for the corresponding irreducible vertices. Following Baym and Kadanoff,\cite{Baym:1962,Kadanoff_Baym:1962} here we use a functional derivative approach to obtain irreducible vertices that satisfy conservation laws. We check that the \textit{f}-sum rule is then satisfied with the Green's function obtained at the second-level of approximation. For the conductivity, we show that, not only the Maki-Thompson-type contributions from spin-density wave (SDW) fluctuations, but also the Aslamasov-Larkin contributions are important. The latter have a dramatic effect in the pseudogap regime.

The paper is structured as follow. The next section contains the details of the methodology and is divided into five subsections: \ref{sec:model}, the model; \ref{sec:linear_resp}, the conductivity in linear response theory; \ref{sec:tpsc}, a derivation of the TPSC approach for the spin and charge response functions and the one-particle self-energy; \ref{sec:cond_tpsc}, the conductivity in the TPSC approach; and finally \ref{sec:algo}, a description of the numerical algorithms that were used to calculate the expression given in subsection \ref{sec:cond_tpsc}. Section \ref{sec:res} presents the results for the system considered, followed by a discussion and a conclusion in sections \ref{sec:discussion} and \ref{sec:conclusion}, respectively. Also, some useful derivations, such as those for the conductivity, the \textit{f}-sum rule, Ward identities, are given in appendices along with details of algorithms that are of more general applicability, such as calculating response functions with Fast-Fourier transforms and cubic splines and analytical continuation of numerical data.

\section{Methodology}\label{sec:method}

We first define the model, recall the conductivity formula and introduce the TPSC method in the functional derivative formalism. This approach allows us, in the fourth subsection, to derive the conductivity formula including vertex corrections. The last subsection briefly describes the numerical algorithms that we implemented. Those are detailed in appendices. In this section, we use units where $e=1$, $\hbar=1$ and $a=1$, $a$ being the lattice parameter. In section \ref{sec:res}, we use both those units and physical units to allow comparison with typical experimental values.

\subsection{Model}\label{sec:model}
Our model is the two-dimensional Hubbard Hamiltonian in the presence of an electromagnetic field that is treated classically. With the usual Peierls substitution, we have
\begin{equation}\label{eq:Hubbard_EM_field}
H(t)=-\sum_{ ij \sigma} t_{ij}  c_{i\sigma}^{\dagger}c_{j\sigma}e^{-i\int_i^j d\mathbf{r}_{ij}\cdot \mathbf{A}(\mathbf{r},t)} + U\sum_{i} n_{i\uparrow}n_{i\downarrow}\,,
\end{equation}
where $\mathbf{r}_{ij}=\mathbf{r}_{i}-\mathbf{r}_{j}$, $t_{ij}$ are the hopping matrix elements between the sites of the lattice,  $c_{j\sigma}$ destroys a particle with spin $\sigma$ at site $j$ and $c_{i\sigma}^{\dagger}$ creates a particle at site $i$, $\mathbf{A}(\mathbf{r},t)$ is the vector potential, $U$ is the on-site repulsion energy and $n_{i\sigma}=c_{i\sigma}^{\dagger}c_{i\sigma}$ is the spin $\sigma$ particle number operator at site $i$. Note that it is perfectly general to use only the vector potential to represent the field, since a scalar potential can always be removed using the proper gauge transformation. The form \eqref{eq:Hubbard_EM_field} for the Hamiltonian is justified by gauge invariance. Further discussion of the Peierls substitution may be found in Ref.~\onlinecite{Paul:2003}.

\subsection{Conductivity in linear response theory}\label{sec:linear_resp}

This derivation of the linear response result will allow us to set the notation. To obtain the conductivity, we first need the expression for the current operator. In the $x$ direction, for example, it is given by
\begin{equation}
j_x^S(\mathbf{r},t)=-\frac{\delta H}{\delta A_x (\mathbf{r},t)}\,,
\end{equation}
where the superscript $S$ indicates that $j_x$ is in Shr\"odinger representation, despite its dependance on $t$. If we apply this definition to Eq.~\eqref{eq:Hubbard_EM_field}, keep terms up to linear order in the vector potential, assuming that $\mathbf{A}(\mathbf{r},t)=A_x(\mathbf{r},t)\hat{x}$, with $A_x(\mathbf{r},t)$ varying slowly on the scale of a lattice spacing so that
\begin{equation}
\int_i^j dx_{ij}A_x(\mathbf{r},t)\approx\frac{x_{ij}}{2}\left(A_x(\mathbf{r}_i,t)+A_x(\mathbf{r}_j,t)\right)\,,
\end{equation}
where $x_{ij}$ is the $x$ component of the vector $\mathbf{r}_j-\mathbf{r}_i$, we obtain,
\begin{multline}\label{eq:jxlt}
j_x^S(\mathbf{r}_l,t)=\frac{i}{2}\sum_{ \delta \sigma} \delta_x t_{\delta} \left(c_{l\sigma}^{\dagger}c_{l-\delta,\sigma}+c_{l+\delta,\sigma}^{\dagger}c_{l,\sigma}\right)\\
-\frac{1}{2}A_x(\mathbf{r}_l,t)\sum_{ \delta \sigma} \delta_x^2 t_{\delta}\left(  c_{l\sigma}^{\dagger}c_{l-\delta,\sigma}
+c_{l+\delta,\sigma}^{\dagger}c_{l\sigma}\right)\,,
\end{multline}
where $\delta_x$ is the projection along $x$ of the vector $\delta$ between neighbors. $t_{\delta}$ is the corresponding hopping matrix element. For a uniform electric field we can take the vector potential independent of position, which means that we need only the $\mathbf{q}=0$ component of the current
\begin{multline}\label{eq:jx_q0}
j_x^S(t)=-\frac{1}{N}\sum_{ \mathbf{k} \sigma} \frac{\partial \epsilon_\mathbf{k}}{\partial k_x} c_{\mathbf{k}\sigma}^{\dagger}c_{\mathbf{k},\sigma}\\
-A_x(t)\frac{1}{N}\sum_{ \mathbf{k} \sigma} \frac{\partial ^2\epsilon_\mathbf{k}}{\partial k_x^2}c_{\mathbf{k}\sigma}^{\dagger}c_{\mathbf{k},\sigma}\,,
\end{multline}
where $\epsilon_\mathbf{k}$ is the dispersion relation and $N$ the number of lattice sites or wave vectors in the Brillouin zone. For the following we need to define the paramagnetic current,
\begin{equation}
j_x^p=-\frac{1}{N}\sum_{ \mathbf{k} \sigma} \frac{\partial \epsilon_\mathbf{k}}{\partial k_x} c_{\mathbf{k}\sigma}^{\dagger}c_{\mathbf{k},\sigma}
\end{equation}
and the diamagnetic current,
\begin{equation}
j_x^d(t)=-A_x(t)\frac{1}{N}\sum_{ \mathbf{k} \sigma} \frac{\partial ^2\epsilon_\mathbf{k}}{\partial k_x^2}c_{\mathbf{k}\sigma}^{\dagger}c_{\mathbf{k},\sigma}\,,
\end{equation}
so that $j_x^S(t)=j_x^p+j_x^d(t)$.

According to linear response theory, the frequency dependent current in response to the field is
\begin{equation}\label{eq:mn_jxw_1}
\mn{j_x(\omega)}=\mn{\hat{j}_x(\omega)}+ \chi_{j_xj_x}(\omega)A_x(\omega)\,
\end{equation}
where $j_x(\omega)$ is the Fourier transform  of
\begin{equation}
j_x(t)=U^\dagger(t,-\infty)j_x^S(t)U(t,-\infty)\,,
\end{equation}
where $U(t,t')$ is the time evolution operator for the Hamiltonian $H(t)$, Eq.\eqref{eq:Hubbard_EM_field},
\begin{equation}\label{eq:chijj}
\chi_{j_xj_x}(\omega)=\frac{i}{N}\int dt\, e^{i\omega(t-t')} \mn{[\hat{j}_x^p(t), \hat{j}_x^p(t')]} \theta(t-t')\,,
\end{equation}
is the \textit{current-current} correlation function and the notation $\hat{O}(t)$ stands for the interaction representation of the operator $O$, namely, $\hat{O}(t)=e^{iH_0t}Oe^{-iH_0t}$, where $H_0$ is the Hamiltonian Eq.\eqref{eq:Hubbard_EM_field} with $ \mathbf{A}=0$. Also, $\mn{\ldots}$ means an equilibrium average, namely, in the system $H_0$.

Since $\mn{\hat{j}_x^p(t)}=\mn{j_x^p}=0$, the equilibrium average of the current operator in the interaction representation is given by the diamagnetic term
\begin{multline}
\mn{\hat{j}_x(t)}=\mn{\hat{j}_x^d(t)}=\mn{j_x^d(t)}\\
=-A_x(t)\frac{1}{N}\sum_{ \mathbf{k} \sigma} \frac{\partial ^2\epsilon_\mathbf{k}}{\partial k_x^2}\mn{c_{\mathbf{k}\sigma}^{\dagger}c_{\mathbf{k},\sigma}}\,.
\end{multline}
Defining
\begin{equation}\label{eq:def_kx0}
\mn{ k_x }= -\frac{1}{N}\sum_{\mathbf{k}}\frac{\partial^2\epsilon_\mathbf{k}}{\partial k_x^2}\mn {n_k}\,,
\end{equation}
where $n_k=c_{\mathbf{k}\sigma}^{\dagger}c_{\mathbf{k},\sigma}$, we have
\begin{equation}
\mn{\hat{j}_x(t)}=\mn{ k_x }A_x(t)
\end{equation}
or, in frequency,
\begin{equation}
\mn{\hat{j}_x(\omega)}=\mn{ k_x }A_x(\omega)\,,
\end{equation}
and Eq.\eqref{eq:mn_jxw_1} therefore becomes
\begin{equation}\label{eq:mn_jxw_2}
\mn{j_x(\omega)}=\left[\mn{ k_x }+ \chi_{j_xj_x}(\omega)\right]A_x(\omega)\,.
\end{equation}
Note that, in the case where the Hamiltonian has only nearest-neighbor hoppings, $\mn{ k_x }$ is proportional to the local kinetic energy, \cite{Scalapino92_1} but in general it cannot be regarded as such as is clear from Eq.\eqref{eq:def_kx0}.

To find the conductivity, it suffices to relate the electric field to the vector potential through
\begin{equation}
E_x(\omega)=i(\omega+i\eta)A_x(\omega)\,,
\end{equation}
where $\eta$ is positive and infinitesimal. Thus, the current is related to the electric field by
\begin{equation}\label{eq:mn_jxw_3}
\mn{j_x(\omega)}=\frac{\mn{ k_x }+ \chi_{j_xj_x}(\omega)}{i(\omega+i\eta)}E_x(\omega)
\end{equation}
and finally, since $\mn{j_x(\omega)}=\sigma_{xx}(\omega)E_x(\omega)$ by definition, the expression for the optical conductivity in linear response theory is
\begin{equation}\label{eq:def_cond}
\sigma_{xx}(\omega)=\frac{\mn{k_x}+\chi_{j_xj_x}(\omega)}{i(\omega+i\eta)}\,.
\end{equation}

In this work, we calculate only the real part of $\sigma(\omega)$, the dissipative part, and the expression that we use in practice is
\begin{equation}\label{eq:Re_sigma}
Re\, \sigma_{xx}(\omega)=\frac{\chi_{j_xj_x}^{\prime\prime}(\omega)}{\omega}\,,
\end{equation}
which is derived in Appendix \ref{sec:Appendix_f_sum_rule} from Eq.\eqref{eq:derivation_Re_sigma_chijj} to Eq.\eqref{eq:Re_sigma_chijj}. If one is interested in the imaginary part, it can be obtained using Kramers-Kr\"onig relations.

\subsection{Two-Particle Self-Consistent approach}\label{sec:tpsc}

In the TPSC approach, one- and two-particle Green's functions for the Hubbard model are calculated in a non-perturbative way. The approach enforces conservation laws, key sum rules and the Pauli principle. It was shown, from benchmarks with quantum Monte Carlo results~\cite{Moukouri:2000,Vilk:1997,Vilk:1994,Veilleux:1995,Vilk:1996,LTP:2006}, to be accurate within a few percent for interaction strengths up to about $U=6t$.  We will derive the TPSC approach below, but the reader can also resort to Refs. \onlinecite{Vilk:1997}, \onlinecite{Allen:2003},\onlinecite{LTP:2006} and \onlinecite{TremblayMancini:2011} for a more detailed discussion of the approach itself and a comparison with other approaches.

In this subsection, we present the key equations for the theory. The following subsection contains details of the derivation. We use the short-hand notation $1=(\mathbf{r}_1,\tau_1)$ for space and imaginary time coordinates and $q=(\mathbf{q},iq_n)$ for reciprocal space and Matsubara frequency coordinates.

The \textit{spin} and \textit{charge} response functions must be computed first from the expressions,
\begin{equation}\label{eq:chi_sp_q}
\begin{split}
\chi_{sp}(q)&=\frac{1}{N}\int d\tau e^{i\omega_n \tau} \mn{T_\tau S^z(\mathbf{q},\tau) S^z(-\mathbf{q})}\\
&= \frac{\chi_0(q)}{1-\frac{U_{sp}}{2}\chi_0(q)}
\end{split}
\end{equation}
and
\begin{equation}\label{eq:chi_ch_q}
\begin{split}
\chi_{ch}(q)&=\frac{1}{N}\int d\tau e^{i\omega_n \tau} \mn{T_\tau n(\mathbf{q},\tau) n(-\mathbf{q})}\,-\,\mn{n}^2\\
&=\frac{\chi_0(q)}{1+\frac{U_{ch}}{2}\chi_0(q)}\,,
\end{split}
\end{equation}
where $T_\tau$ is the \textit{imaginary time}-ordering operator and $\chi_0(q)$ is the Lindhard function, given by
\begin{equation}\label{eq:Lindhard}
\chi_{0}(q)=-2\frac{T}{N}\sum_k G^{(1)}(k+q)G^{(1)}(k)\,.
\end{equation}
Here, $G^{(1)}$ corresponds numerically to a non-interacting Green's function because the initial approximation for the self-energy, that is given below in Eq.\eqref{eq:ansatz_self}, is constant and the chemical potential is adjusted to obtain the correct filling. This point will become clear in the next subsection. Note that we assume here that the system is paramagnetic so that the spin index will often be omitted and the sum over it replaced by a factor of two, as in Eq.\eqref{eq:Lindhard}. In Eq.\eqref{eq:chi_sp_q} and Eq.\eqref{eq:chi_ch_q}, the parameters $U_{sp}$ and $U_{ch}$ are the spin and charge irreducible vertices, respectively. First, $U_{sp}$ is defined by
\begin{equation}
U_{sp}=U\frac{\left\langle n_\uparrow(1)n_{\downarrow}(1)\right\rangle }{\left\langle n_\uparrow(1)\right\rangle \left\langle n_{\downarrow}(1)\right\rangle }
\end{equation}
(this definition is used for hole doping, electron doping will be discussed below), so that it can be determined from the fluctuation-dissipation theorem
\begin{equation}\label{eq:sum_rule_Usp}
\begin{split}
\frac{T}{N}\sum_q \chi_{sp}(q)&=\mn{S^zS^z}\\
&=\mn{n}-2\mn{n_\uparrow n_\downarrow}\\
&=\mn{n}-\frac{U_{sp}}{U}\frac{\mn{n}^2}{2}\,,
\end{split}
\end{equation}
where we have used the Pauli principle $\mn{n_{\sigma}^2}=\mn{n_\sigma}$. Note that all quantities on right-hand side of Eq.\eqref{eq:sum_rule_Usp} are local ones. Then, once double occupancy is known, $U_{ch}$ can be determined from
\begin{equation}\label{eq:sum_rule_Uch}
\frac{T}{N}\sum_q \chi_{ch}(q)=\mn{n}+2\mn{n_\uparrow n_\downarrow}-\mn{n}^2\,.
\end{equation}
We also call Eq.\eqref{eq:sum_rule_Usp} and Eq.\eqref{eq:sum_rule_Uch} the \textit{local spin} and \textit{local charge} sum rules.

Finally, with the response functions \eqref{eq:chi_sp_q} and \eqref{eq:chi_ch_q} we can obtain the one-particle self-energy,
\begin{multline}\label{eq:self_2_1}
\Sigma_\sigma^{(2)}(k)=Un_{-\sigma}\\
+\frac{U}{8}\frac{T}{N}\sum_{q} \left[ 3U_{sp}\chi_{sp}(q)+U_{ch}\chi_{ch}(q)\right] G_\sigma^{(1)}(k+q)
\end{multline}
whose form will be explained below. Expressions \eqref{eq:chi_sp_q}, \eqref{eq:chi_ch_q} and \eqref{eq:self_2_1} are the basic TPSC equations.

\subsubsection{First step: spin and charge susceptibilities}

First let us derive expressions \eqref{eq:chi_sp_q} and \eqref{eq:chi_ch_q} for the spin and charge susceptibilities. In the following, we use Einstein's convention for the sums (or integrals), namely that an index appearing twice or more in an expression is summed over lattice sites and integrated over imaginary time. An overbar helps clarify which indices are involved. The approach follows the Martin-Schwinger techniques~\cite{MartinSchwinger:1959} described in Kadanoff and Baym's book~\cite{Kadanoff_Baym:1962}.

It is convenient to introduce a ``source field'' $\phi_{\sigma}(1,2)$ that couples to one-particle excitations in the system. It allows us to easily obtain correlation functions from functional derivatives. The source field can be set to zero at the end of the calculation. The source-field dependent Green's function in the grand canonical ensemble is then given by
\begin{equation}\label{eq:def_Green_phi}
\begin{split}
G_\sigma(1,2;\{\phi\})&=-\frac{Tr\left[ e^{-\beta K } T_\tau e^{-c_{\bar{\sigma}}^\dagger(\bar{1})\phi_{\bar{\sigma}}(\bar{1},\bar{2})c_{\bar{\sigma}}(\bar{2})} c_{\sigma}(1) c_{\sigma}^\dagger(2)\right] }{Tr\left[ e^{-\beta K } T_\tau e^{-c_{\bar{\sigma}}^\dagger(\bar{1})\phi_{\bar{\sigma}}(\bar{1},\bar{2})c_{\bar{\sigma}}(\bar{2})}\right] }\\
&=-\mn{T_\tau c_{\sigma}(1) c_{\sigma}^\dagger(2)}_\phi
\end{split}
\end{equation}
where $K=H-\mu N$, $\mu$ being the chemical potential and $N$ the number operator. We have also used the notation $\mn{\ldots}_\phi$ which means that the average is taken with the source field turned on. Response functions can be obtained from functional derivatives of $G_\sigma$ with respect to $\phi_{\sigma'}$ since
\begin{multline}\label{eq:dQdphi}
\frac{\delta G_\sigma(1,2;\{\phi\})}{\delta \phi_{\sigma'}(3,4)}=G_{\sigma'}(4,3;\{\phi\})G_{\sigma}(1,2;\{\phi\})\\
-\mn{T_\tau c_{\sigma'}^\dagger(3) c_{\sigma'}(4) c_{\sigma}^\dagger(2) c_{\sigma}(1)}_\phi \,.
\end{multline}
We also have
\begin{equation}\label{eq:chisp_def}
\begin{split}
\chi_{sp}(1,2)&=\mn{T_\tau S^z(1)S^z(2)}\\
&=\mn{T_\tau\left[ n_\uparrow(1)-n_\downarrow(1)\right]\left[ n_\uparrow(2)-n_\downarrow(2)\right]}\\
&=2\left(\mn{T_\tau  n_\uparrow(1)n_\uparrow(2)}-\mn{T_\tau  n_\uparrow(1)n_\downarrow(2)}\right)\,,
\end{split}
\end{equation}
where we have used spin rotation symmetry to obtain the last line from the previous one. For the charge response function, the corresponding expression is
\begin{equation}\label{eq:chich_def}
\begin{split}
\chi_{ch}(1,2)&=\mn{n(1)n(2)}-\mn{n(1)}\mn{n(2)}\\
&=2\left(\mn{T_\tau  n_\uparrow(1)n_\uparrow(2)}+\mn{T_\tau  n_\uparrow(1)n_\downarrow(2)}\right)\\
&\qquad\qquad-\mn{n(1)}\mn{n(2)}\,.
\end{split}
\end{equation}
According to Eq.~\eqref{eq:dQdphi}, the last two results Eqs.~\eqref{eq:chisp_def} and \eqref{eq:chich_def} can be written as
\begin{multline}\label{eq:chisp_ch_def}
\chi_{ch/sp}(1,2)=\\
-2\left(\frac{\delta G_\uparrow(1,1^+;\{\phi\})}{\delta \phi_{\uparrow}(2^+,2)} \pm \frac{\delta G_\uparrow(1,1^+;\{\phi\})}{\delta \phi_{\downarrow}(2^+,2)}\right)\Bigg|_{\{\phi\}=0}\,,
\end{multline}
where the expressions with the plus and minus sign correspond respectively to the charge and spin response functions. Here, $1^+=(\mathbf{r}_1,\tau_1+\epsilon)$, where $\epsilon$ is positive and infinitesimal. For the remainder of this section, we implicitly assume that derivatives with respect to $\phi$ are evaluated at $\{\phi\}=0$.

To obtain integral equations for the response functions, one begins with
\begin{equation}
G_\sigma(1,\bar{1})G_\sigma^{-1}(\bar{1},2)=\delta(1-2)\,.
\end{equation}
Taking the functional derivative of this equation with respect to $\phi_{\sigma'}(3,4)$, taking $2\rightarrow \bar{2}$, multiplying on the right by $G_\sigma(\bar{2},2)$ and summing over $\bar{2}$, we obtain
\begin{equation}\label{eq:dGdphi}
\frac{\delta G_\sigma(1,2)}{\delta \phi_{\sigma'}(3,4)}=-G_\sigma(1,\bar{1})\frac{\delta G_\sigma^{-1}(\bar{1},\bar{2})}{\delta \phi_{\sigma'}(3,4)}G_\sigma(\bar{2},2)\,.
\end{equation}
On the other hand, Dyson's equation in the presence of the field $\phi$ reads
\begin{equation}\label{eq:Dyson_phi}
G_\sigma^{-1}(1,2)=G_\sigma^{(0)-1}(1,2)-\phi_{\sigma}(1,2)-\Sigma_\sigma(1,2)\,,
\end{equation}
where $G_\sigma^{(0)}$ is the non interacting Green's function and $\Sigma_\sigma$ is the self-energy, so that
\begin{equation}
\frac{\delta G_\sigma^{-1}(1,2)}{\delta \phi_{\sigma'}(3,4)}=-\delta(1-3)\delta(2-4)\delta_{\sigma\sigma'}-\frac{\delta\Sigma_\sigma(1,2)}{\delta \phi_{\sigma'}(3,4)}\,,
\end{equation}
and therefore, from \eqref{eq:dGdphi},
\begin{multline}\label{eq:Bethe-Salpeter0}
\frac{\delta G_\sigma(1,2)}{\delta \phi_{\sigma'}(3,4)}=G_\sigma(1,3)G_\sigma(4,2)\delta_{\sigma\sigma'}\\
+ G_\sigma(1,\bar{1})\frac{\delta\Sigma_\sigma(\bar{1},\bar{2})}{\delta \phi_{\sigma'}(3,4)}G_\sigma(\bar{2},2)\,.
\end{multline}
Following Luttinger and Ward\cite{Luttinger_Ward:1960}, $\Sigma_\sigma$ is a functional of $G_\sigma$ and $G_{-\sigma}$, we find, applying the chain rule, that
\begin{multline}\label{eq:Bethe-Salpeter1}
\frac{\delta G_\sigma(1,2)}{\delta \phi_{\sigma'}(3,4)}=G_\sigma(1,3)G_\sigma(4,2)\delta_{\sigma\sigma'}\\
+ G_\sigma(1,\bar{1})\frac{\delta\Sigma_\sigma(\bar{1},\bar{2})}{\delta G_{\bar{\sigma}}(\bar{3},\bar{4})}\frac{\delta G_{\bar{\sigma}}(\bar{3},\bar{4})}{\delta \phi_{\sigma'}(3,4)}G_\sigma(\bar{2},2)\,.
\end{multline}
This is the analog of the Bethe-Salpeter equation for the \textit{particle-hole} channel. Defining the \textit{particle-hole} irreducible vertex
\begin{equation}
\Gamma_{\sigma\sigma'}(1,2;3,4)=\frac{\delta\Sigma_\sigma(1,2)}{\delta G_{\sigma'}(3,4)}\,,
\end{equation}
Eq. \eqref{eq:Bethe-Salpeter1} reads
\begin{multline}\label{eq:Bethe-Salpeter}
\frac{\delta G_\sigma(1,2)}{\delta \phi_{\sigma'}(3,4)}=G_\sigma(1,3)G_\sigma(4,2)\delta_{\sigma\sigma'}\\
+ G_\sigma(1,\bar{1})\Gamma_{\sigma\bar{\sigma}}(\bar{1},\bar{2};\bar{3},\bar{4})\frac{\delta G_{\bar{\sigma}}(\bar{3},\bar{4})}{\delta \phi_{\sigma'}(3,4)}G_\sigma(\bar{2},2)\,.
\end{multline}
The spin and charge response functions in Eq.~\eqref{eq:chisp_ch_def} are special cases of the more general functions
\begin{equation}\label{eq:chipm_def}
\chi_{\pm}(1,2;3,4)=-2\left(\frac{\delta G_\uparrow(1,2;\{\phi\})}{\delta \phi_{\uparrow}(3,4)} \pm \frac{\delta G_\uparrow(1,2;\{\phi\})}{\delta \phi_{\downarrow}(3,4)}\right)\,.
\end{equation}
It is straightforward to show, from the Bethe-Salpeter equation Eq.~\eqref{eq:Bethe-Salpeter} and spin-rotational invariance, that
\begin{multline}\label{eq:chi_plus}
\chi_{\pm}(1,2 ; 3,4)=-2G(1,3)G(4,2)\\
\pm \Gamma_{ch/sp}(\bar{1},\bar{2};\bar{3},\bar{4})G(1,\bar{1})G(\bar{2},2)\,\chi_{\pm}(\bar{3},\bar{4} \vert 3,4)
\end{multline}
where
\begin{equation}\label{eq:def_gam_ch_sp}
\Gamma_{ch/sp}(\bar{1},\bar{2};\bar{3},\bar{4})=\frac{\delta \Sigma_\uparrow(\bar{1},\bar{2})}{\delta G_{\downarrow}(\bar{3},\bar{4})}\pm \frac{\delta \Sigma_\uparrow(\bar{1},\bar{2})}{\delta G_{\uparrow}(\bar{3},\bar{4})}
\end{equation}
and $G=G_\uparrow=G_\downarrow$.

Up to now, all the results are exact. The first step in the TPSC method is to obtain a first approximation for the self-energy that will be used to obtain irreducible vertices. First, let us rewrite Dyson's equation Eq.~\eqref{eq:Dyson_phi} with zero source field in the form,
\begin{equation}\label{eq:Dyson}
G_\sigma^{(0)-1}(1,\bar{3})G_\sigma(\bar{3},2)=\delta(1-2)+\Sigma_\sigma(1,\bar{3})G_\sigma(\bar{3},2)\,.
\end{equation}
Then, note that the equation of motion for the Green's function reads
\begin{multline}\label{eq:eq_tau_G}
\sum_{l}\left[\left(-\frac{\partial}{\partial \tau}+\mu\right)\delta_{il} + t_{il}\right] G_{\sigma}(l-j,\tau)=\delta(\tau)\delta_{ij}\\
-U\langle T_{\tau}n_{i\tilde{\sigma}}(\tau)c_{i\sigma}(\tau) c_{j\sigma}^{\dagger}\rangle\,.
\end{multline}
or, in compact form,
\begin{equation}\label{eq:eqn_mvnt}
G_\sigma^{(0)-1}(1,\bar{3})G_\sigma(\bar{3},2)=\delta(1-2)-U\mn{ T_{\tau}n_{\tilde{\sigma}}(1)c_{\sigma}(1) c_{\sigma}^{\dagger}(2)}\,,
\end{equation}
where $\tilde{\sigma}=-\sigma$. By comparing Eqs.\eqref{eq:Dyson} and \eqref{eq:eqn_mvnt} one concludes that
\begin{equation}\label{eq:SigmaG_id}
\Sigma_\sigma(1,\bar{3})G_\sigma(\bar{3},2)=-U\mn{ T_{\tau}n_{\tilde{\sigma}}(1)c_{\sigma}(1) c_{\sigma}^{\dagger}(2)}\,.
\end{equation}

Now, comes our first approximation. We replace the last expression Eq.~\eqref{eq:SigmaG_id} by
\begin{equation}\label{eq:ansatz_SigmaG}
\Sigma_\sigma(1,\bar{1})G_\sigma(\bar{1},2)\approx Ug_{\uparrow\downarrow}(1)G_{\tilde{\sigma}}(1,1^+)G_\sigma(1,2)\,,
\end{equation}
where, for the hole-doped case,
\begin{equation}\label{eq:def_gup_down}
g_{\uparrow\downarrow}(1)=\frac{\mn{ n_\uparrow(1)n_{\downarrow}(1)} }{\mn{ n_\uparrow(1)\right\rangle \left\langle n_{\downarrow}(1)} }\,.
\end{equation}
For the electron-doped case, one replaces $n_\sigma$ in this expression by  $1-n_\sigma$. Thus, when the lattice is bipartite, particle-hole symetry of the phase diagram is preserved. It also gives a better agreement with quantum Monte Carlo results when the lattice is not bipartite. Those two different approximation are equivalent to assume that the approximation \eqref{eq:ansatz_SigmaG} is made for holes when hole density is smaller than particle density and for particles otherwise. Now, substituting the definition \eqref{eq:def_gup_down} in Eq.~\eqref{eq:ansatz_SigmaG}, it is clear that one recovers the exact result for the self-energy Eq.~\eqref{eq:SigmaG_id} in the special case $2=1^+$. The approximation~\cite{Hedeyati:1989,Vilk:1994} is justified if it is correct to assume that the four-point correlation function factorizes only when points $1$ and $2$ are different.

Within this approximation, the self-energy can be obtained by multiplying Eq.~\eqref{eq:ansatz_SigmaG} by $G_\sigma^{-1}(2,3)$ from the right and summing over $2$, to obtain
\begin{equation}\label{eq:ansatz_self}
\Sigma_\sigma^{(1)}(1,2)= Ug_{\uparrow\downarrow}(1)G_{\tilde{\sigma}}(1,1^+)\delta(1-2)\,,
\end{equation}
which is the self-energy ansatz that is used to define the Green's function $G^{(1)}$ in the Lindhard function \eqref{eq:Lindhard}, from which are defined the TPSC susceptibilities \eqref{eq:chi_sp_q} and \eqref{eq:chi_ch_q}. Note that once the source field $\phi$ is turned off and translational invariance is restored, $\Sigma_\sigma^{(1)}(1,2)$ becomes independent of position and its Fourier transform is simply a constant that can be absorbed in the chemical potential in $G_\sigma^{(1)}(k)$, so that this Green's function is in practice a non-interacting one.

Given the self-energy, as well as the result
\begin{equation}
\frac{\delta g_{\uparrow\downarrow}(1)}{{\delta G_{\sigma}(2,3)}}=\frac{\delta g_{\uparrow\downarrow}(1)}{{\delta G_{-\sigma}(2,3)}}\,,
\end{equation}
valid in the paramagnetic phase, and the definition of the spin vertex, Eq.(\ref{eq:def_gam_ch_sp}), we obtain the spin vertex%
\begin{equation}\label{eq:gam_sp}
\Gamma_{sp}(1,2;3,4)=U_{sp}\delta(1-3)\delta(1^{+}-4)\delta(1^{-}-2)
\end{equation}
where $U_{sp}=Ug_{\uparrow\downarrow}(1)$. It is possible to obtain an analytical expression for $\delta g_{\uparrow\downarrow}(1)/\delta G_{\sigma}(3,4)$ to compute the charge vertex from the definition \eqref{eq:def_gam_ch_sp} and the ansatz \eqref{eq:ansatz_self}, but to this day the most successful approach has been to approximate this functional derivative as local, i.e. proportional to $\delta(1-3)\delta(1-4)$, which leads to
\begin{equation}\label{eq:gam_ch_1}
\Gamma_{ch}(1,2;3,4)=U_{ch}\delta(1-3)\delta(1^{+}-4)\delta(1^{-}-2).
\end{equation}
Substituting in the Bethe-Salpeter equation for the susceptibilities, Eq.\eqref{eq:chi_plus}, we find the corresponding TPSC expressions for the spin \eqref{eq:chi_sp_q} and charge \eqref{eq:chi_ch_q} susceptibilities. They suffice to determine also $U_{ch}$.

\subsubsection{Second step: improved self-energy}

Collective modes are generally less influenced by details of the single-particle properties than the other way around since collective modes depend more strongly on general principles like conservation laws. We thus wish now to obtain an improved approximation for the self-energy that takes advantage of the fact that we have found accurate approximations for the low-frequency spin and charge fluctuations. We begin from the general definition of the self-energy Eq.\eqref{eq:SigmaG_id}  obtained from Dyson's equation.

We start with the longitudinal channel ($\phi$ diagonal in spin indices) and use the corresponding expression for the correlation function in terms of response function Eq.\eqref{eq:dQdphi}. In that case, the right-hand side of the general definition of the self-energy Eq.\eqref{eq:SigmaG_id} may be written as
\begin{multline}
\Sigma_{\sigma}\left(  1,\bar{1}\right)  G_{\sigma}\left(  \bar{1},2\right)  =\\
-U\left[\frac{\delta G_{\sigma}\left(  1,2\right)  }{\delta\phi_{-\sigma}\left(  1^{+},1\right)  }-G_{-\sigma}\left(  1,1^{+}\right)  G_{\sigma}\left(  1,2\right)\right]  .
\end{multline}
The last term is the Hartree-Fock contribution, which gives the exact result for the left-hand side in the limit $\omega\rightarrow\infty$.\cite{Vilk:1997} The $\delta G_{\sigma}/\delta\phi_{-\sigma}$ term is thus a contribution to lower frequencies and it comes from the spin and charge fluctuations. Right-multiplying the last equation by $G^{-1}$, replacing the lower energy part $\delta G_{\sigma}/\delta\phi_{-\sigma}$ by its general expression in terms of irreducible vertices, Eq.(\ref{eq:Bethe-Salpeter1}), and taking $\Sigma$ and $G$ on the right-hand side to be the first level approximations $\Sigma^{(1)}$ and $G^{(1)}$, respectively, we find
\begin{multline}\label{SigmaLongExact}
\Sigma_{\sigma}^{\left(  2\right)  }\left(  1,2\right)    =UG_{-\sigma}^{\left(  1\right)  }\left(  1,1^{+}\right)  \delta\left(  1-2\right)\\
-UG_{\sigma}^{\left(  1\right)  }\left(  1,\bar{3}\right)   \frac{\delta\Sigma_{\sigma}^{\left(  1\right)  }\left(  \bar{3},2\right)  }{\delta G_{\bar{\sigma}}^{\left(  1\right)}\left(  \bar{4},\bar{5}\right)  }  \frac{\delta G_{\bar{\sigma}}^{\left(  1\right)  }\left(\bar{4},\bar{5}\right)  }{\delta\phi_{-\sigma}\left(1^{+},1\right)  } \,.
\end{multline}
If we expand the sum over spins on the right-hand side to express the irreducible vertices in terms of their spin and charge versions Eqs.(\ref{eq:def_gam_ch_sp}) we find, after using the TPSC vertices, Eqs.(\ref{eq:gam_sp},\ref{eq:gam_ch_1}),
\begin{multline}\label{eq:self_long}
\Sigma_{\sigma}^{l(2)}(k)=Un_{-\sigma}\\
+\frac{U}{4}\frac{T}{N}\sum_{q}\left[U_{sp}\chi_{sp}(q)+U_{ch}\chi_{ch}(q)\right]  G_{\sigma}^{(1)}(k+q)\,.
\end{multline}

There is, however, an ambiguity in obtaining the self-energy formula. Indeed, we can obtain an expression for the self-energy by using a transverse source field $\phi_{-\sigma\sigma}$ in Eq.\eqref{eq:def_Green_phi}. By taking functional derivatives of $G$ with respect to this $\phi$, we first obtain the transverse spin correlation functions $\chi_{+-}(1,2)=\langle T_{\tau} S_+(1)S_-(2)\rangle$ and $\chi_{-+}$. Then, using a derivation analogous to that for the longitudinal case, \cite{Allen:2003} we find
\begin{equation}\label{eq:self_transv}
\Sigma_\sigma^{t(2)}(k)=Un_{-\sigma}+\frac{U}{2}\frac{T}{N}\sum_k U_{sp}\chi_{sp}(q) G_{-\sigma}^{(1)}(k+q)\,.
\end{equation}
During the derivation, $\chi_{+-}(1,2)=\chi_{-+}(1,2)=\frac{1}{2}\chi_{sp}(1,2)$ was used, taking spin rotational invariance into account.

The two previous expressions for the self-energy clearly show that our approximations for the fully reducible vertex does not preserve crossing symmetry, that is the symmetry under the exchange of two particles or two holes. To improve our approximation and restore crossing symmetry,\cite{Moukouri:2000} we average the two expressions \eqref{eq:self_long} and \eqref{eq:self_transv}, which gives the final result Eq.\eqref{eq:self_2_1} that we use in the rest of this paper. It turns out that this ``symmetric'' expression of the self energy gives a better agreement with quantum Monte Carlo results.\cite{Moukouri:2000} In addition, one verifies numerically that the exact sum rule (Appendix A in Ref. \onlinecite{Vilk:1997})
\begin{equation}
-\int \frac{d\omega ^{\prime }}{\pi }\text{Im}\,\Sigma _{\sigma}^{R}\left( \mathbf{k,}\omega ^{\prime }\right) =U^{2}n_{-\sigma }\left(1-n_{-\sigma }\right)\,,
\end{equation}
determining the high-frequency behavior of $\Sigma$, is satisfied to a higher degree of accuracy with the symmetrized self-energy expression Eq.\eqref{eq:self_2_1}.

\subsubsection{Internal accuracy checks}\label{Sec Internal accuracy}

Our final expression for the self-energy $\Sigma ^{\left( 2\right) },$ Eq.\eqref{eq:self_2_1}, is in principle an improvement over the constant self-energy entering the calculation of the susceptibilities. But clearly the approach is not one-particle self-consistent. All the Green's functions entering the right-hand side of Eq.\eqref{eq:self_2_1} are evaluated with $G^{\left( 1\right) }$, which has a constant self-energy. The advantage is that all quantities, including the vertices, on the right-hand side of Eq.\eqref{eq:self_2_1} are computed at the same level of approximation. In fact, one can miss some important physics if this is not the case \cite{Vilk:1997}.

Apart from comparisons with quantum Monte Carlo (QMC) calculations, we can check the accuracy in other ways. For example, the \textit{f}-sum rule,
\begin{equation}
\int \frac{d\omega}{\pi} \omega\chi_{ch}^{\prime\prime}(\mathbf{q},\omega)=\frac{1}{N}\sum_{\mathbf{k}\sigma}(\epsilon_{\mathbf{k}+\mathbf{q}}+\epsilon_{\mathbf{k}-\mathbf{q}}-2\epsilon_{\mathbf{k}})\mn{n_{\mathbf{k}\sigma}}\,,
\end{equation}
is exactly satisfied at the first level of approximation (i.e. with $\mn{n_{\mathbf{k}\sigma}}^{\left( 1\right) }$ on the right-hand side) and the charge susceptibility obtained with $U_{ch}$. Suppose that on the right-hand side of that equation, one uses $\mn{n_{\mathbf{k}\sigma}}$ obtained from $G^{\left( 2\right) }$ instead of the Fermi function. In cases where the agreement with QMC calculations is good, one should find that the right-hand side does not change by more than a few percent.

When we are in the Fermi liquid regime, another way to verify the accuracy of the approach is to verify if the Fermi surface obtained from $G^{\left(2\right) }$ satisfies Luttinger's theorem very closely.

Finally, the consistency relation between one- and two-particle quantities ($\Sigma$ and $\mn{n_\uparrow n_\downarrow}$), Eq. \eqref{eq:SigmaG_id}, should be satisfied exactly in the Hubbard model. In the special case where $2=1^+$, this relation can be written as
\begin{equation}\label{eq:trSigmaG_general}
\frac{1}{2}\mathrm{Tr}\left( \Sigma G\right)=\lim_{\tau\rightarrow 0^-}\frac{T}{N}\sum_k e^{-ik_n\tau} \Sigma_{\sigma}(k)G_{\sigma}(k) =U\left\langle n_{\uparrow}n_{\downarrow }\right\rangle\,.
\end{equation}
In standard many-body books \cite{Mahan:2000}, this expression is encountered in the calculation of the free energy through a coupling-constant integration. In the TPSC approach, it is not difficult to show (Appendix B of Ref. \onlinecite{Vilk:1997}) that the following equation
\begin{equation}\label{Consistency 1-2}
\frac{1}{2}\mathrm{Tr}\left( \Sigma ^{\left( 2\right) }G^{\left( 1\right)}\right) =U\left\langle n_{\uparrow }n_{\downarrow }\right\rangle
\end{equation}
is satisfied exactly with the self-consistent $U\left\langle n_{\uparrow}n_{\downarrow }\right\rangle $ obtained from the sum rule \eqref{eq:sum_rule_Usp}. An internal accuracy check consists in verifying by how much $\frac{1}{2}\mathrm{Tr}\left( \Sigma ^{\left( 2\right) }G^{\left( 2\right) }\right) $ differs from $\frac{1}{2}\mathrm{Tr}\left( \Sigma ^{\left( 2\right)}G^{\left( 1\right) }\right) .$ Again, in regimes where we have agreement with Quantum Monte Carlo calculations, the difference is only a few percent.

The above relation between $\Sigma $ and $\left\langle n_{\uparrow}n_{\downarrow }\right\rangle $ gives us another way to justify our expression for $\Sigma ^{\left( 2\right) }$, Eq.\eqref{eq:self_2_1}. Suppose one starts from \eqref{eq:self_long} to obtain a self-energy that contains only the longitudinal spin fluctuations and the charge fluctuations, as was done in the first papers on the TPSC approach\cite{Vilk:1994}. One finds that the spin part and the charge part each contribute an amount $U\left\langle n_{\uparrow }n_{\downarrow}\right\rangle /2$ to the consistency relation Eq.(\ref{Consistency 1-2}). Similarly, if we work only in the transverse spin channel \cite{Moukouri:2000,Allen:2003}, we find that each of the two transverse spin components also contributes $U\left\langle n_{\uparrow }n_{\downarrow }\right\rangle /2$ to $\frac{1}{2}\mathrm{Tr}\left( \Sigma ^{\left( 2\right) }G^{\left(1\right) }\right)$. Hence, averaging the two expressions also preserves rotational invariance since each spin component contributes equally to Eq.\eqref{Consistency 1-2}. Note that Eq.\eqref{eq:self_2_1} for $\Sigma ^{\left( 2\right) }$ is different from so-called Berk-Schrieffer type expressions \cite{Berk:1966} that contains only bare vertices and do not satisfy (Appendix E in Ref.~ \onlinecite{Vilk:1997}) the consistency condition between one- and two-particle properties, Eq.\eqref{eq:trSigmaG_general}.

\subsection{Conductivity in the Two-Particle Self-Consistent approach}\label{sec:cond_tpsc}

To compute Re~$\sigma(\omega)$ from the Kubo formula Eq.\eqref{eq:Re_sigma},  we have to obtain the \textit{current-current} correlation function given by Eq. \eqref{eq:chijj}, which contains the $\mathbf{q}=0$ component of the paramagnetic current. From the general expression for the current Eq.\eqref{eq:jxlt}, we have
\begin{equation}\label{eq:j_param_q0}
j_x^p=\sum_l j_x^p(\mathbf{r}_l)=i\sum_{ \delta \sigma} \delta_x t_{\delta} \sum_l c_{l\sigma}^{\dagger}c_{l-\delta,\sigma}\,.
\end{equation}
Since, the actual calculation is done in Matsubara frequency instead of the real-frequency definition of $\chi_{j_xj_x}$, Eq.~\eqref{eq:chijj}, we use
\begin{equation}\label{eq:chijj_Matsubara}
\chi_{j_xj_x}(iq_n)=\frac{1}{N}\int dt\, e^{iq_n(\tau-\tau')} \mn{ T_\tau \hat{j}_x^p(\tau) \hat{j}_x^p(\tau')}\,,
\end{equation}
where $\hat{j}_x^p(\tau)=e^{\tau H_0}j_x^pe^{-\tau H_0}$ and $q_n=2n\pi T$, with $n$ integer, is a bosonic Matsubara frequency. Substituting \eqref{eq:j_param_q0} in this expression gives
\begin{multline}
\mn{ T_\tau \hat{j}_x^p(\tau) \hat{j}_x^p(\tau')}=-\sum_{il \delta_1\delta_2 \sigma_1\sigma_2} \delta_{x1}\delta_{x2} t_{\delta_1}t_{\delta_2}\\  \times\mn{T_\tau c_{i\sigma}^{\dagger}(\tau)c_{i-\delta_1,\sigma}(\tau)c_{l\sigma}^{\dagger}(\tau')c_{l-\delta_2,\sigma}(\tau')}\,.
\end{multline}
If we substitute the functional derivative expression for the correlation function, Eq.\eqref{eq:dQdphi}, in this equation, we obtain
\begin{equation}
\mn{ T_\tau \hat{j}_x^p(\tau_1) \hat{j}_x^p(\tau_2)}=
\sum_{\mathbf{r}_1\mathbf{r}_2\delta_1\delta_2\sigma_1\sigma_2}\delta_{x1}\delta_{x2}\, t_{\delta 1} t_{\delta 2} \;\frac{\delta G_{\sigma_2}(2',2)}{\delta \phi_{\sigma_1}(1,1')}
\end{equation}
where have used the notation $1=(\mathbf{r}_1,\tau_1)$ and $1'=(\mathbf{r}_1-\delta_1,\tau_1)$. Note that the first term on the right-hand side of Eq.~\eqref{eq:dQdphi} does not contribute to the sum since $\mn{ j_x^p}=0$. Now, summing over spin indices and using spin rotational invariance, we obtain
\begin{equation}\label{eq:chi_jj_12}
\mn{ T_\tau \hat{j}_x^p(\tau_1) \hat{j}_x^p(\tau_2)}= -\sum_{\mathbf{r}_1\mathbf{r}_2\delta_1\delta_2}\delta_{x1}\delta_{x2}\, t_{\delta 1} t_{\delta 2}\; \chi_+(2',2 ;\, 1,1')\,,
\end{equation}
where we have used the definition Eq.~\eqref{eq:chipm_def} for the generalized susceptibility $\chi_+$.

The most general way of thinking about the last result is that it comes from a functional derivative of the current with respect to an applied vector potential representing the electric field. The remaining of the derivation, that we give in the rest of this subsection, is based on the idea that we should evaluate this functional derivative in a systematic way to obtain a conserving approximation, as shown by Baym \cite{Baym:1962}. We will keep working with the source field $\phi_{\sigma}$ but it is useful to remember that within simple prefactors, it is equivalent to working with the vector potential as the source field.

The susceptibility $\chi_+$ is defined in terms of a functional derivative in Eq.~\eqref{eq:chipm_def}. That functional derivative leads to a Bethe-Salpeter like equation \eqref{eq:Bethe-Salpeter0} that contains two different types of terms
\begin{multline}\label{eq:chi_plus_2_0}
\chi_+(1,2\vert 3,4)=-2G(1,3)G(4,2)\\
-2 G(1,\bar{1})\left(\frac{\delta \Sigma_{\sigma}(\bar{1},\bar{2})}{\delta \phi_{\sigma}(3,4)}+\frac{\delta \Sigma_{\sigma}(\bar{1},\bar{2})}{\delta \phi_{-\sigma}(3,4)}\right)G(\bar{2},2)\,.
\end{multline}
The product $GG$ comes from the explicit dependence of $G^{-1}$ on the source field while the last one comes from the implicit dependence of the self-energy on that source field. The product $GG$ is what leads to the "bubble" in standard conductivity calculations. The other term is the vertex correction.

Up to now, everything is exact in the present subsection. To work within the TPSC approach, it suffices to use everywhere the results obtained at the second level of approximation, namely $G_\sigma^{(2)}$ and $\Sigma_{\sigma}^{(2)}$ since at the first level of approximation the conductivity is infinite. Our explicit formula for $\Sigma_{\sigma}^{(2)}$ Eq.~\eqref{eq:self_2_1} is a functional of $G_\sigma^{(1)}$. We can thus write
\begin{equation}
\frac{\delta \Sigma_{\sigma}^{(2)}(\bar{1},\bar{2})}{\delta \phi_{\sigma'}(3,4)}=\frac{\delta \Sigma_{\sigma}^{(2)}(\bar{1},\bar{2})}{\delta G_{\bar{\sigma}}^{(1)}(\bar{3},\bar{4})} \frac{\delta G_{\bar{\sigma}}^{(1)}(\bar{3},\bar{4})}{\delta \phi_{\sigma'}(3,4)}\,.
\end{equation}
Expanding the sum over spin, using the chain rule, spin rotation invariance and the definition of $\chi_+$ Eq.~\eqref{eq:chipm_def} we obtain
\begin{multline}
\frac{\delta \Sigma_{\sigma}^{(2)}(\bar{1},\bar{2})}{\delta \phi_{\sigma}(3,4)}+\frac{\delta \Sigma_{\sigma}^{(2)}(\bar{1},\bar{2})}{\delta \phi_{-\sigma}(3,4)}=\\
-\frac{1}{2}\,\left(\frac{\delta \Sigma_{\sigma}^{(2)}(\bar{1},\bar{2})}{\delta G_{\sigma}^{(1)}(\bar{3},\bar{4})} + \frac{\delta \Sigma_{\sigma}^{(2)}(\bar{1},\bar{2})}{\delta G_{-\sigma}^{(1)}(\bar{3},\bar{4})}\right)\,\chi_+^{(1)}(\bar{3},\bar{4}\vert 3,4)
\end{multline}
where $\chi_+^{(1)}$ is computed with $G_\sigma^{(1)}$. Substituting in the exact expression for $\chi_+$ Eq.~\eqref{eq:chi_plus_2_0}, we find
\begin{multline}\label{eq:chi_plus_2_1}
\chi_+^{(2)}(1,2\vert 3,4)=-2G^{(2)}(1,3)G^{(2)}(4,2)\\
+G^{(2)}(1,\bar{1})G^{(2)}(\bar{2},2)\left(\frac{\delta \Sigma_{\sigma}^{(2)}(\bar{1},\bar{2})}{\delta G_{\sigma}^{(1)}(\bar{3},\bar{4})} + \frac{\delta \Sigma_{\sigma}^{(2)}(\bar{1},\bar{2})}{\delta G_{-\sigma}^{(1)}(\bar{3},\bar{4})}\right)\,\\
\times\chi_+^{(1)}(\bar{3},\bar{4}\vert 3,4)\,.
\end{multline}

If $\Sigma_{\sigma}^{(2)}$ had been a functional of $G^{(2)}$ we would have had an infinite series to sum. Instead, here the series ends as we now show. First note that, from Eq.\eqref{eq:chi_plus} and the approximation \eqref{eq:gam_ch_1} for the charge vertex,
\begin{equation}\label{eq:chi_plus_1_0}
\begin{split}
&\chi_+^{(1)}(\bar{3},\bar{4}\vert 3,4)=-2G^{(1)}(\bar{3},3)G^{(1)}(4,\bar{4})\\
&\qquad +G^{(1)}(\bar{3},\bar{5})G^{(1)}(\bar{6},\bar{4})\Gamma_{ch}^{(1)}(\bar{5},\bar{6},\bar{7},\bar{8})\chi_+^{(1)}(\bar{7},\bar{8}\vert 3,4)\\
&\qquad=-2G^{(1)}(\bar{3},3)G^{(1)}(4,\bar{4})\\
&\qquad + U_{ch}G^{(1)}(\bar{3},\bar{5})G^{(1)}(\bar{5},\bar{4})\chi_+^{(1)}(\bar{5},\bar{5}^+\vert 3,4)\,.
\end{split}
\end{equation}
From this expression, only the first term has to be included in our calculation because the second term will not contribute to the $\mathbf{q}=0$ \textit{current-current} correlation function. This is because $\Gamma_{ch}^{(1)}$ is local in space and local vertex corrections do not contribute to the uniform conductivity. To see why, one can represent the correlation function Eq.~\eqref{eq:chi_jj_12} as a series of diagrams with current vertices at their ends and vertex functions inserted between them. Whenever a vertex function in a diagram does not depend on $\mathbf{q}$, a bubble is closed with the local vertex on one end and the current vertex on the other. Since a current vertex is an odd function in space and the product of the Green's function is even because the bubble does not carry any momentum, the integral vanishes.

Inserting the first term of $\chi_+^{(1)}$, Eq.~\eqref{eq:chi_plus_1_0}, into $\chi_+^{(2)}$, Eq.~\eqref{eq:chi_plus_2_1}, we obtain
\begin{multline}\label{eq:chi_plus_2_2}
\chi_+^{(2)}(1,2\vert 3,4)=-2G^{(2)}(1,3)G^{(2)}(4,2)\\
-2G^{(2)}(1,\bar{1})G^{(2)}(\bar{2},2)\left(\frac{\delta \Sigma_{\sigma}^{(2)}(\bar{1},\bar{2})}{\delta G_{\sigma}^{(1)}(\bar{3},\bar{4})} + \frac{\delta \Sigma_{\sigma}^{(2)}(\bar{1},\bar{2})}{\delta G_{-\sigma}^{(1)}(\bar{3},\bar{4})}\right)\,\\
\times G^{(1)}(\bar{3},3)G^{(1)}(4,\bar{4})\,.
\end{multline}
The last step is in principle straightforward, but very tedious. We must obtain an expression for the functional derivatives in parenthesis in this expression. The self-energy Eq.~\eqref{eq:self_2_1} in real space is
\begin{multline}\label{eq:self_2_re}
\Sigma_\sigma^{(2)}(1,2)=UG_{-\sigma}^{(1)}(1,1^+)\delta(1-2)\\
+\frac{U}{8}\left[ 3 U_{sp}\chi_{sp}(2,1)+U_{ch}\chi_{ch}(2,1)\right] G_\sigma^{(1)}(1,2)\,.
\end{multline}
so that
\begin{multline}\label{eq:dSigma_dG}
\frac{\delta \Sigma_{\sigma}^{(2)}(1,2)}{\delta G_{\sigma'}^{(1)}(3,4)}=U\delta(1-2)\delta(1-3)\delta(1^+-4)\delta_{-\sigma,\sigma'}\\
+\frac{U}{8}\left[ 3 U_{sp}\chi_{sp}(2,1)+U_{ch}\chi_{ch}(2,1)\right]\delta(1-3)\delta(2-4)\delta_{\sigma\sigma'}\\
+\frac{U}{8}G_\sigma^{(1)}(1,2)\left[ 3 U_{sp}\frac{\delta \chi_{sp}(2,1)}{\delta G_{\sigma'}^{(1)}(3,4)}+U_{ch}\frac{\delta \chi_{ch}(2,1)}{\delta G_{\sigma'}^{(1)}(3,4)}\right]\,.
\end{multline}
In this expression, the terms involving the functional derivatives of $\Gamma_{sp}$ and $\Gamma_{ch}$ have been omitted because they do not contribute to the conductivity. Fundamentally, what is needed is $\frac{\delta \Gamma_{sp/ch}}{\delta A_x}$, where $A_x$ is the vector potential. Since $\Gamma_{sp}$ and $\Gamma_{ch}$ are local, this is the correlation function of a local operator, that is thus even under parity, and the current operator, that is odd. The $\mathbf{q}=0$ component of this correlation function that enters the \textit{current-current} correlation function, Eq.\eqref{eq:chijj_Matsubara}, therefore vanishes.

Now, we need an explicit expression for $\delta \chi_{ch/sp}(2,1)/\delta G_{\sigma'}^{(1)}(3,4)$ to know the vertex correction Eq.~\eqref{eq:dSigma_dG} completely. Let us start with $\chi_{ch}$. One can obtain an expression for this function by taking $2=1^+$ and $3=4^+$ in the expressions for $\chi_{+/-}$ Eq.~\eqref{eq:chi_plus}. But this expression has been simplified using spin rotational invariance. If we separate the spin contributions, we have
\begin{equation}\label{eq:chi_ch_2}
\begin{split}
\chi_{ch}(2,1)&=-G_\sigma(2,1^+)G_\sigma(1,2^+)-G_{-\sigma}(2,1^+)G_{-\sigma}(1,2^+)\\
&+ \frac{1}{2}U_{ch}G_\sigma(2,\bar{2}^+)G_\sigma(\bar{2},2^+)\,\chi_{ch}(\bar{2}, 1)\\
&+\frac{1}{2}U_{ch}G_{-\sigma}(2,\bar{2}^+)G_{-\sigma}(\bar{2},2^+)\,\chi_{ch}(\bar{2}, 1)
\end{split}
\end{equation}
so that
\begin{widetext}
\begin{multline}\label{eq:dchi_ch_dG_0}
\frac{\delta \chi_{ch}(2,1)}{\delta G_{\sigma'}(3,4)}=-\delta(2-3)\delta(1^+-4)G_{\sigma'}(1,2^+)-G_{\sigma'}(2,1^+)\delta(1-3)\delta(2^+-4)\\
+\frac{1}{2}U_{ch}\delta(2-3)G_{\sigma'}(4,2)\,\chi_{ch}(4, 1)+\frac{1}{2}U_{ch}\delta(2-4)G_{\sigma'}(2,3)\,\chi_{ch}(3, 1)\\
+ \frac{1}{2}U_{ch}G_\sigma(2,\bar{2}^+)G_\sigma(\bar{2},2^+)\,\frac{\delta \chi_{ch}(\bar{2}, 1)}{\delta G_{\sigma'}(3,4)}+ \frac{1}{2}U_{ch}G_{-\sigma}(2,\bar{2}^+)G_{-\sigma}(\bar{2},2^+)\,\frac{\delta \chi_{ch}(\bar{2}, 1)}{\delta G_{\sigma'}(3,4)}\,,
\end{multline}
where, again, the functional derivatives of $U_{ch}$ have been omitted, for the same reason as in \eqref{eq:dSigma_dG}. Using again spin rotational invariance, this gives
\begin{multline}\label{eq:dchi_ch_dG_1}
\frac{\delta \chi_{ch}(2,1)}{\delta G_{\sigma'}(3,4)}=-\delta(2-3)\delta(1^+-4)G(1,2^+)-G(2,1^+)\delta(1-3)\delta(2^+-4)\\
+\frac{1}{2}U_{ch}\delta(2-3)G(4,2)\,\chi_{ch}(4, 1)+\frac{1}{2}U_{ch}\delta(2-4)G(2,3)\,\chi_{ch}(3, 1)+ U_{ch}G(2,\bar{2}^+)G(\bar{2},2^+)\,\frac{\delta \chi_{ch}(\bar{2}, 1)}{\delta G_{\sigma'}(3,4)}\,.
\end{multline}
By Fourier transforming this equation with respect to $2$, we obtain an algebraic equation that is trivial to solve. Fourier transforming back the result, we obtain
\begin{multline}\label{eq:dchichdG}
\frac{\delta \chi_{ch}(2,1)}{\delta G_{\sigma'}(3,4)}=\frac{1}{1 + \frac{U_{ch}}{2} \chi_0} (2,3)\,G(4,3)\Big[ -\delta(1-4)+\frac{1}{2}U_{ch}\,\chi_{ch}(4, 1)\Big]\\
 +\frac{1}{1 + \frac{U_{ch}}{2} \chi_0} (2,4) \,G(4,3)\Big[-\delta(1-3)+\frac{1}{2}U_{ch}\,\chi_{ch}(3, 1)\Big]\,,
\end{multline}
where
\begin{equation}
\frac{1}{1 + \frac{U_{ch}}{2} \chi_0} (1,2)=\frac{T}{N}\sum_q e^{iq\cdot (1-2)}\frac{1}{1 + \frac{U_{ch}}{2} \chi_0(q)}\,.
\end{equation}
The expression for $\delta \chi_{sp}(2,1)/\delta G_{\sigma'}(3,4)$ is obtained by simply replacing $U_{ch}$ by $-U_{sp}$ and $\chi_{ch}$ by $\chi_{sp}$ in the right-hand side of expression \eqref{eq:dchichdG},
\begin{multline}\label{eq:dchispdG}
\frac{\delta \chi_{sp}(2,1)}{\delta G_{\sigma'}(3,4)}=\frac{1}{1 - \frac{U_{sp}}{2} \chi_0} (2,3)\,G(4,3)\Big[ -\delta(1-4)-\frac{1}{2}U_{sp}\,\chi_{sp}(4, 1)\Big]\\
 +\frac{1}{1 - \frac{U_{sp}}{2} \chi_0} (2,4)\,G(4,3) \Big[-\delta(1-3)-\frac{1}{2}U_{sp}\,\chi_{sp}(3, 1)\Big]\,.
\end{multline}
Substituting Eqs. \eqref{eq:dchichdG} and \eqref{eq:dchispdG} in our expression for the vertex, Eq.~\eqref{eq:dSigma_dG}, we obtain
\begin{small}
\begin{multline}
\frac{\delta \Sigma_{\sigma}^{(2)}(1,2)}{\delta G_{\sigma'}^{(1)}(3,4)}=U\delta(1-2)\delta(1-3)\delta(1^+-4)\delta_{-\sigma,\sigma'}
+\frac{U}{8}\left[ 3 U_{sp}\chi_{sp}(2,1)+U_{ch}\chi_{ch}(2,1)\right]\delta(1-3)\delta(2-4)\delta_{\sigma\sigma'}\\
-\frac{U}{8}G^{(1)}(1,2)G^{(1)}(4,3)\left[ 3 U_{sp}\left(\frac{1}{1 - \frac{U_{sp}}{2} \chi_0} (2,3)\Big[ \delta(1-4)+\frac{1}{2}U_{sp}\,\chi_{sp}(4, 1)\Big]\right.+\frac{1}{1 - \frac{U_{sp}}{2} \chi_0} (2,4)\, \Big[\delta(1-3)+\frac{1}{2}U_{sp}\,\chi_{sp}(3, 1)\Big]\right)\\
+ \left.U_{ch}\left(\frac{1}{1 + \frac{U_{ch}}{2} \chi_0} (2,3)\,\Big[ \delta(1-4)-\frac{1}{2}U_{ch}\,\chi_{ch}(4, 1)\Big] +\frac{1}{1 + \frac{U_{ch}}{2} \chi_0} (2,4) \,\Big[\delta(1-3)-\frac{1}{2}U_{ch}\,\chi_{ch}(3, 1)\Big]\right)\right]
\end{multline}
\end{small}
and therefore,
\begin{small}
\begin{multline}\label{eq:dSigma2dG1_1}
\frac{\delta \Sigma_{\sigma}^{(2)}(\bar{1},\bar{2})}{\delta G_{\sigma}^{(1)}(\bar{3},\bar{4})} + \frac{\delta \Sigma_{\sigma}^{(2)}(\bar{1},\bar{2})}{\delta G_{-\sigma}^{(1)}(\bar{3},\bar{4})}=
U\delta(\bar{1}-\bar{2})\delta(\bar{1}-\bar{3})\delta(\bar{1}^+-\bar{4})+\frac{U}{8}\left[ 3 U_{sp}\chi_{sp}(\bar{2},\bar{1})+U_{ch}\chi_{ch}(\bar{2},\bar{1})\right]\delta(\bar{1}-\bar{3})\delta(\bar{2}-\bar{4})\\
-\frac{U}{4}G^{(1)}(\bar{1},\bar{2})G^{(1)}(\bar{4},\bar{3})\left[ 3 U_{sp}\left(\frac{1}{1 - \frac{U_{sp}}{2} \chi_0} (\bar{2},\bar{3})\,\Big[ \delta(\bar{1}-\bar{4})+\frac{1}{2}U_{sp}\,\chi_{sp}(\bar{4}, \bar{1})\Big]\right. +\frac{1}{1 - \frac{U_{sp}}{2} \chi_0} (\bar{2},\bar{4})\, \Big[\delta(\bar{1}-\bar{3})+\frac{1}{2}U_{sp}\,\chi_{sp}(\bar{3}, \bar{1})\Big]\right)\\
+ U_{ch}\left(\frac{1}{1 + \frac{U_{ch}}{2} \chi_0} (\bar{2},\bar{3})\,\Big[ \delta(\bar{1}-\bar{4})-\frac{1}{2}U_{ch}\,\chi_{ch}(\bar{4}, \bar{1})\Big]\left. +\frac{1}{1 + \frac{U_{ch}}{2} \chi_0} (\bar{2},\bar{4}) \,\Big[\delta(\bar{1}-\bar{3})-\frac{1}{2}U_{ch}\,\chi_{ch}(\bar{3}, \bar{1})\Big]\right)\right]\,.
\end{multline}
\end{small}

We now have everything to write down an explicit form for the susceptibility $\chi_+^{(2)}$ in \eqref{eq:chi_plus_2_2},
\begin{multline}\label{eq:chi_plus_2}
\chi_+^{(2)}(1,2\vert 3,4)=-2G^{(2)}(1,3)G^{(2)}(4,2)
-2UG^{(2)}(1,\bar{1})G^{(2)}(\bar{1},2)G^{(1)}(\bar{1},3)G^{(1)}(4,\bar{1}^+)\\
-\frac{U}{4}G^{(2)}(1,\bar{1})G^{(2)}(\bar{2},2)\left[ 3 U_{sp}\chi_{sp}(\bar{2},\bar{1})+U_{ch}\chi_{ch}(\bar{2},\bar{1})\right]G^{(1)}(\bar{1},3)G^{(1)}(4,\bar{2})\\
-\frac{U}{2}G^{(2)}(1,\bar{1})G^{(2)}(\bar{2},2)G^{(1)}(\bar{1},\bar{2})G^{(1)}(\bar{4},\bar{3})G^{(1)}(\bar{3},3)G^{(1)}(4,\bar{4})\\
\times\Bigg[ 3 U_{sp}\Bigg(\frac{1}{1 - \frac{U_{sp}}{2} \chi_0} (\bar{2},\bar{3})\,\Big[ -\delta(\bar{1}-\bar{4})-\frac{1}{2}U_{sp}\,\chi_{sp}(\bar{4}, \bar{1})\Big]
+\frac{1}{1 - \frac{U_{sp}}{2} \chi_0} (\bar{2},\bar{4})\, \Big[-\delta(\bar{1}-\bar{3})-\frac{1}{2}U_{sp}\,\chi_{sp}(\bar{3}, \bar{1})\Big]\Bigg)\\
+ U_{ch}\Bigg(\frac{1}{1 + \frac{U_{ch}}{2} \chi_0} (\bar{2},\bar{3})\,\Big[ -\delta(\bar{1}-\bar{4})+\frac{1}{2}U_{ch}\,\chi_{ch}(\bar{4}, \bar{1})\Big]
+\frac{1}{1 + \frac{U_{ch}}{2} \chi_0} (\bar{2},\bar{4}) \,\Big[-\delta(\bar{1}-\bar{3})+\frac{1}{2}U_{ch}\,\chi_{ch}(\bar{3}, \bar{1})\Big]\Bigg)\Bigg]\,.
\end{multline}
To evaluate the current-current correlation function entering the conductivity, it is clearly necessary to go to Fourier space. Inserting then this last result in Eq.~\eqref{eq:chi_jj_12} for the current-current correlation function and using
\begin{equation}
\delta_x t_\delta=\frac{i}{N}\sum_{\mathbf{k}}e^{i\mathbf{k}\cdot\delta}\frac{\partial \epsilon_\mathbf{k}}{\partial k_x}
\end{equation}
and
\begin{equation}
1+\frac{U_{sp}}{2}\,\chi_{sp}(q)=\frac{1}{1 - \frac{U_{sp}}{2} \chi_0(q)}\,,\qquad 1-\frac{U_{ch}}{2}\,\chi_{ch}(q)=\frac{1}{1 + \frac{U_{ch}}{2} \chi_0(q)}\,,
\end{equation}
we finally obtain for the Fourier-Matsubara transformed expression at $\mathbf{q}=0$,
\begin{multline}\label{eq:chi_jj_qn1}
\chi_{j_xj_x}(iq_n)
=\frac{-2T}{N}\sum_k \left(\frac{\partial \epsilon_\mathbf{k}}{\partial k_x}(k)\right)^2 G^{(2)}(k)G^{(2)}(k+iq_n)\\
-\frac{U}{4}\left(\frac{T}{N}\right)^2\sum_{k_1k_2} G^{(2)}(k_1)G^{(2)}(k_1+iq_n)G^{(1)}(k_2)G^{(1)}(k_2+iq_n)\frac{\partial \epsilon_\mathbf{k}}{\partial k_x}(k_1)\frac{\partial \epsilon_\mathbf{k}}{\partial k_x}(k_2) \left[3U_{sp}\chi_{sp}(k_2-k_1)+U_{ch}\chi_{ch}(k_2-k_1)\right]\\
+\frac{U}{2}\left(\frac{T}{N}\right)^3\sum_{k_1,k_2,q_1}\frac{\partial \epsilon_\mathbf{k}}{\partial k_x}(k_1)\frac{\partial \epsilon_\mathbf{k}}{\partial k_x}(k_2)G^{(2)}(k_1)G^{(2)}(k_1+iq_n)  G^{(1)}(k_2)G^{(1)}(k_2+iq_n)\left[G^{(1)}(k_2+q_1+iq_n)+G^{(1)}(k_2-q_1)\right]\\
\times G^{(1)}(k_1+q_1+iq_n)\Bigg(3 U_{sp}\frac{1}{1 - \frac{U_{sp}}{2} \chi_0(q_1)} \,\frac{1}{1 - \frac{U_{sp}}{2} \chi_0(q_1+iq_n)}+U_{ch}\frac{1}{1 + \frac{U_{ch}}{2} \chi_0(q_1)} \,\frac{1}{1 + \frac{U_{ch}}{2} \chi_0(q_1+iq_n)}\Bigg)
\end{multline}
\end{widetext}
In this expression, we use the compact notation $k+iq_n=(\mathbf{k},ik_m+iq_n)$. Note that the second term on the right-hand side of Eq.~\eqref{eq:chi_plus_2} does not contribute to $\chi_{j_xj_x}$ because it has a local vertex.

\begin{figure}[h]
\includegraphics[width=0.45\textwidth]{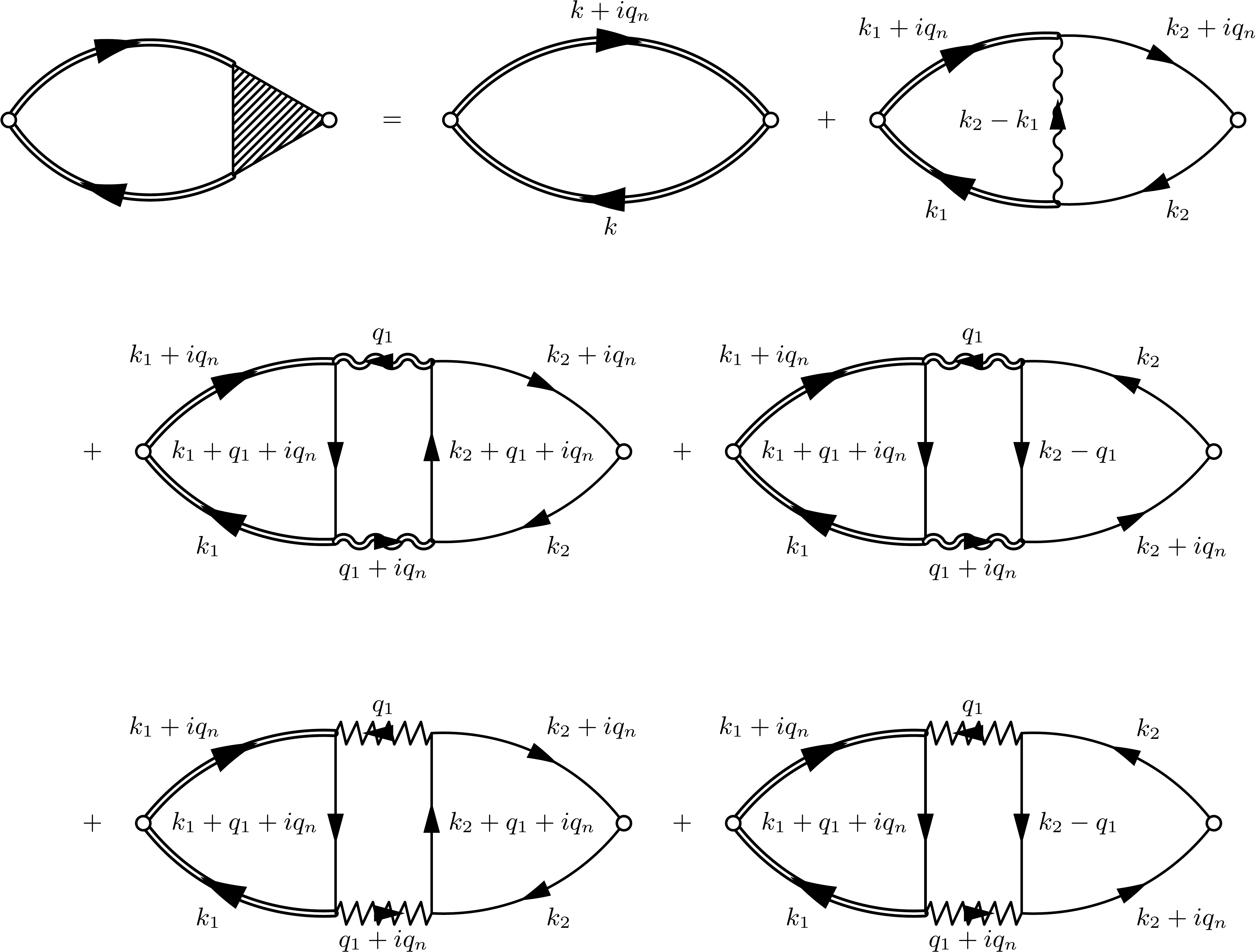}
\caption{\label{fig:chijj} Schematic representation of the various terms in the \textit{current-current} correlation function Eq.\eqref{eq:chi_jj_qn1}. Simple fermion lines represent the Green's function $G^{(1)}$ (in practice it has the same form as the bare Green's function), double fermion lines are dressed Green's function $G^{(2)}$, the simple wiggly line in the last diagram on the first line is the function $[3U_{sp}\chi_{sp}+U_{ch}\chi_{ch}]$, the double wiggly lines in the third and fourth diagrams on the right-hand side are the function $1/(1-U_{sp}\chi_0/2)$ and the zigzag lines in the last two diagrams are the function $1/(1+U_{ch}\chi_0/2)$. Those diagrams serve only as a graphical representation of Eq.\eqref{eq:chi_jj_qn1}, they are not obtained in a perturbative way.}
\end{figure}
Eq.~\eqref{eq:chi_jj_qn1} is our final result for the current-current correlation function in Matsubara space, including both bubble and vertex corrections. It is useful for the intuition and to verify the Fourier transforms to represent it schematically, as we have done in Fig. \ref{fig:chijj}. We stress that this representation is not the result of a perturbative calculation. It exists merely because we are working with Green's functions and response functions. This representation helps noticing that an analogy can be made between the different contributions to $\chi_{j_xj_x}$ and the diagrams considered in the theory of paraconductivity.\cite{Larkin:2001} In the latter theory, one considers the effect of the superconducting fluctuations on the conductivity of the normal state, while in our case, the bosons exchanged are spin and charge fluctuations. The diagram with a single boson propagator on the first line of  Fig.\ref{fig:chijj} is the analog of the Maki-Thompson (MT)  contribution,\cite{Maki:1968:1,Maki:1968:2,Maki:1969,Thompson:1970} while the diagrams on the second and third lines are analogs of the Aslamasov-Larkin (AL) contributions.\cite{Aslamasov:1968} However, it is important to note that, because our approach is not perturbative, some electron propagators are at the second level of approximation and some are at the first level, and the boson propagators are the susceptibilities computed with the renormalized spin and charge irreducible vertices.

In the system considered in this paper, the relevant collectives modes are magnetic fluctuations with a wave vector close to $\mathbf{Q}=(\pi,\pi)$. First, when those fluctuations become strong, they scatter quasiparticles, which has a dramatic effect on the single particle spectrum and therefore the conductivity obtained from the bubble alone.  But the magnetic fluctuations also lead to important vertex corrections because they correlate the regions of the Fermi surface that are connected by wave vectors close to $\mathbf{Q}$. For the MT diagram in Fig. \ref{fig:chijj}, in the DC limit, the exchange of such a fluctuation between the particle and the hole created by the field causes the pair to be scattered from a region of the Fermi surface where it carries a positive current along the direction of the field, to a region where it carries a negative current. This correlation between currents flowing in oposite directions degrades the DC conductivity. In the case of the AL diagrams in Fig. \ref{fig:chijj}, a particle-hole pair creates another one via two magnetic fluctuations. If they have a large correlation length, this will tend to correlate currents on large distances. However, the particle-hole pair created by the fluctuations can carry a current that is either positive or negative. Keeping in mind that, when quasiparticles create or absorb magnetic fluctuations, their velocity along the direction of the field changes sign, one finds that the first diagram in the second line of  Fig. \ref{fig:chijj} correlates currents flowing in the same direction, while the second diagram correlates currents in opposite directions. In addition, the former contribution is large if the single particle spectral density is large below the Fermi level in regions connected by wave vectors around $\mathbf{Q}$, while the latter is large if the spectral density is large above the Fermi level. When both processes are summed up, there will therefore be a net effect on the conductivity only if there is an asymmetry in the relevant parts of the density of states.

\subsection{Calculation algorithms}\label{sec:algo}

Let us recall that we have used the short-hand notation $k=(\mathbf{k},ik_n)$, with $\mathbf{k}=(k_x,k_y)$, in the general expression for the current-current correlation function Eq.~\eqref{eq:chi_jj_qn1}. Thus, each sum over a tri-vector $k$ is over the two-dimensional Brillouin zone and over Matsubara frequencies. The second and third sums are therefore six- and nine-dimensional, respectively. The six-dimensional sum would be extremely long to do in a direct way for relevant system sizes and temperatures. For example, for a finite system of $512\times 512$ sites with about 4000 frequencies, which would allow to go down to temperatures about $T=0.01t$ (without finite size effects), it would take of the order of thirty years to compute one frequency $iq_n$, if $10^9$ terms are summed per second. Of course, this is using pure brute force, when all wave vectors and frequencies are summed, which is not necessary in practice. In the case of the nine-dimensional sum, it would take about 40 billion years to compute a single frequency in the same conditions. If one was to keep the direct summation approach, but optimizing the procedure by using a very efficient adaptive scheme, it would still be extremely hard to do the six-dimensional sum to obtain, for example, 100 frequencies in a reasonnable time, namely, of the order of a few days. In the case of the nine-dimensional sum, it is obviously impossible to do this way.

One thus has to resort to a numerical approach completely different from direct summation to succeed in calculating all the terms of Eq.\eqref{eq:chi_jj_qn1} for a useful number of frequencies $iq_n$. The main tool that we use to make this calculation possible is the \textit{fast Fourier transform} (FFT), which changes the scaling of Fourier transforms (FTs) from $N^2$ to $N\log N$. But other mathematical and numerical tricks are also necessary to make the calculation both fast and precise. Precision is critical here since we have to numerically perform the analytical continuation of the computed Matsubara current-current correlation function and numerical analytical continuation is inherently an ill-conditionned problem. For this analytical continuation procedure, that produces the real frequency optical conductivity from the Matsubara current-current correlation function, a maximum entropy algorithm is also developped to maximize the accuracy of the result. The calculation algorithms for $\chi_{j_xj_x}(iq_n)$ and our analytical continuation algorithm are summarized in the next two subsections and detailed in appendices \ref{sec:FFT_splines_HF_exp} to \ref{sec:analytical_cont_app}.

\subsubsection{Fast Fourier transforms, cubic splines and asymptotic expansions}

The key property that allows us to compute Eq.\eqref{eq:chi_jj_qn1} is that some of its sums are convolutions. Since the convolution of two functions can be written as the Fourier transform of the product of the of their Fourier transforms, we can use fast Fourier transforms (FFT) to do those convolutions in a very efficient way. However, FFTs are discrete transform, while some of the transforms we have to do are continuous ones. In the case of the spatial Fourier transforms, we can use a finite system with periodic boundary conditions, so that all the transforms are discrete and FFTs can be used directly. Since we work at finite temperature in two dimensions, all correlation lengths are finite and we can use a system large enough to reach interesting regimes. The system size we use in this work is $512\times 512$ and the lowest temperature we reach is $0.008t$, which corresponds to $32K$ if $t=0.35eV$. At this temperature, the thermal De Broglie wavelength is about $100$ in lattice units, so that no finite size effect are seen unless the magnetic correlation length becomes large.

In the case of imaginary time, if it is discretized, then the Fourier transform will be periodic in Matsubara frequencies, which is unphysical. In fact, the lowest frequencies will be acceptable, though not very precise, but the precision will decrease rapidly with increasing frequency, so that the high frequencies will be completely wrong. To overcome this problem, we use a cubic spline to interpolate the function between the discrete imaginary time points and we do a continuous Fourier transform on the spline. This technique has already been used in the context of dynamical mean field theory calculations.\cite{Oudovenko:2002} However, to our knowledge, it has not been pointed out that, because the spline is only twice continuously differentiable, after integrating by parts three times the Fourier integral, we are left with an expression containing a discrete Fourier transform (DFT) that can be done with a FFT. Using the more general formula derived in appendix \ref{sec:FT_spline}, the final formula for the Fourier transform of the cubic spline of an imaginary time function $g(\tau)$ is
\begin{equation}\label{eq:TF_spline_tau}
\begin{split}
g(i\omega_n)&=\int_{0}^{\beta} d\tau \,g(\tau) e^{i\omega_n\tau}\\
&\approx \sum_{j=1}^N \int_{\tau_{j-1}}^{\tau_j} d\tau \,S_j(\tau) e^{i\omega_n\tau}\\
&=-\frac{g(0)-e^{i\omega_n\beta}g(\beta)}{i\omega_n}+\frac{S_1'(0)-e^{i\omega_n\beta}S_N'(\beta)}{(i\omega_n)^2}\\
&\qquad-\frac{S_1''(0)-e^{i\omega_n\beta}S_N''(\beta)}{(i\omega_n)^3}\\
&\qquad\qquad+\frac{1-e^{i\omega_n\Delta \tau}}{(i\omega_n)^4}\sum_{j=0}^{N-1} S_{j+1}^{(3)} e^{i\omega_n\tau_{j}}\,,
\end{split}
\end{equation}
where $N$ is the number of intervals in the discretized imaginary time between $0$ and $\beta$, $\Delta \tau$ is the size of an interval, $S_j(\tau)$ is the cubic polynomial in the $j^{th}$ interval, $S_j'(\tau)$, $S_j''(\tau)$ and $S_{j}^{(3)}$ are, respectively, the first, second and the third derivative of $S_j(\tau)$. If $g$ is a fermionic function, $e^{i\omega_n\beta}=-1$, while $e^{i\omega_n\beta}=1$ if it is bosonic.

When using formula \eqref{eq:TF_spline_tau}, we work explicitly with the high frequency expansion up to the $1/(i\omega_n)^3$ term, which can be very useful, as will be explained shortly. Up to which term this expansion will be exact depends on how the spline is defined. As shown in appendix \ref{sec:FT_spline}, there are two conditions defining the spline that can be chosen depending on the information available. If the derivatives at the boundaries $g'(0)$ and $g'(\beta)$ are known, one can fix $S_1'(0)$ and $S_N'(\beta)$ to those values. That is what we do in our calculations. Otherwise, one has to use exact results for the second and third high frequency coefficients of $g$, i.e., the second and third moments of its spectral function. Another important point about Eq.\eqref{eq:TF_spline_tau} is that the last term containing the DFT only contributes to the low frequencies because of its factor $1/(i\omega_n)^4$, therefore the error at high frequency that comes from the discretization of $\tau$ rapidly vanishes.

Given the above considerations, it should now be clear that we can compute convolutions both in wave vector and Matsubara frequency using fast Fourier transforms. Before the calculation of $\chi_{j_xj_x}$, Eq.\eqref{eq:chi_jj_qn1}, this technique is used to compute the Lindhard function Eq.\eqref{eq:Lindhard} and then, the self-energy Eq.\eqref{eq:self_2_1}. The computation of those functions with FFT is relatively straightforward to implement, although some care must be taken in the definition of the splines. Those details are given in appendices \ref{sec:Lindhard} and \ref{sec:self-energy}. In the case of Eq.\eqref{eq:chi_jj_qn1}, we only seek the $\mathbf{q}=0$ component of $\chi_{j_xj_x}(\mathbf{q},iq_n)$ so that, in each term, the last sum over the Brillouin zone to be performed is a simple sum and FFTs are not useful there. Regarding the other sums  it is not obvious which one can be put in the form of a convolution, except for the bubble which has the same form as the
Lindhard function. In the case of the second term of Eq.\eqref{eq:chi_jj_qn1}, represented by the second diagram (Maki-Thompson) on the right-hand side of Fig.\ref{fig:chijj}, it is possible to put all those sums in the form of convolution, and thus to compute all of them with FFTs. This means that, all the external frequencies $iq_n$ can be obtained at once, when the last FFT is performed. The scaling of this calculation with system size and inverse temperature is thus the same as for the bubble. That is not the case however of the last term of Eq.\eqref{eq:chi_jj_qn1}, represented by the last four (Aslamasov-Larkin) diagrams in Fig.\ref{fig:chijj}. For this term, each frequency has to be calculated separately and it is the calculation time of only one frequency that has the same scaling as the bubble. Calculating this term for 100 values of $iq_n$ therefore takes hundreds of times the calculation time of all frequencies of the bubble. The calculation of the bubble is nevertheless very quick, of the order of a few minutes at most for our $512\times 512$ system at low temperature.

Note however that a quite large amount of analytical work is needed to put the last two terms of Eq.\eqref{eq:chi_jj_qn1} in a form suitable for computation with FFTs. This work involves some transformations and a certain number of sums over Matsubara frequencies must be performed exactly. The details of those transformations and analytical calculations are given in appendix \ref{sec:chijj}.

Still, it is impossible to calculate the third term of Eq.\eqref{eq:chi_jj_qn1} for thousands of frequencies, namely the number of frequencies below the cutoff used in the calculation. Therefore, to reduce the number of frequencies to be calculated, we have used a non-uniform Matsubara frequency grid in which the frequency spacing increases with frequency magnitude. We give the definition of this grid in appendix \ref{eq:grid_Matsubara}. We have verified that the function $\chi_{j_xj_x}(iq_n)$ evaluated on this grid has enough information for the analytically continued conductivity Re~$\sigma(\omega)$ to be converged.

Using the formula \eqref{eq:TF_spline_tau} for our functions in Matsubara frequencies has the advantage that we always have at our disposal their high frequency expansion. Thus if we want the inverse Fourier transform (from Matsubara frequencies to imaginary time) of a more complex function, for example the product of two functions of the form \eqref{eq:TF_spline_tau}, we can always use the asymptotic form to perform the sum over Matsubara frequencies up to infinity using contour integrals in the complex plane and the residue theorem. In a lot of cases, this is necessary since the sum does not converge otherwise. As will be explained in the next subsection, precision is important in our calculations since we have to extract Re~$\sigma(\omega)$ from the Matsubara function $\chi_{j_xj_x}(iq_n)$ and analytical continuation is a very ill-conditioned problem.

Finally, as one will notice from appendix \ref{sec:chijj}, the calculation of Eq.\eqref{eq:chi_jj_qn1}, especially the second and third terms, contains many steps. The debugging part of the work is therefore considerable. To make sure the formula \eqref{eq:chi_jj_qn1} is implemented correctly, we have compared its brute force implementation, term by term, with its fast one given in appendix  \ref{sec:chijj}. Here, brute force means that it is coded exactly as written in \eqref{eq:chi_jj_qn1}, with one exception that will be explained shortly. Therefore, we compare the results of two calculations that are completely different numerically, but mathematically equivalent. Of course, those verifications can only be done for very small system at very high temperature, since otherwise the brute force calculation is impossible to do in a reasonable time. But even then, the third term of \eqref{eq:chi_jj_qn1} takes too much time to compute with brute force. In that case, the procedure is to first verify that the sum over $k_2$ for some random values of $q_n$ and $q_1$ gives the same result with the two implementations and then to replace this sum in the brute force version by its fast implementation. After that step we compare this modified brute force implementation with the fast implementation of the whole third term.

\subsubsection{Analytical continuation}

Expression \eqref{eq:chi_jj_qn1} gives the \textit{current-current} correlation function in Matsubara frequencies while we need it in real frequency to compute the conductivity from Eq.\eqref{eq:def_cond} or Eq.\eqref{eq:Re_sigma}. Thus we need a reliable analytical continuation method. To do so, we use a maximum entropy approach. This kind of analytical continuation procedure is often used to extract real frequency results from imaginary time quantum Monte Carlo data using some information known \textit{a priori} such as sum rules and a default model which contains some known properties of the expected function\cite{Jarrell1996133}. This information is included in the algorithm in the form of constraints or in the entropy definition. This kind of approach is well suited for quantum Monte Carlo since the results are usually in discretized imaginary time and the amount of noise can be important, so that any additional information is welcome. In our case, the original data is in Matsubara frequencies and the noise level is much lower since it comes only from finite precision rounding errors accumulating throughout the calculation. However, our calculation of $\chi_{j_xj_x}$ contains many steps so that the final result may have an amount of noise such that Pad\'e approximants will not work. Those approximants are very sensitive to noise \cite{Beach_Gooding_2000} and their reliability is not very good except at very low temperature. In any case, we have noticed that they give unstable results as a function of temperature for our data while our maximum entropy procedure is in general stable.

We give here the main features of our approach. The details are given in appendix \ref{sec:analytical_cont_app}. As usual we minimize the function $\chi^2-\alpha S$, where $S$ is the entropy, $\alpha$, a weighting parameter for $S$ while $\chi^2$ is the quadratic error between $\chi_{j_xj_x}(iq_n)$ and the quantity $\bar{\chi}_{jj}(iq_n)$ computed from the spectral representation of $\chi_{j_xj_x}(iq_n)$ with a trial real frequency conductivity Re~$\sigma(\omega)$. First, an accurate numerical integration scheme has to be chosen to compute $\bar{\chi}_{jj}(iq_n)$ from Re~$\sigma(\omega)$ known on a fixed discrete grid in real frequency $\omega$. We use a cubic spline to approximate Re~$\sigma(\omega)$. Since we work in Matsubara frequencies instead of imaginary time, the spectral form is simple and can be integrated analytically if Re~$\sigma(\omega)$ is approximated by a piecewise cubic polynomial function. Also, because we want to integrate Re~$\sigma(\omega)$ in the interval $[0,\infty $, we integrate the low frequency part with respect to $\omega$ and the high frequency part with respect to $1/\omega$. To do so, we use a spline cubic in $\omega$ for low frequencies and cubic in $1/\omega$ for high frequencies. As for the choice of grid, we take it to be uniform in $\omega$  for the low frequency part and uniform in $1/\omega$ for high frequencies. This choice of grid ensures that the matrix for the spline linear system is well conditioned, and keeps the number of values Re~$\sigma(\omega_j)$ reasonable, so that the minimization procedure is not too heavy. The integration using the spline turns out to be very accurate compared to a simple piecewise linear approximation. In our tests with a well defined analytical form for Re~$\sigma(\omega)$ for which $\chi_{j_xj_x}(iq_n)$ could be computed very accurately with an adaptive integration routine, the relative precision was typically five orders of magnitude smaller with the spline than the piecewise linear approximation. The reason why we have to use a very accurate integration method is that the relative difference between $\bar{\chi}_{jj}(iq_n)$ and $\chi_{j_xj_x}(iq_n)$ has to be very small, typically $<10^{-5}$ in our calculation, for Re~$\sigma(\omega)$ to be converged in the optimization procedure. The precision of the numerical integral for a given Re~$\sigma(\omega)$ has clearly to be smaller than this required relative error on $\bar{\chi}_{jj}(iq_n)$ for the result to be reliable.

The fact that the spline is integrated analytically has the great advantage that low temperatures are not more difficult to handle, while it is the case with standard numerical integration because the integrand becomes sharper as temperature decreases. Note that all those precision issues are important if one is interested in quantitative results. For instance, we want to obtain the resistivity as a function of temperature, but since the results at different temperatures are numerically completely independent, the quantitative aspect becomes crucial. If one is only interested in the shape of Re~$\sigma(\omega)$ at a given temperature, simpler and cruder approximations can be sufficient.

\section{Numerical results}\label{sec:res}

Before we show the results of this work, let us recall some important former TPSC results. First, the theory respects the Mermin-Wagner theorem, so that no phase transition occur at finite temperature. However, with proper values of $U$ and hopping parameters, antiferromagnetic correlations are present up to very high temperatures around half-filling. For example, with $U=6t$ and nearest neighbor hopping only, for dopings smaller than $p_c=0.205$ a crossover to a renormalized classical regime appears. This regime appears when $k_BT\gg \hbar\omega_{sp}$, where $\omega_{sp}$ is the characteristic frequency of the antiferromagnetic fluctuations, i.e., the frequency at which the imaginary part of the spin correlation function $\chi_{sp}''(\omega)$ is maximum. In this regime, the antiferromagnetic correlation length has the form $\xi_{sp}\propto\exp(C/T)$, where $C$ has a very weak temperature dependence. Therefore, at a certain temperature $T^*$, $\xi_{sp}$ becomes larger than the single-particle thermal De Broglie wavelength $\xi_{th}=\hbar v_F/(\pi k_BT)$. When this happens, the parts of the Fermi surface that are connected by the antiferromagnetic wave vector, called the hot spots, are strongly scattered by the magnetic fluctuations and eventually destroyed, producing a gap in those regions of the Brillouin zone.\cite{Vilk:1996,Vilk:1997} However, before the correlation length becomes infinite, which is the case only at $T=0$ in two dimensions, there is still spectral weight at the Fermi level and thus no real gap exits, but what is observed instead is a pseudogap, namely, a depression of the density of states at the Fermi level. Therefore, the crossover temperature to the renormalized classical regime $T^*$ can also be called a pseudogap temperature. When $T=0$, long-range spin-density wave (SDW) order exists for $p<p_c$ and thus $p_c$ is an quantum critical point (QCP). Depending on band parameters and doping, this SDW state can be commensurate or incommensurate. Usually, it is commensurate close to half-filling
 and a transition to incommensurate appears at a certain doping.\cite{Roy:2008}

Benchmarks of TPSC calculations were made against quantum Monte Carlo (QMC) results for quantities such as the spin- an charge structure factors,\cite{Vilk:1994} the quasiparticle renormalization factor and the imaginary time Green's function,\cite{Vilk:1996} the finite frequency spin susceptibility, the double occupancy and the one-particle occupation number,\cite{Vilk:1997} and finally, the one-particle the spectral weight \cite{Moukouri:2000}. Those benchmarks were made in the weak to intermediate coupling regime for a large range of doping around half-filling and for temperatures where no finite size effect are seen in the QMC results and, at finite doping, when the sign problem is not too strong. In general, a good quantitative agreement is obtained for all quantities at a coupling $U=4t$. For some quantities such as the spin structure factor, the agreement is almost perfect above $T^*$ for couplings up to $U=8t$. Since the TPSC approach has some mean field aspects\cite{Dare:1996} coming from the ansatz Eq.\eqref{eq:ansatz_self}, it slightly overestimates $T^*$ (see Fig. 7 of ref. \onlinecite{Vilk:1997}), but the qualitative behavior just below $T^*$, when the spin correlation length grows exponentially, is still very well reproduced. The spectral weight at half-filing in this regime is also very well reproduced by TPSC calculations.\cite{Moukouri:2000} This is the regime where precursors of the antiferromagnetic bands are formed and a pseudogap appears in the spectral weight.

The results we present in this section are for the one-band Hubbard model with nearest-neighbor hopping only. All numerical examples are for $U=6t$ and various dopings and temperature. We begin by showing the accuracy with which the \textit{f}-sum rule is satisfied. We then give a few typical examples of the frequency dependent conductivity. The last subsection will focus on the temperature dependent resistivity for various dopings.

\subsection{f-sum-rule}\label{sec:f-sum-rule_res}

\begin{figure}
\includegraphics[width=0.4\textwidth]{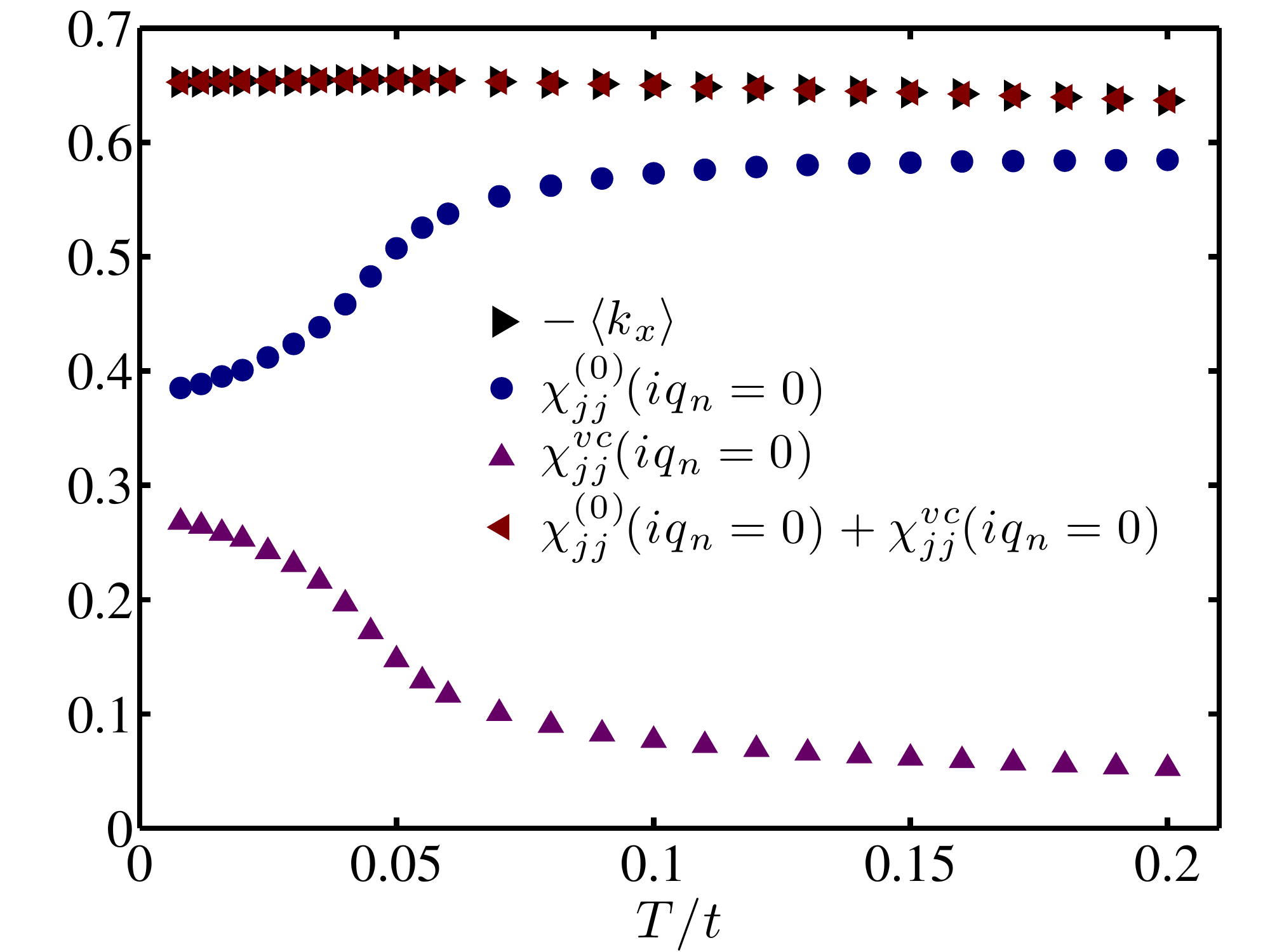}
\caption{\label{fig:f-sum-rule} Contributions to the zero Matsubara frequency value of the \textit{current-current} correlation function compared to the sum-rule value $-\mn{k_x}$ for $p=0.17$.}
\end{figure}

Although our expression for the conductivity Eq. \eqref{eq:chi_jj_qn1} was obtained from functional derivative methods that lead to results that satisfy conservation laws, \cite{Baym:1962} usually this method is applied to perturbative one-particle self-consistent schemes. In the TPSC equations, all the functional dependence on vector potential is in $G^{(1)}$. One may question whether this preserves conservation laws. The full Ward identity is derived in Appendix~\ref{sec:Ward} where one also finds comments on why it cannot be used to find the vertex corrections in the limiting case we are interested in. Given the difficulty of computing the current-current correlation function alone for only the wave vector $\mathbf{q}=0$, it will be clear why the full Ward identity cannot be verified. As a test of particle conservation, we focus instead on how accurately the \textit{f}-sum rule, Eq.\eqref{eq:sum_rule_chijj_Matsubara} derived in Appendix \ref{sec:Appendix_f_sum_rule},
\begin{equation}\label{eq:f_sum_rule}
\chi_{j_xj_x}(iq_n=0)=\frac{1}{N}\sum_{\mathbf{k}\sigma}\frac{\partial^2 \epsilon_{\mathbf{k}}}{\partial k_x^2}\mn{n_{\mathbf{k}\sigma}}\,,
\end{equation}
is satisfied numerically. In this equation, the occupation probability $\mn{n_{\mathbf{k}\sigma}}$ is computed with $G^{(2)}$, while the left-hand side is the zero-Matsubara frequency \textit{current-current} correlation function $\chi_{j_xj_x}$, Eq.\eqref{eq:chi_jj_qn1}, obtained with the functional derivative approach that gives us $\partial \Sigma^{(2)}/\partial G^{(1)}$ as irreducible vertex.

Typically, using $500t$ as the cutoff Matsubara frequency, i.e. about 60 times the bandwidth, the above equation is satisfied to a relative accuracy of $10^{-7}$. By increasing the cutoff, the accuracy can be increased at will. We have reached an accuracy of $10^{-10}$. The separate contributions of the different terms of expression \eqref{eq:chi_jj_qn1}, also represented schematically in Fig.\ref{fig:chijj}, are shown in Fig.\ref{fig:f-sum-rule} as a function of temperature for $17\%$ doping, i.e. on the left side of the quantum critical point. As one would expect, the bubble contribution is dominant at high temperature, although the first vertex correction is not negligible. At low temperature, in the renormalized classical regime, the two terms give comparable contributions. The contribution of the third term in Eq.\eqref{eq:chi_jj_qn1}, the Aslamasov-Larkin-like diagrams in Fig.\ref{fig:chijj}, vanishes at all temperatures. As will be seen in the next subsections, despite this vanishing contribution of this term to $\chi_{j_xj_x}(iq_n=0)$, i.e. to the integral of the real part of the conductivity, this term contributes in a non-trivial way both to the DC and the finite frequency conductivity.

\subsection{Optical conductivity}

\setlength{\unitlength}{1in}

\begin{figure}
\includegraphics[width=0.233\textwidth]{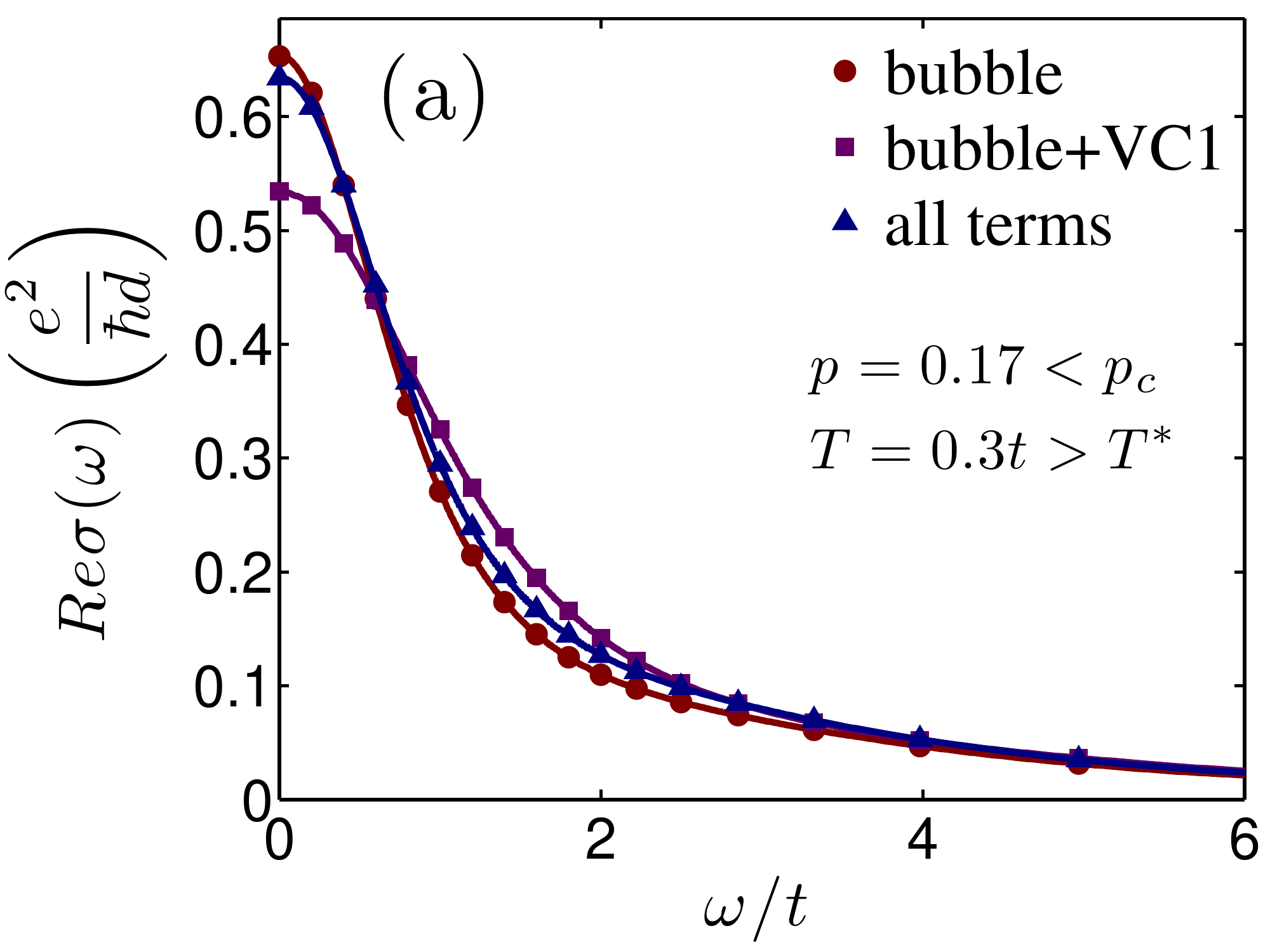}
\includegraphics[width=0.233\textwidth]{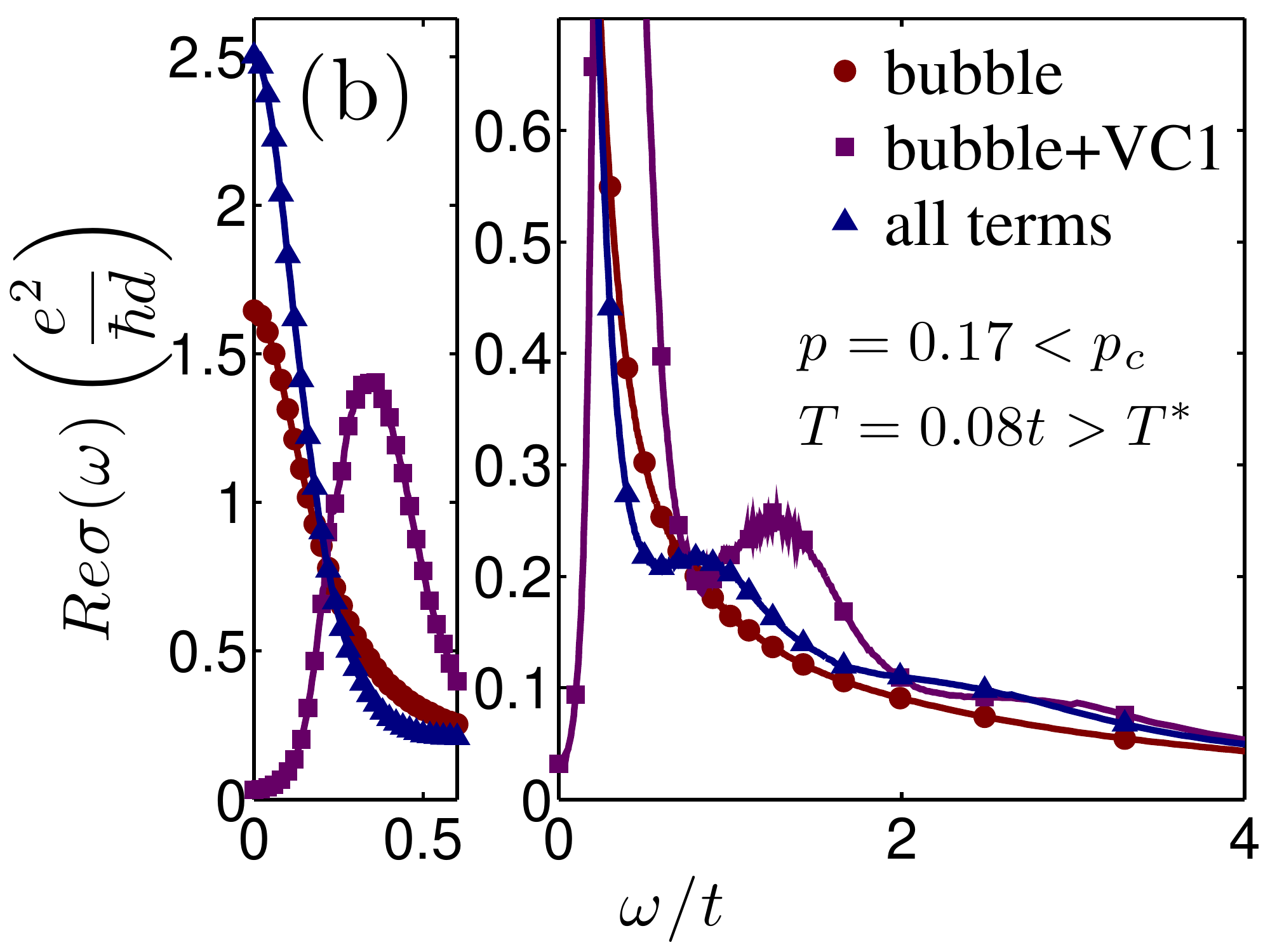}\\
\includegraphics[width=0.23\textwidth]{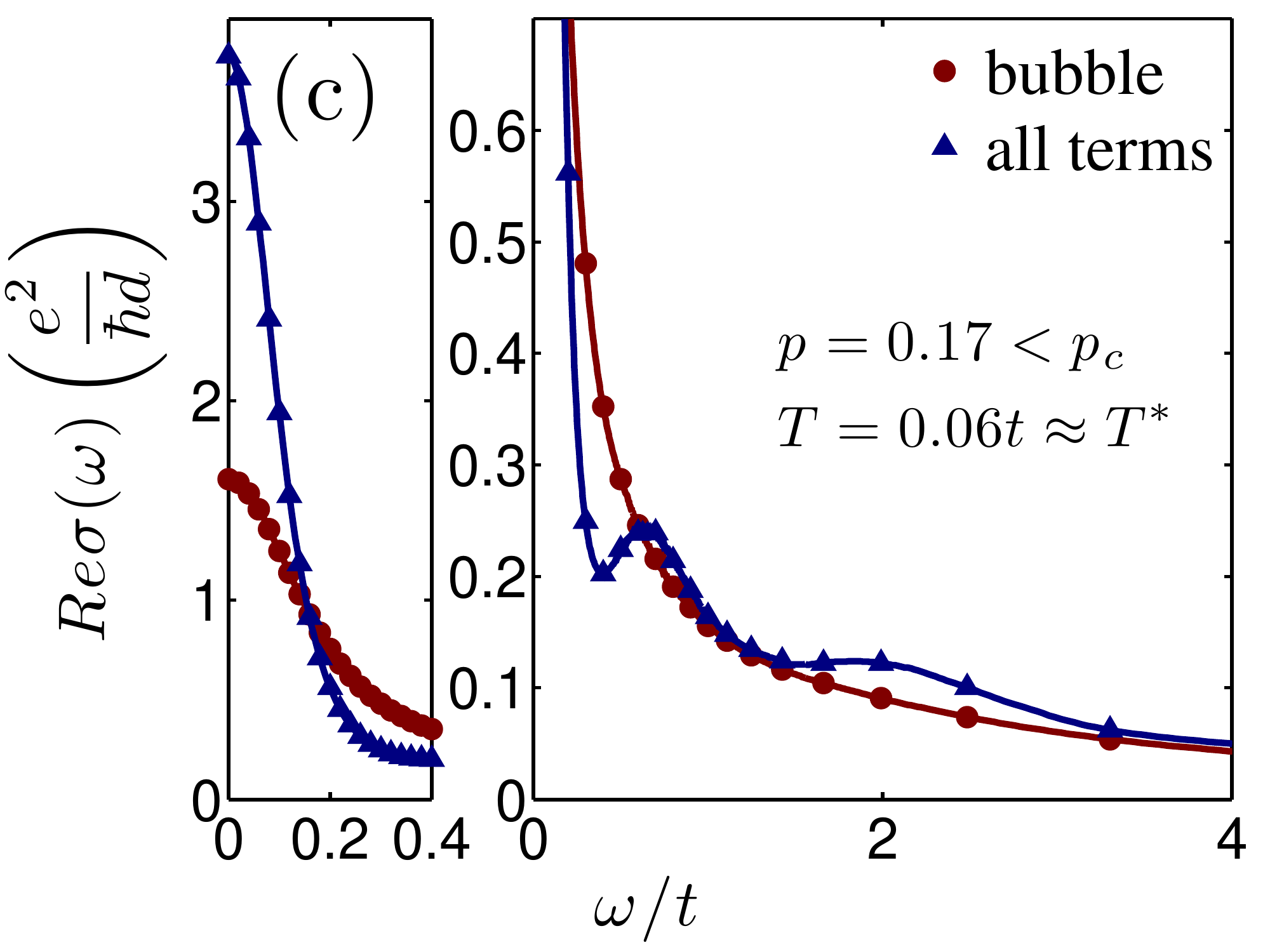}
\includegraphics[width=0.23\textwidth]{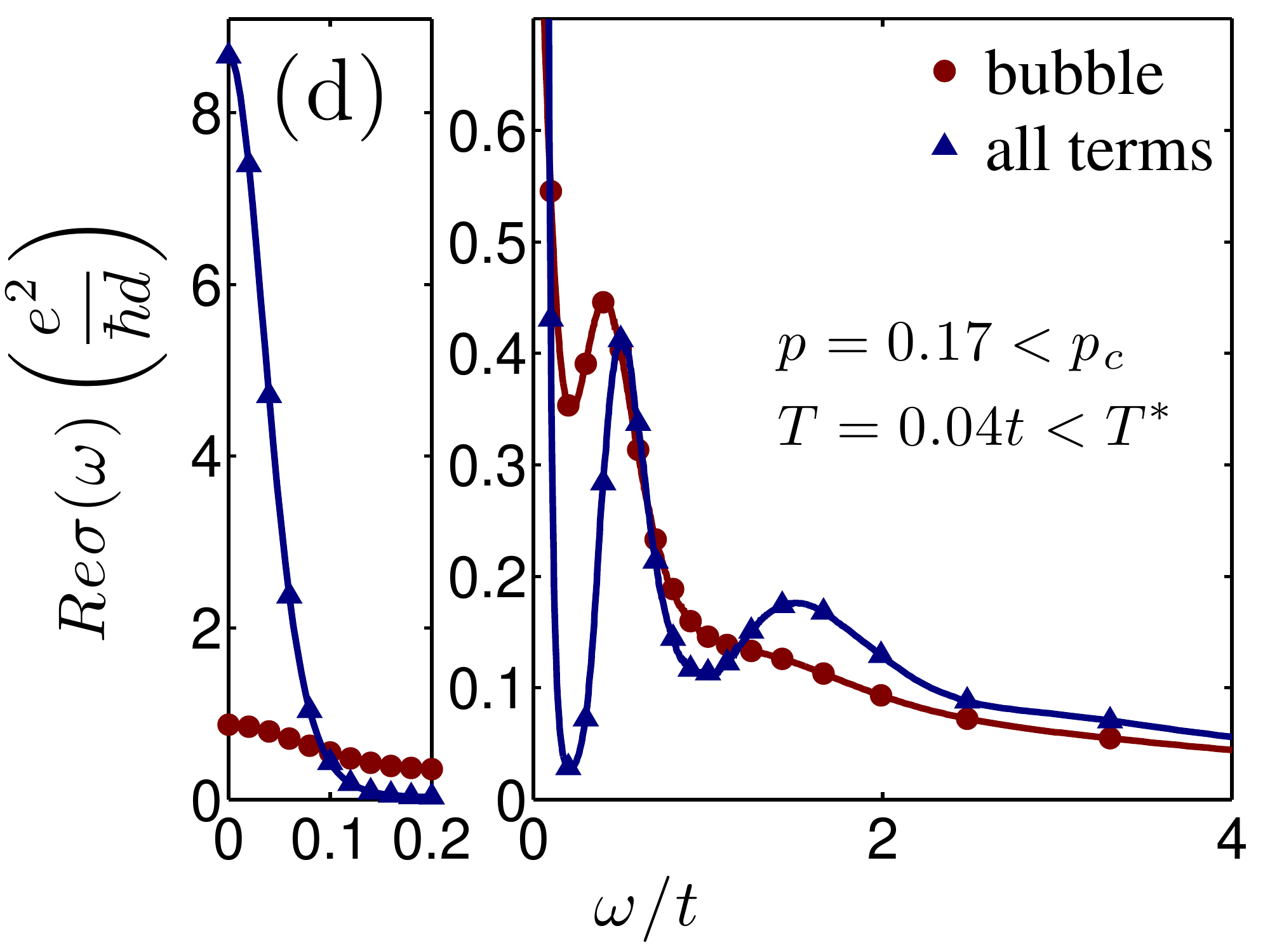}\\
\includegraphics[width=0.23\textwidth]{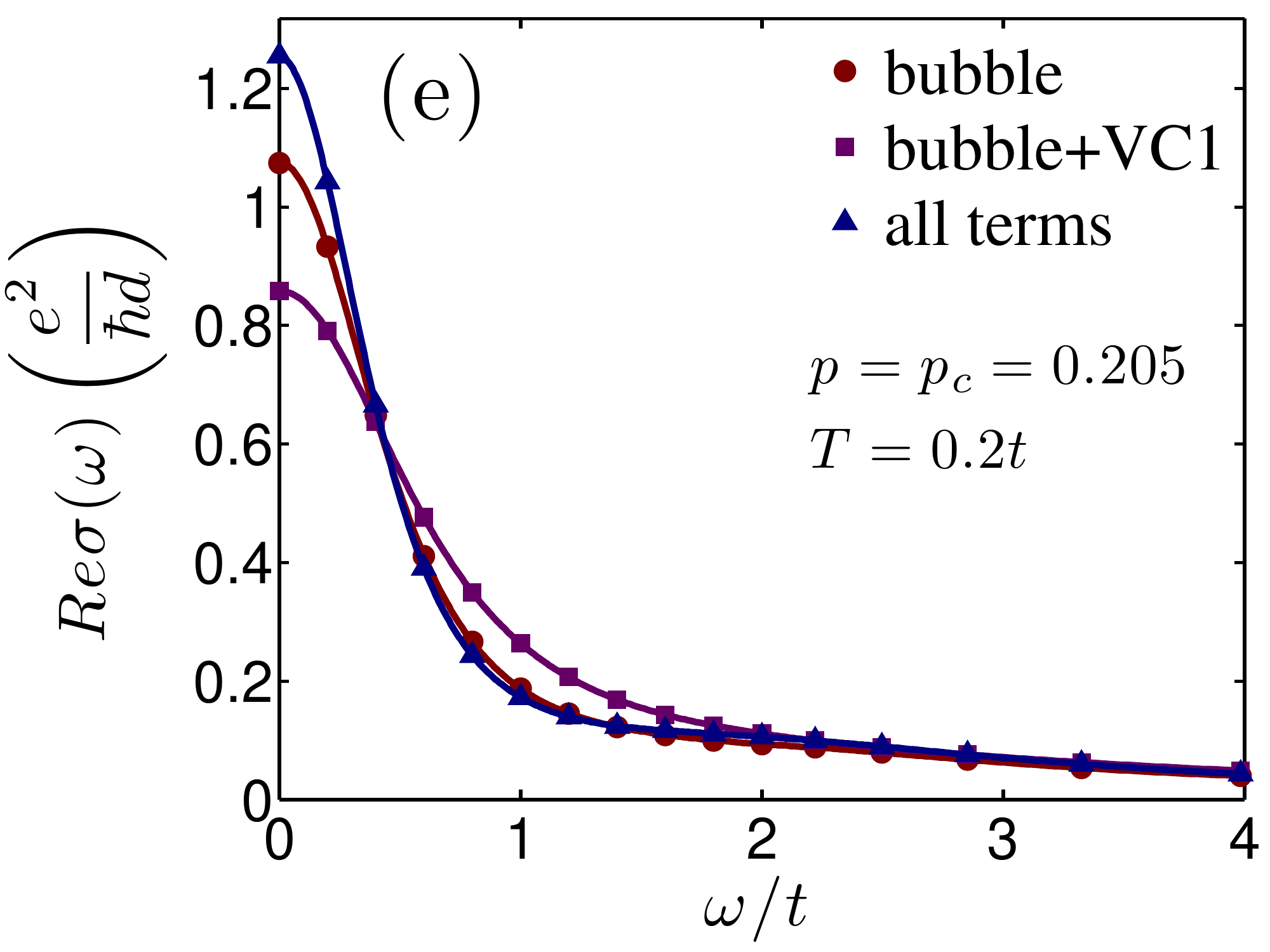}
\includegraphics[width=0.23\textwidth]{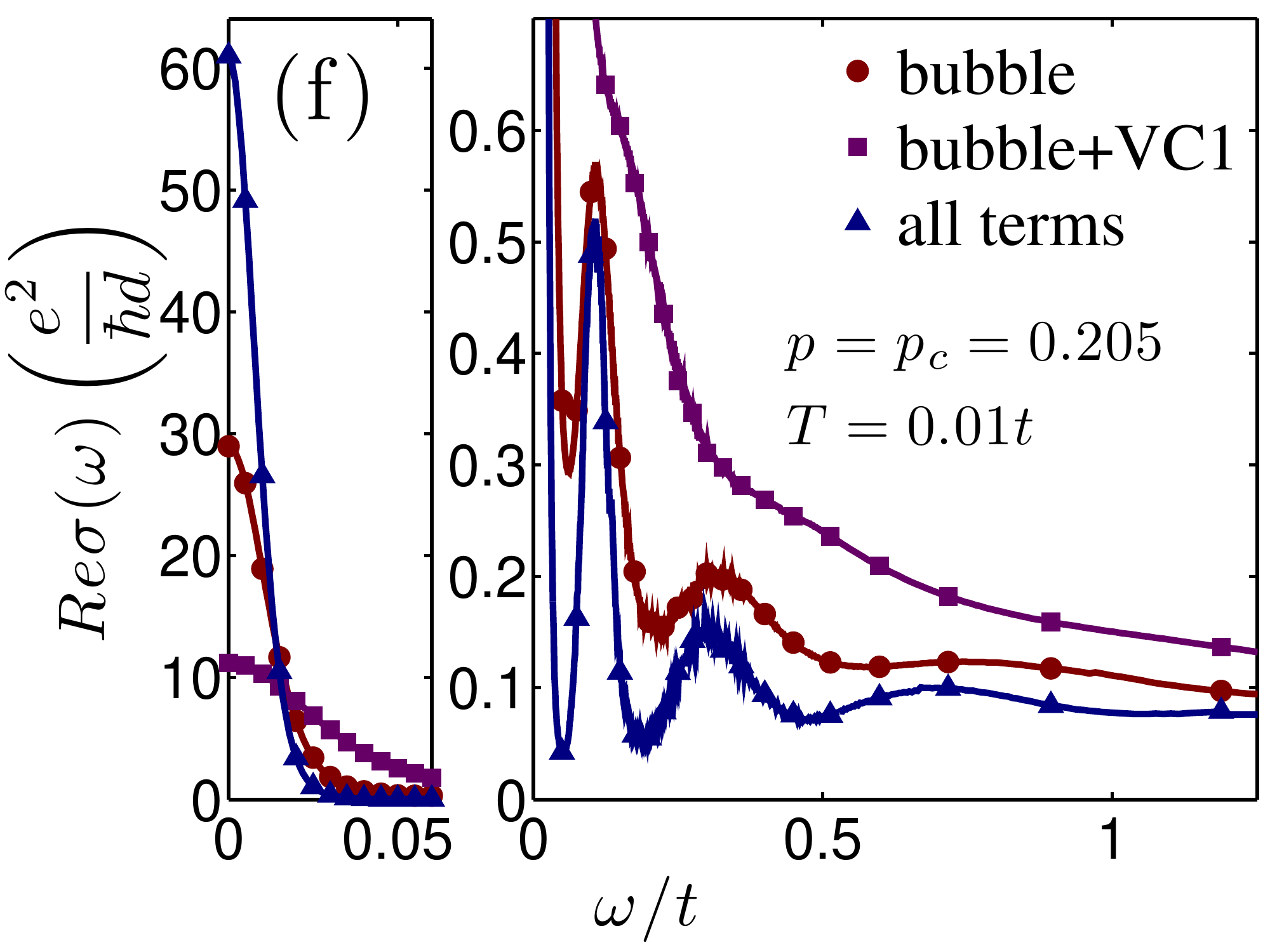}\\
\includegraphics[width=0.23\textwidth]{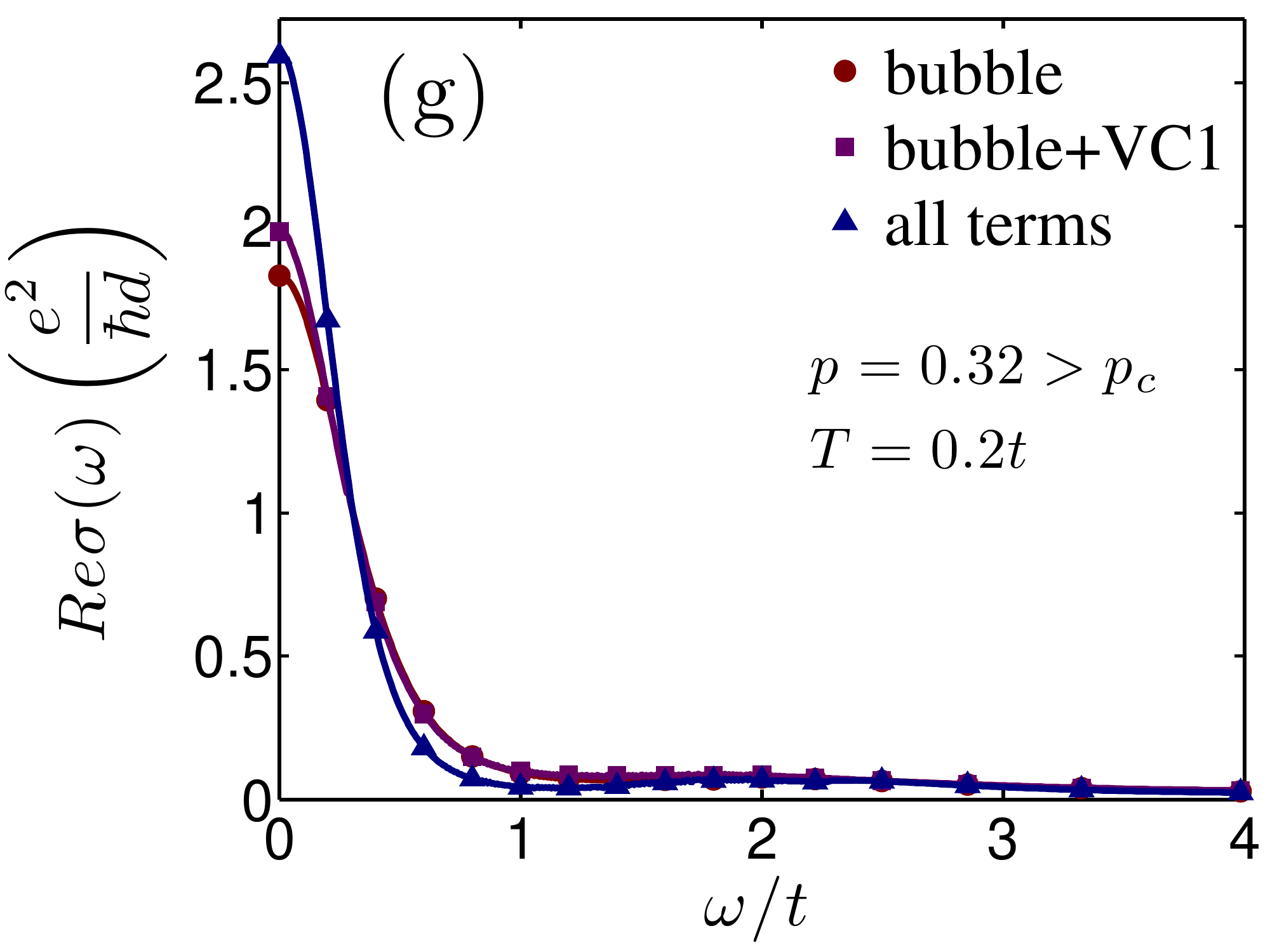}
\includegraphics[width=0.23\textwidth]{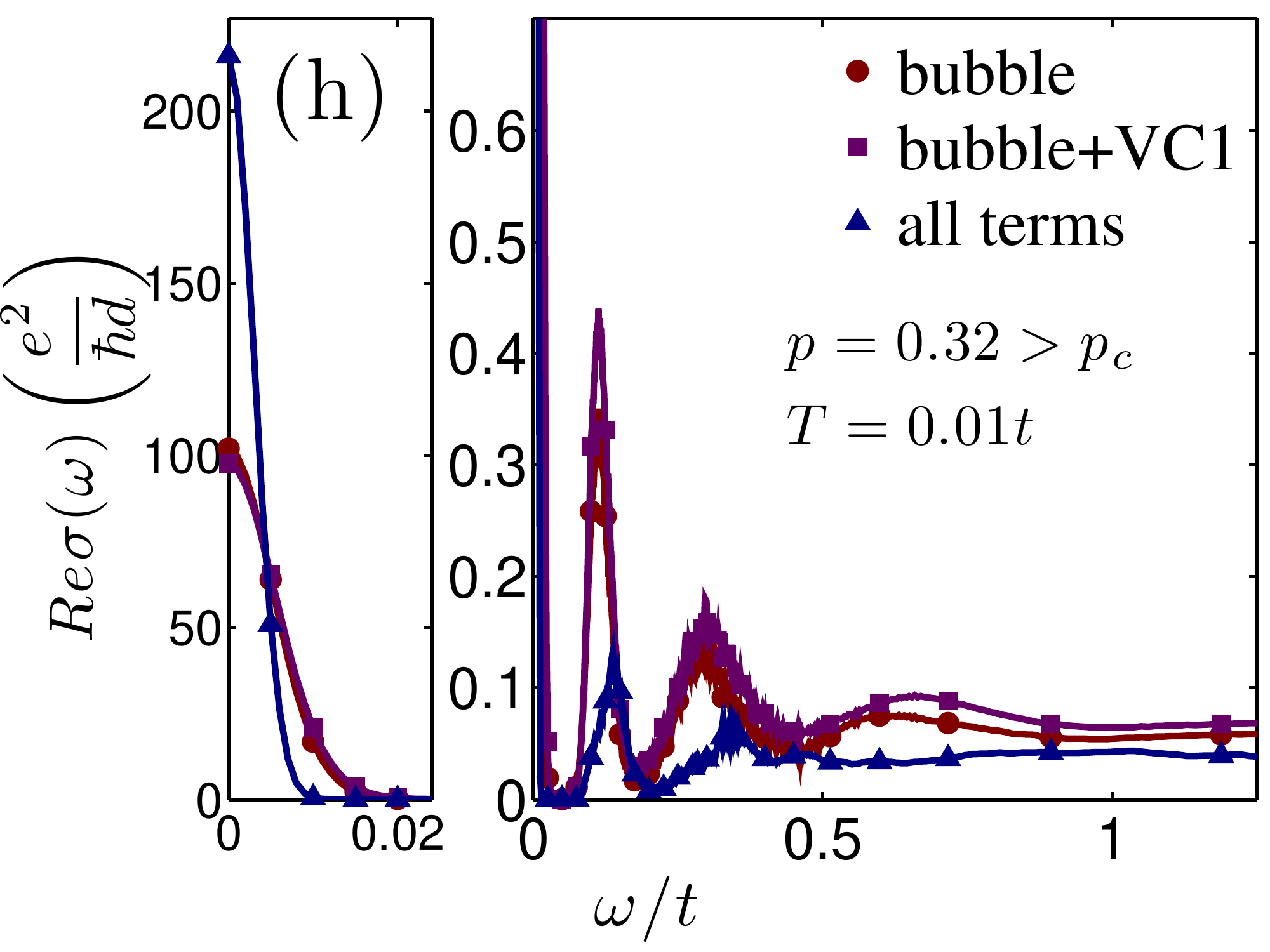}
\caption{\label{fig:opt_cond} Optical conductivity with and without vertex corrections at different dopings and temperatures. From (a) to (d) the doping is $p=0.17<p_c$ and the temperatures are (a) $T=0.3t>T^*$, (b)  $T=0.08t>T^*$, (c) $T=0.06\approx T^*$ and (d) $T=0.04t<T^*$. The other dopings are $p=0.205=p_c$ at temperatures (e) $T=0.2t$ and (f) $T=0.01t$ and $p=0.32>p_c$ at (g) $T=0.2t$ and (h) $T=0.2t$. All panels on the right-hand side contain a blow up of the low-frequency region of the rightmost plot. The symbols are only shown for a small fraction of the total number of points in the grids.}
\end{figure}
The optical conductivity and the effect of the vertex corrections are different depending on which side of the quantum critical point the system is and on the temperature. On the left-hand side of the critical point, the conductivity changes qualitatively as the temperature approaches the crossover temperature to the renormalized classical regime $T^*$. Here $T^*$ is defined as the temperature at which $\xi_{sp}=\xi_{th}$. As shown in Fig.\ref{fig:opt_cond}(a), at high temperature it has a Lorentzian-like shape at low frequency, whether we look at the bubble alone, the bubble with the first vertex correction or with both vertex corrections. In this region of the phase diagram, the low frequency conductivity is smaller if the vertex corrections are included. Then, as seen in the left part of Fig.\ref{fig:opt_cond}(b), as $T$ is lowered toward $T^*$ but still above it, the effect of the first vertex correction is to strongly decrease the low frequency conductivity, while the second correction does not just compensate this effect, but make the total even higher than the bubble alone. At higher frequencies, in the right part of Fig.\ref{fig:opt_cond}(b), a hump, not present without vertex corrections, appears in the total conductivity. Note that at this temperature, the spin fluctuation frequency $\omega_{sp}$ is about $T/2$. When $T$ is around $T^*$, in Fig.\ref{fig:opt_cond}(c), the hump is more pronounced and finally, when $T<T^*$, in Fig.\ref{fig:opt_cond}(d), it becomes a very distinct peak. At this temperature, a hump similar to the one seen with vertex corrections at higher temperature appears in the bubble term alone. At this temperature, $T\approx 600\omega_{sp}$ and $\xi_{sp}=147\approx 6\xi_{th}$. If we were to compare our result to experiments, assuming an energy scale $t=350meV$, the hump seen around $0.5t$ would correspond to the feature observed in the mid-infrared frequency range in the optical conductivity of electron-doped  cuprates.\cite{Onose:2004,Zimmers:2005} Note that another clear hump appears in the conductivity with vertex corrections in Fig.\ref{fig:opt_cond}(d). At this doping and temperature, the spin fluctuations are incommensurate, hence their effect on the spectral weight at finite energy can be quite complex. This additional structure in the conductivity may be a consequence of this incommensurability.

At the critical doping, Figs. \ref{fig:opt_cond}(e) and  \ref{fig:opt_cond}(f), the low frequency conductivity is lower than the bubble result when only the first vertex correction is taken into account, while it is higher with both corrections. This effect is much more pronounced at low temperature in Fig. \ref{fig:opt_cond}(f). New secondary peaks also appear at low temperature. Those peaks are at frequencies considerably smaller than the peak seen at $p=0.17$ below $T^*$ and, as will be clear from the DC resistivity results in the next subsection, there is no pseudogap regime at this doping. Those peaks may nevertheless be caused by correlations that are present far beyond the critical doping, at least at finite temperature. The effect of those correlations are also clear in the resistivity results in the following subsection.

Finally, at a doping higher than the critical doping, in Figs. \ref{fig:opt_cond}(g) and \ref{fig:opt_cond}(h), the conductivity with the first vertex correction can be almost the same as the bubble result at low frequency both at high and low temperature, although this correction does not vanish in Matsubara frequency (at $T=0.01t$, $\chi_{j_xj_x}^{vc1}(0)$ is about $7\%$ of $\chi_{j_xj_x}^{(0)}(0)$ and $\chi_{j_xj_x}^{vc1}(i2\pi T)$ is about $14\%$ of $\chi_{j_xj_x}^{(0)}(i2\pi T)$). Adding the other correction makes the low frequency conductivity increase substantially. However, while the finite frequency conductivity stays finite when the doping is closer to $p_c$ or smaller, at low temperature, in Fig.\ref{fig:opt_cond}(h), it vanishes completely just after the main low frequency peak when both vertex corrections are included. This peak is also sharper and higher at high doping (not shown). As will be confirmed in the next subsection, this means that the system becomes closer to a Fermi liquid, although it has not yet reached this regime at this doping. When it does, the DC conductivity will be inversely proportional to $T^2$ at low temperature so that, to conserve the weight which is roughly constant with respect to temperature, i.e. to respect the \textit{f}-sum-rule, the width has to be proportional to $T^2$. At zero temperature, since there is no impurity scattering in our model, the low frequency part becomes a delta function. Note that an absoption band remains at finite frequency at the highest doping since there is still an incoherent part in the one particle spectrum. However, one must also note that the highest values of the conductivity in that band are about a thousand times smaller than the DC value and that the total optical weight of this band is about ten times smaller than the weight of the low frequency peak.

The frequency region where the effect of vertex corrections is important becomes smaller as the doping increases. For $p=0.17$, the difference between the different results vanishes around $\omega=4t$, at the critical point, this happens around $\omega=2t$ and at $p=0.32$, around  $\omega=1.5t$.

As the doping increases, the conductivity at low temperature becomes extremely sharp and it becomes very hard to do the maximum entropy analytical continuation. That is because a very fine grid must be used at low frequency, while one still needs a cutoff larger that the bandwidth. This makes the number of points in the real frequency grid explode, as well as the time for the optimization process.

\subsection{DC resistivity as a function of temperature and doping close to the quantum critical point}\label{sec:RC_res}

\begin{figure}
\includegraphics[width=0.35\textwidth]{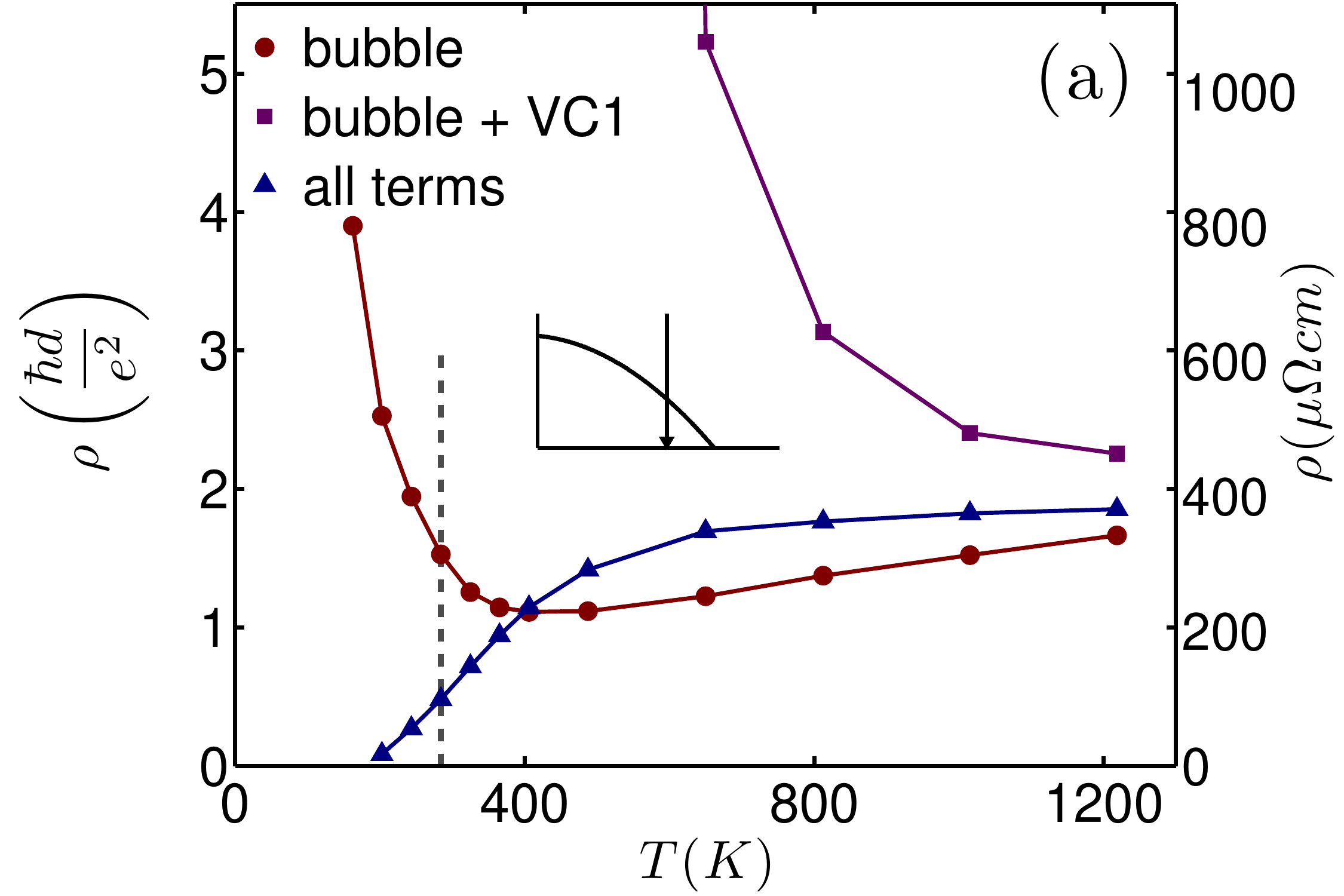}\\
\includegraphics[width=0.35\textwidth]{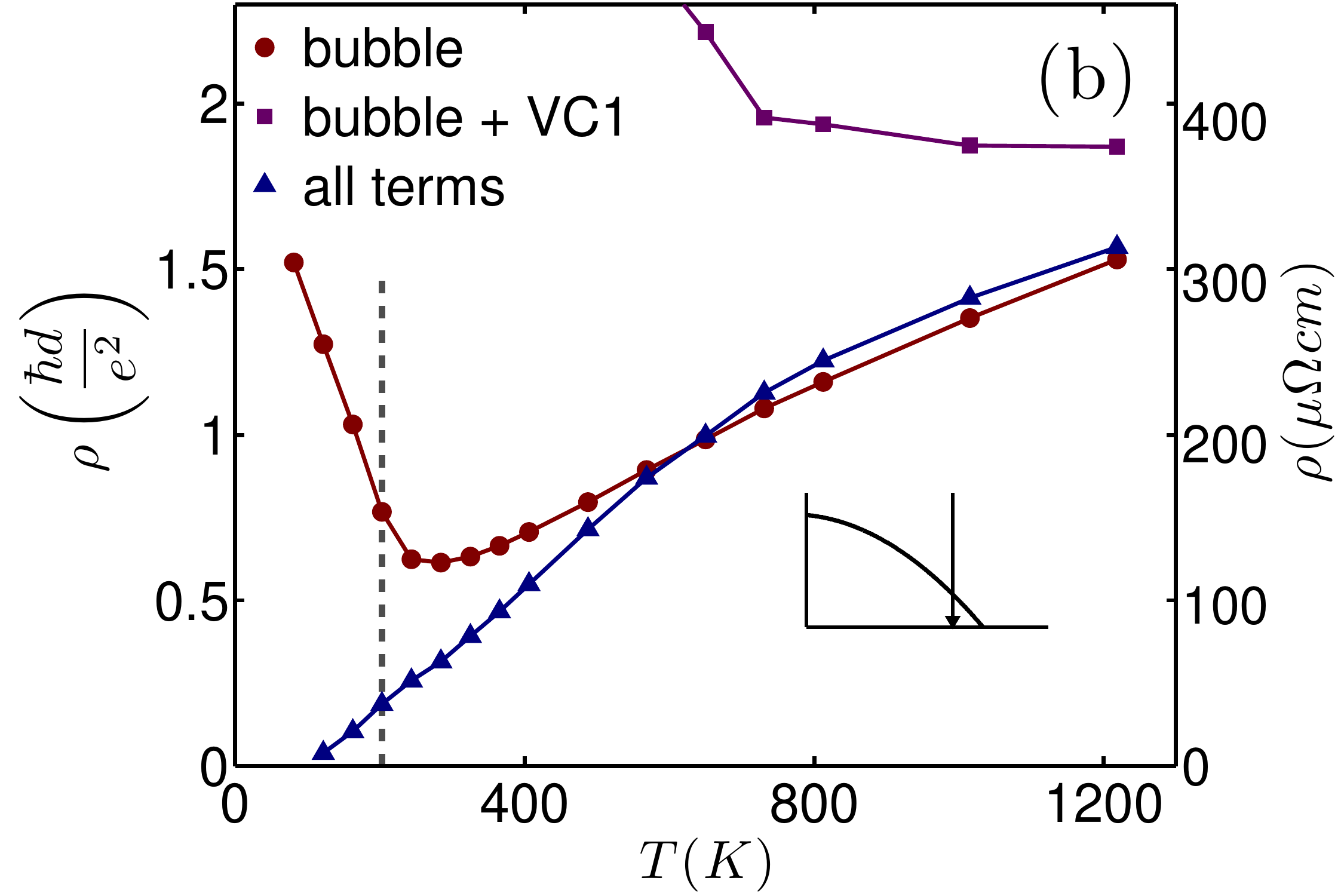}\\
\includegraphics[width=0.355\textwidth]{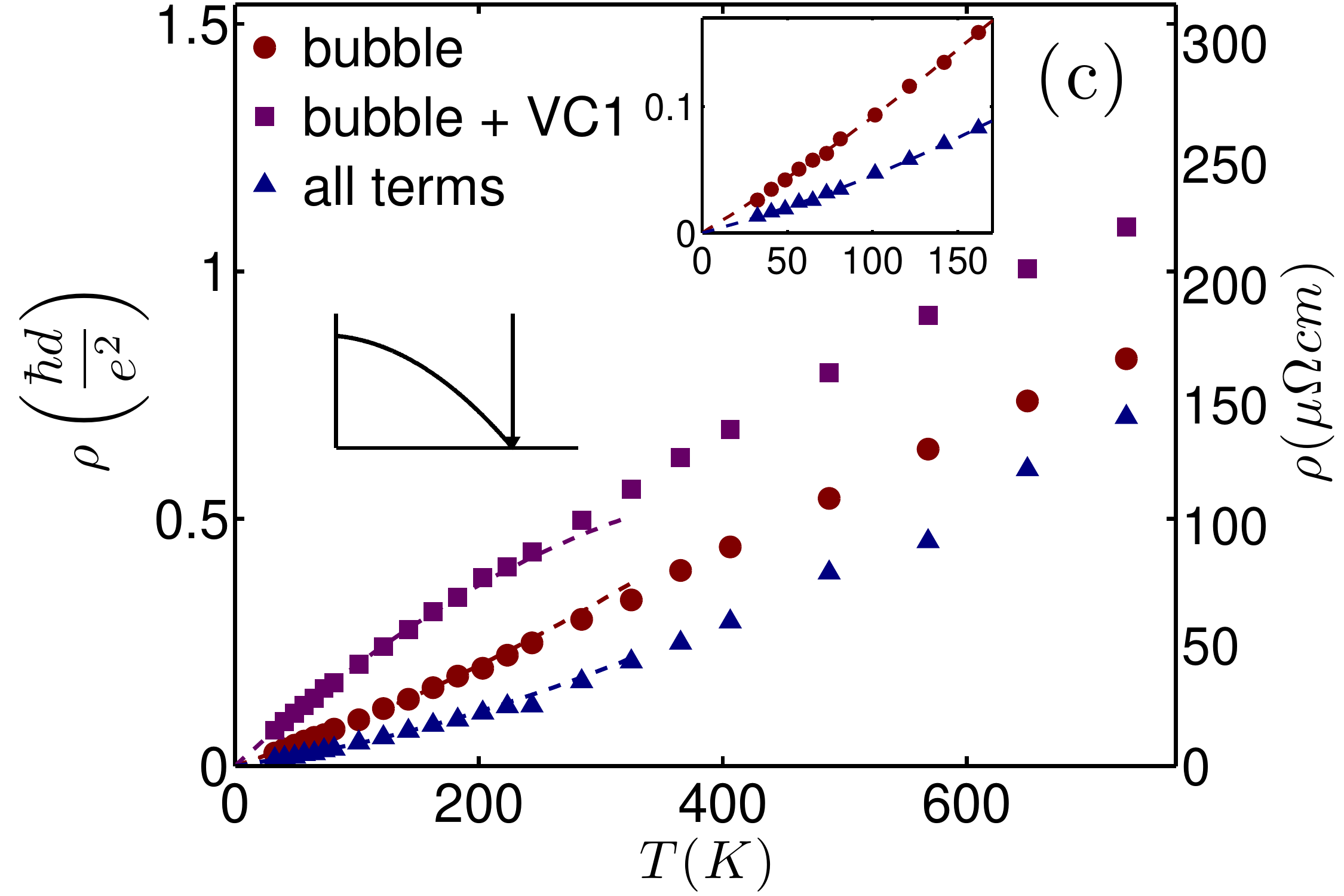}\\
\includegraphics[width=0.355\textwidth]{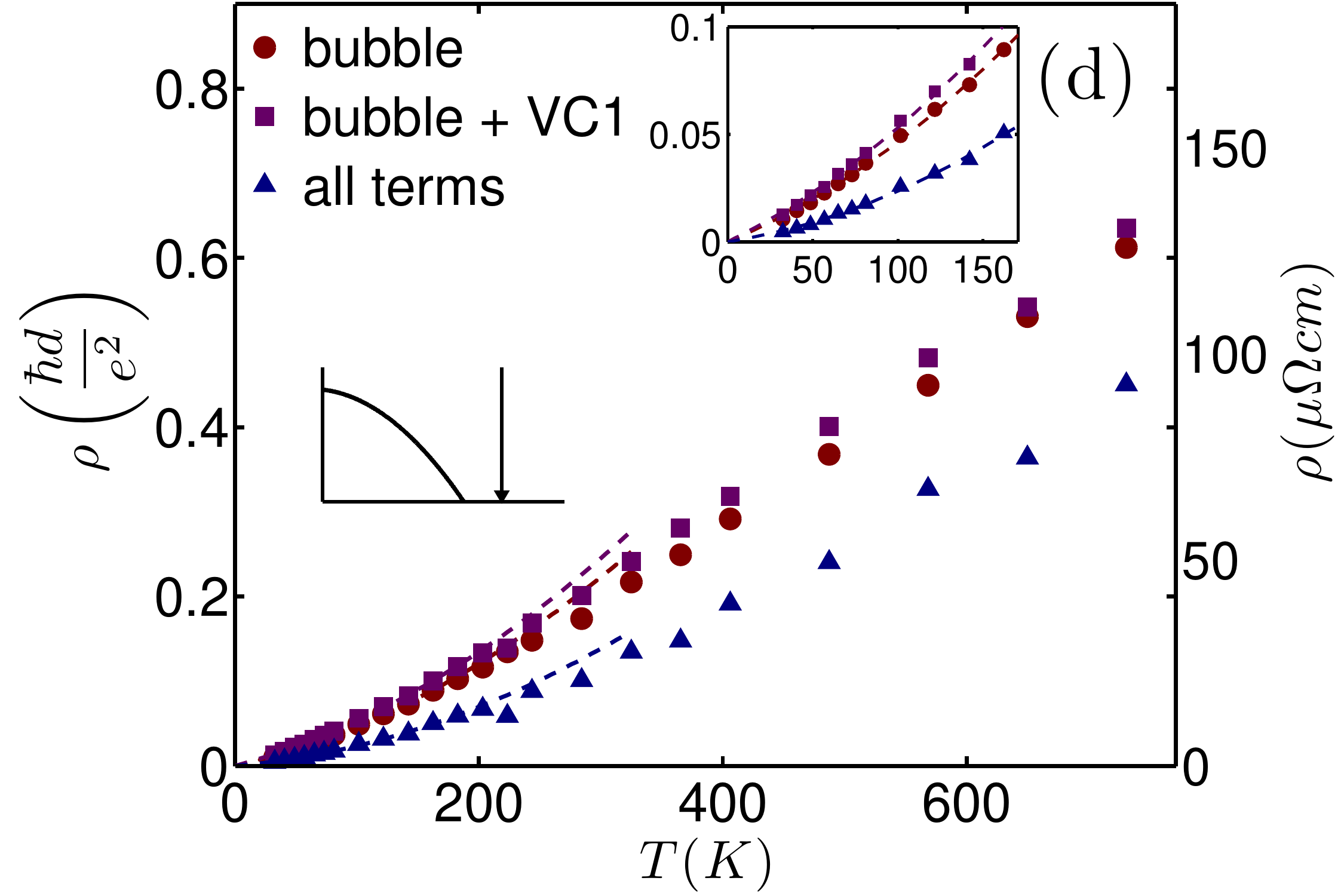}\\
\includegraphics[width=0.355\textwidth]{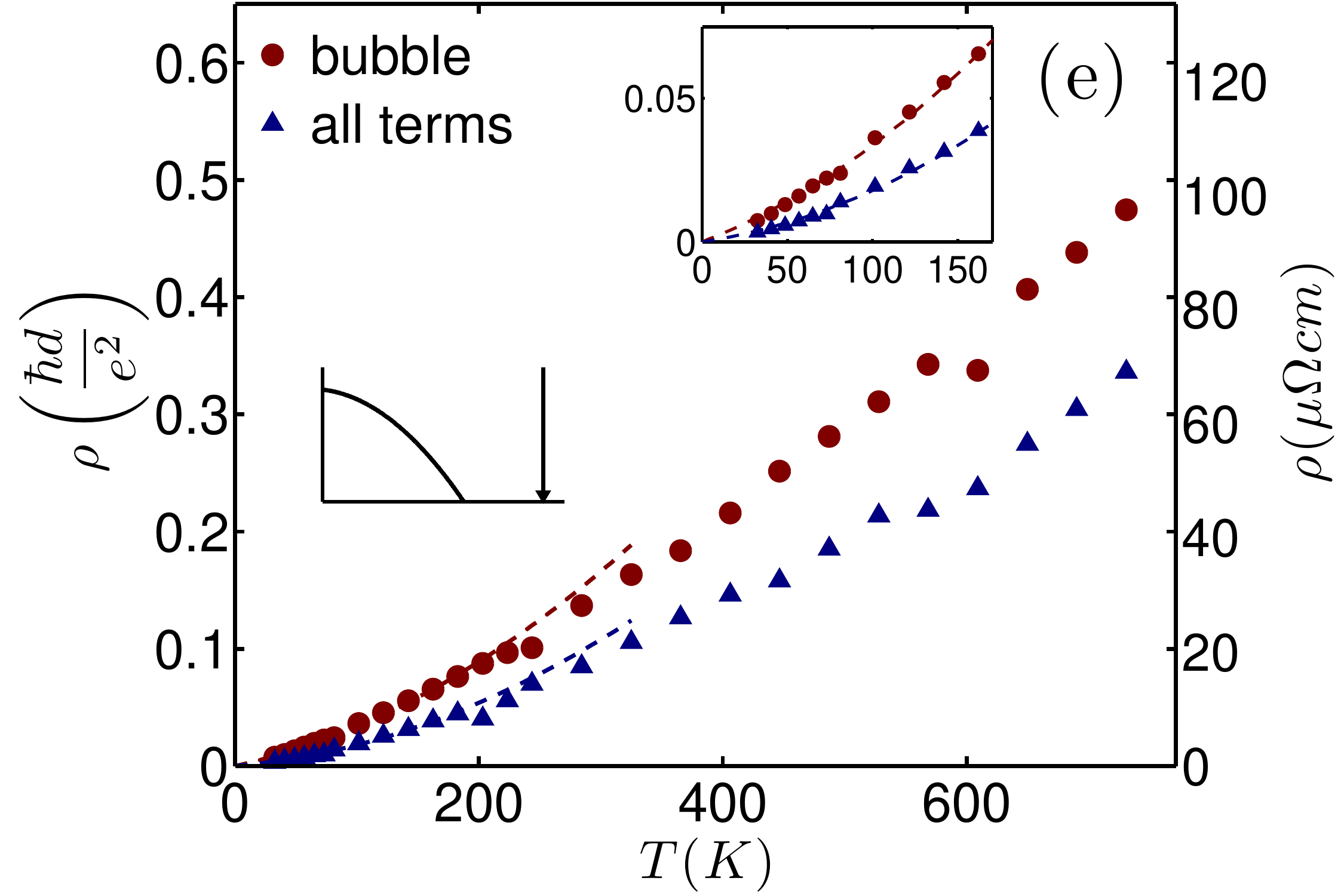}
\caption{\label{fig:dc_resistivity} Resistivity as a function of temperature for a) $p=0.15$ b) $p=0.17$ c) $p=0.205$ (critical doping) d) $p=0.26$ and e) $p=0.32$. The dashed vertical line in (a) and (b) indicates the temperature at which the internal accuracy starts failling as $T$ decreases. Dashed curves in (c), (d) and (e) are the results of fits of the low $T$ resistivity  to the form $AT+BT^2$.}
\end{figure}
Figure \ref{fig:dc_resistivity} shows five interesting doping regimes for the DC resistivity. The left vertical axis is in units of the Ioffe-Regel maximum metallic resistivity. The right vertical axis translates the result in micro-ohm-centimeters by taking $d=5 {\AA}$ as the interplane lattice constant. For the temperature scale, we use $t=350 meV$. The resistivity without vertex corrections (bubble) appears as red circles, with the first correction, namely the second term in expression \eqref{eq:chi_jj_qn1} or the Maki-Thompson-like diagram in figure \ref{fig:chijj}, in purple squares (bubble+VC1) and the total resistivity appears as blue triangles. The insets of the last three figures display the low temperature behavior. The position of each doping is indicated by an arrow for each figure on a schematic ``phase diagrams'' with the quantum critical point and the crossover line.

At the smallest doping, $p=0.15$ in Fig. \ref{fig:dc_resistivity}(a), one observes at high temperature the expected Ioffe-Regel maximum metallic resistivity saturation. The value is about $2 \hbar d/e^2$. The resistivity is higher when vertex corrections are included at high temperature. Without vertex corrections, it increases with decreasing temperature below the crossover temperature to the renormalized classical regime $T^*$, which is about $400K$ ($0.1t$) at this doping. The effect of vertex corrections is dramatic at low temperature, essentially changing the resistivity from insulating to metallic. An important point is that this effect can only be obtain when both corrections are included. When only the first correction is included, the scattering effect of magnetic correlations is largely overestimated and the resistive behavior thus amplified. At $p=0.17$, closer to the quantum critical point, Fig. \ref{fig:dc_resistivity}(b) exhibits essentially the same behavior except that, given the overall smaller resistivity, the saturation at high temperature occurs beyond the range displayed. At low temperature in Figs. \ref{fig:dc_resistivity}(a) and (b), the resistivity with vertex corrections seems to extrapolate to negative values. However, as indicated by the vertical dashed lines in those figures, this happens in a region where the TPSC result is no longer quantitatively reliable because the internal accuracy check starts to fail when the magnetic correlation length becomes too large compared to the thermal De Broglie wavelength. The results at those temperatures and dopings, namely, deep in the peudogap regime, can thus at most be considered as qualitative tendencies.

At the quantum critical point, $p=0.205$, one observes in Fig. \ref{fig:dc_resistivity}(c) that, without vertex corrections, the resistivity is quite linear at low temperature, as found previously in spin fluctuation theories.\cite{Moriya:1990,Moriya:2000} When all the corrections are included, the most obvious effect is that the resistivity decreases at all temperatures. The linear behavior remains, at low temperature, but the vertex corrections tend to reduce the linear contribution and a $T^2$ behavior appears at a lower temperature than without vertex corrections. Note that, with only the first correction, there is a change in curvature. Also, the resistivity is larger than the bubble result at all temperatures instead of smaller as found when all vertex corrections are included.

As the doping becomes larger than the quantum critical doping, figures \ref{fig:dc_resistivity}(d) and \ref{fig:dc_resistivity}(e) show that a linear $T$ behavior is still present at the lowest temperatures, but gradually disappears as the doping increases and the Fermi liquid-like $T^2$ behavior becomes dominant. The resistivity with the first vertex correction only is omitted in figure \ref{fig:dc_resistivity}(e) because it is almost equal to the bubble result at all temperatures.

\begin{figure}
\includegraphics[width=0.35\textwidth]{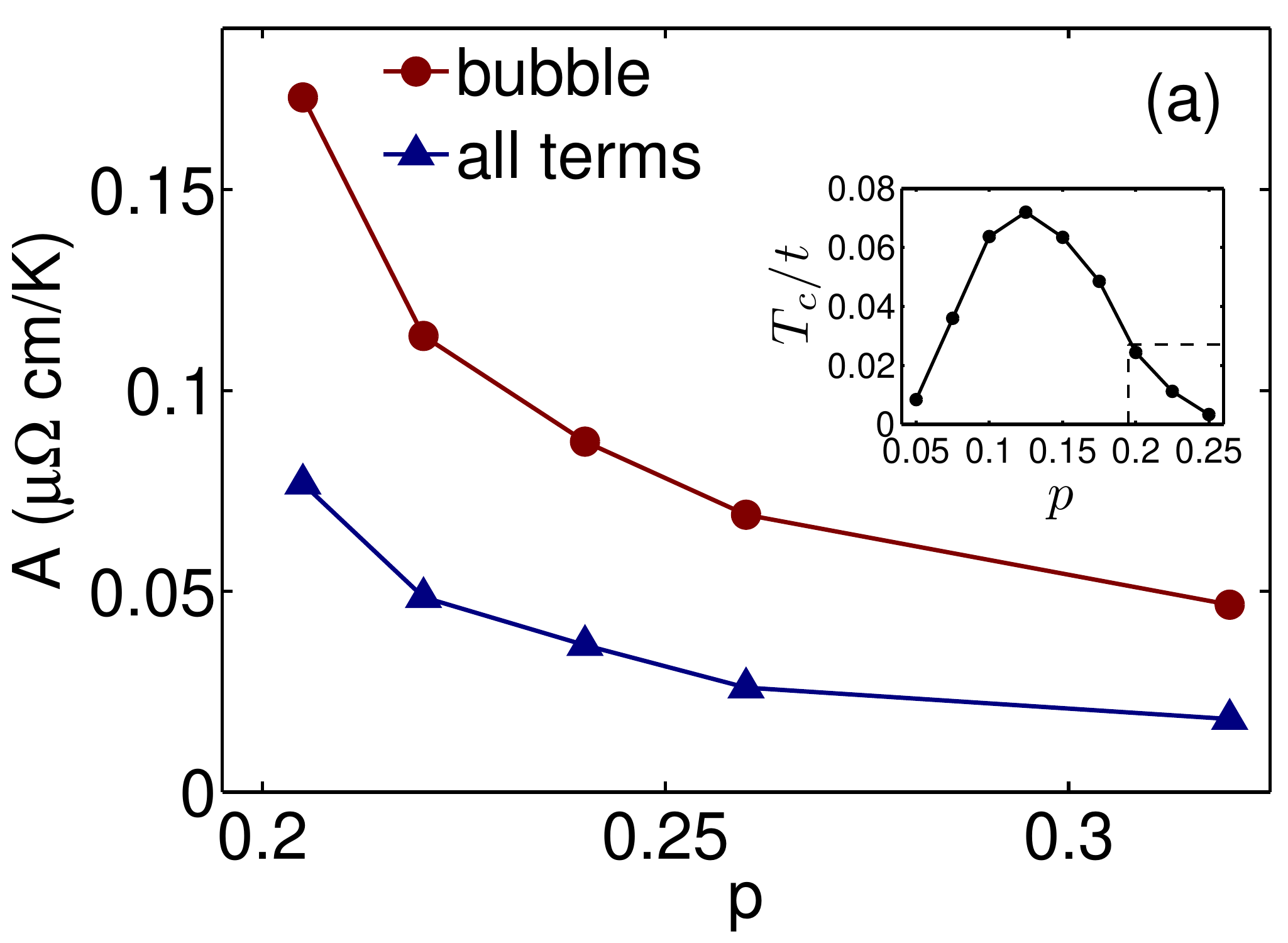}\\
\includegraphics[width=0.35\textwidth]{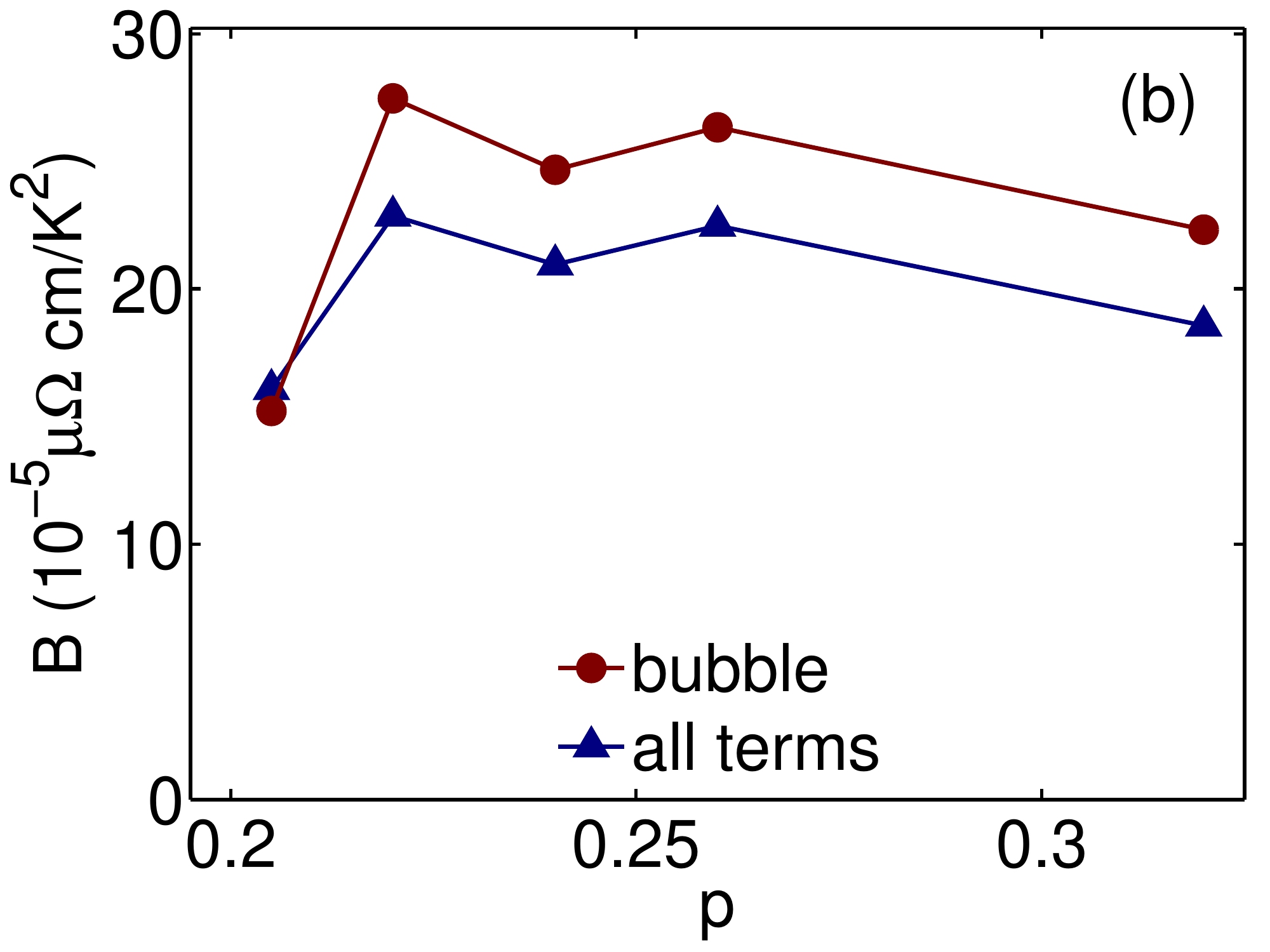}
\caption{\label{fig:at+bt2} Coefficients a) $A$ and b) $B$ in the fit of the form $AT+BT^2$ to the resistivity, as a function of doping, starting at the QCP. The inset in (a) shows the superconducting transition temperature as a function of doping estimated with the TPSC approach, from Ref. \onlinecite{Kyung:2003}. The doping region relevant for our fits is indicated with dashed lines in this inset.}
\end{figure}
The result of fits of the temperature dependence of the resistivity to the functional form $AT+BT^2$ over the range $0.008t<T<0.05t$ ($30 K < T < 200 K$) is illustrated in Figs.~\ref{fig:at+bt2}a and \ref{fig:at+bt2}b. Those fits were done using two other resistivity curves in addition to those shown in figure \ref{fig:dc_resistivity}, namely, at $p=0.22$ and $p=0.24$. The linear coefficient $A$ decreases as one moves away from the quantum critical point. This decrease is correlated with the superconducting transition temperature, shown in the inset, estimated from TPSC calculations of the $d_{x^2-y^2}$-wave susceptibility\cite{Kyung:2003}. As for the coefficient $B$, it seems to have a rapid increase as we move away from the quantum critical point (QCP) and then to remain roughly constant as the doping increases. Note however that this coefficient is hard to obtain precisely close to the QCP. For example, if we change the number of points used in the fit, the value of $B$ in this doping region can change by about $20\%$. This is because at low temperature, the region of interest here, the quadratic term has a very small contribution compared to the linear term close to the QCP. Also, there is always some noise in the resistivity, a consequence of the fact that those results are obtained by analytical continuation of Matsubara response functions, a procedure very sensitive to finite precision noise in these response functions. Therefore, the value of $B$ for the two smallest dopings in figure \ref{fig:at+bt2}(b) are rough estimations. However, as the doping increases, the quadratic contribution becomes more important and thus $B$ is more accurate.

\section{Discussion}\label{sec:discussion}

Starting with the self-energy at the second level of approximation, \eqref{eq:self_2_re} the only non-straightforward part of the derivation is the neglect of the functional derivatives of the irreducibles vertices $\Gamma_{sp}$ and $\Gamma_{ch}$ in the derivative of $\Sigma^{(2)}$ with respect to the field, or with respect to $G^{(1)}$. However, as was explained in subsection \ref{sec:cond_tpsc}, following equation \eqref{eq:dSigma_dG}, the vanishing of those contributions to the conductivity is exact. Therefore, starting from the approximation \eqref{eq:self_2_1} for the self-energy that contains the local spin and charge irreducible vertices $\Gamma_{sp}$ and $\Gamma_{ch}$, Eqs. \eqref{eq:gam_sp} and \eqref{eq:gam_ch_1}, the rest of the calculation is exact.

In fact, based purely on our numerical results, with the Green's function at the second level of approximation and the irreducible vertex generated from the corresponding self-energy, we would have strong reasons to suggest that the \textit{f}-sum rule is satisfied exactly in our approach. At all dopings and temperatures, that sum rule is satisfied with very high accuracy and the precision increases with the cutoff Matsubara frequency. Since the \textit{f}-sum rule is a consequence of particle conservation (see appendix \ref{sec:Appendix_f_sum_rule}), this means that our approach is consistent with particle number conservation. This result is a consequence of our use of functional derivatives to calculate the Green's function and correlation functions, as in conserving approximations.\cite{Baym:1962} However, unlike in those approaches, there is no self-consistency at the one particle level, i.e., our self-energy $\Sigma^{(2)}$ is calculated with $G^{(1)}$ and not $G^{(2)}$. Thus one-particle self-consistency is not necessary to ensure particle conservation.

The saturation of the resistivity to the Ioffe-Regel maximum value at high temperature is a general constraint, this time based on general physical considerations, that is satisfied by our approach and not by Boltzmann-based approaches. At high doping the Fermi-liquid $T^2$ resistivity is also recovered, as expected.

From the fact that the first vertex correction, the Maki-Thompson-like diagram in Fig.\ref{fig:chijj}, is sufficient to respect the sum rule, one would be tempted to assume that the other type of vertex corrections will not contribute, as was done in previous calculations.\cite{Kontani:2007} It is clear from our results that this is not the case. In the pseudogap regime, the Aslamasov-Larkin-like contribution has a drastic effect on the DC resistivity, making the system metallic instead of insulating. If only the first correction is included, we would wrongly assume that the system is even more insulating than without any correction. On the right-hand side of the quantum critical point, it is also very important to include both vertex corrections since their total effect is to reduce the resistivity, while it would increase, for a large range of doping, if only the first correction was taken into account. Also, when both corrections are included, the $T^2$ Fermi liquid resistivity is recovered at a lower doping. The importance of including both corrections is also very clear in the optical conductivity, especially on the left of the quantum critical point, where the result is qualitatively very different with only the first vertex correction.

One may ask whether it is enough to include only those two vertex corrections. To answer that question we recall that our expression for the \textit{current-current} response function \eqref{eq:chi_jj_qn1} is not derived perturbatively, but using functional derivatives. Since the calculation of $\chi_{j_xj_x}$ is exact starting from the TPSC self-energy, Eq.\eqref{eq:self_2_1}, as long as the TPSC approach is valid, all the terms needed are the ones that we have used. The region of the phase diagram where the TPSC result breaks down is deep in the pseudogap regime, when some parts of the Fermi surface are destroyed and thus the one-particle spectrum is dramatically different from the non-interacting one. Therefore, in this regime, the TPSC results should be regarded as more qualitative than quantitative.

Since we have not used realistic band parameters, we cannot directly compare our results with any real material. However, some tendencies can help us understand experiments and provide some hints on what would be interesting to investigate using more realistic band structure.

First, there is a pseudogap in the TPSC approach. For electron-doped cuprates, it was shown that the pseudogap has properties predicted by this approach, namely it appears when the antiferromagnetic correlation length becomes larger than the single particle De Broglie wavelength, or the mean free path.\cite{Motoyama:2007} The hump around $\omega=0.5t$, seen in figure \ref{fig:opt_cond}(b), (c) and (d), thus corresponds to the analogous feature seen, for example, in $Nd_{2-x}Ce_xCuO_4$ in the mid-infrared energy range.\cite{Onose:2004,Zimmers:2005} If we use $t=0.35eV$, the location in energy of this structure is also in the correct energy range. Those optical conductivity results and previous comparison of TPSC spectral weights with photoemission experiments on electron-doped cuprates add to the evidence that those materials are well described by the accurate solutions of the Hubbard model provided by the TPSC approach at intermediate coupling.\cite{Kyung:2004,Hankevych:2005}

An interesting aspect of our results is the linear low-temperature resistivity observed at the quantum critical doping. As mentioned in section \ref{sec:RC_res}, linear $T$ resistivity has already been obtained in spin-fluctuation theory within the so-called \textit{self-consistent renormalization} approach.\cite{Moriya:1990,Moriya:2000} This is discussed in the review Ref.\onlinecite{RoschRMP:2007}. However, this theory is based on the variational approach to the Boltzmann equation. It has been shown with a better variational \textit{ansatz} that the resistivity should be in $T^2$.\cite{Hlubina:1995} That variational \textit{ansatz} is better because it takes into account that hot regions on the Fermi surface that are strongly influenced by spin scattering should be short-circuited by the cold regions. Indeed, it is the conductivities of the different Fermi surface regions that are integrated and not the resistivities. Thus, like parallel resistors, the total resistivity depends only on the more conductive regions. In our results, vertex corrections tend to make the resistivity more quadratic, but a linear term remains at low temperature. This result tells us that, at least for the parameters used in this work, the region of the Fermi surface with the lowest resistivity has linear $T$ resitivity. This is coherent with our observation, from spectral density calculations (not shown), that the whole Fermi surface seems incoherent, or hot, over a range of dopings that extends beyond the QCP at the temperatures considered.\cite{BergeronPhD:2011}

Another important question is whether, in the quantum critical region above the QCP, the whole Fermi surface is always incoherent, as suggested by some experiments.\cite{Kaminski:2003,Hussey:2008} This is possible because what characterizes this region is that the magnetic fluctuations are strong, but not yet strong enough to destroy any part of the Fermi surface. In other words, instead of having spin excitations that are well defined, they are broad in energy and wave vector, which means that they affect large parts of the Fermi surface and possibly all of it. To understand the transport, it is important to make the distinction between this regime and the pseudogap regime, where the well defined hot spots appear. This distinction is not made in Ref.\onlinecite{Hlubina:1995}. In the present work, the incoherence of the whole Fermi surface may come however from the fact that we have included only nearest-neighbor hoppings and that large parts of the Fermi surface are almost nested. More calculations of the conductivity using second and third nearest-neighbor hoppings will be necessary to verify if this global incoherence of the Fermi surface in the quantum critical region is universal. Note that this is a subtle question that cannot be answered using approaches with adjustable parameters, like those of Refs. \onlinecite{Moriya:1990} and \onlinecite{Hlubina:1995}, which can give qualitatively different results depending on how those parameters are chosen.

Another interesting point about the linear term in the resistivity is its correlation with the superconducting transition temperature $T_c$. This correlation seems to be present in all the unconventional superconductors. It has been observed in the cuprates, the pnictides, and the organics \cite{Cooper:2009,Doiron-Leyraud:2009_1,Doiron-Leyraud:2009, Abdel-Jawad_9957290}. In those materials, it seems that the linear coefficient $A$ disappears exactly at the end of the superconducting dome on the overdoped side. Which strongly suggests a common origin for linear resistivity and superconductivity. In the Hubbard model, in the weak to intermediate coupling regime, the TPSC method finds that superconductivity and linear resistivity also have the same origin, namely the interaction of quasiparticles with antiferromagnetic fluctuations. However, from Fig.\ref{fig:at+bt2}, it does not seem that $A$ disappears completely with $T_c$, which vanishes at around $p=0.25$ as shown in the inset of the figure. Note that there is some uncertainty in the values of $A$ that could come from a small systematic error in the resistivity. This is possible because it is difficult to obtain very precise analytically continued results. Nevertheless, it is clear that the most important drop in $A$ is before $T_c$ vanishes.

One striking result we obtain in the pseudogap regime on the left side of the QCP is the change of the resistivity from an insulating to a metallic behavior when the vertex corrections are taken into account. This is an effect of the Aslamasov-Larkin-like contribution which, as mentioned at the end of subsection \ref{sec:cond_tpsc}, gives a positive contribution to the conductivity when the density of states below the Fermi level is larger than above. In the present case the Fermi level is just above the Van Hove singularity. The density of states is therefore extremely asymmetric at the Fermi level and that explains why the AL term is so large that it counters both the effect of the one-particle self-energy that makes the bubble result insulating and the effect of the Maki-Thompson-like term that tends to make the system even more insulating. From preliminary results with second and third nearest neighbor hoppings $t'$ and $t''$, we notice that the second vertex corrections cannot always compensate the first correction and the resistivity for $p<p_c$ can be higher than the bubble result even with both vertex corrections.  This happens when the Fermi level is farther from the Van Hove singularity, so that the density of states is much less asymmetric around $\omega=0$ and thus the AL term is weaker. Those results tell us that the behavior observed in the pseudogap regime on the left side of the QCP in Figs. \ref{fig:dc_resistivity}(a) and \ref{fig:dc_resistivity}(b) is not universal. However, on the right-hand side of the QCP, the results should be qualitatively the same since eventually Fermi-liquid physics dominate. That is effectively what we observe so far.

Another interesting question is the effect of disorder on the importance of vertex corrections. In our case, as mentioned in section \ref{sec:cond_tpsc}, if the vertices do not depend on wave vector, i.e. they are isotropic, they have no effect on the conductivity. Adding disorder would probably make the vertices more isotropic, and therefore their effect should be smaller. It would be very interesting to see if, by adding enough disorder, we could reduce the effect of vertex corrections to the point where the resistivity recovers its insulating behavior in the pseudogap region. This would also provide one possible explanation for the fact that, in less clean cuprates, such as La$_{2-x}$Sr$_x$Cu$_2$O$_4$ the resistivity increases below $T^*$, while it increases in cleaner systems such as YBa$_2$Cu$_3$O$_{7-y}$. That contrasting behavior can be seen in Ref.~\onlinecite{Ando:2004} when, for the latter compound, we take the pseudogap line to end at optimal doping. However, from our preliminary results with hoppings $t'$ and $t''$ relevant for cuprates, we have noticed that, on the hole-doped side, the qualitative behavior with and without vertex corrections can be inverted with respect to the results of Figs. \ref{fig:dc_resistivity}(a) and \ref{fig:dc_resistivity}(b). That is, the resistivity without vertex corrections decreases below $T^*$,  while it increases with vertex corrections. By changing the band parameters and the doping we can therefore change dramatically the transport properties in the pseudogap regime.

\section{Conclusion}\label{sec:conclusion}

To satisfy current conservation, conductivity calculations must include vertex corrections that are consistent with the self-energy. A systematic way of achieving this proceeds with functional derivatives with respect to the vector potential. We have shown how this approach can be generalized to the non-perturbative TPSC approach. The various terms of the resulting algebraic expression, Eq.\eqref{eq:chi_jj_qn1}, have the physical interpretation illustrated schematically in Fig.\ref{fig:chijj}. One type of vertex correction has the structure of the Maki-Thompson term in fluctuation superconductivity, while the other has the structure of the Aslamasov-Larkin term. These diagrams contain many elements that are not computed perturbatively, as for example the irreducible spin and charge vertices, and the vertex corrections. With this approach, the \textit{f}-sum-rule is satisfied in principle exactly. We verified the agreement to the accuracy of the numerical calculations (a typical relative precision of $10^{-7}$).

The numerical evaluation of the conductivity with vertex corrections can be done only if FFT and other advanced numerical algorithms are employed. Brute force calculations are impossible with any kind of computing resource presently available. Analytical continuation is performed with a specialized maximum entropy technique that we have described in detail in appendices, along with all other algorithms.

We have shown that our approach allows us to compute the optical conductivity and DC resistivity of the nearest-neighbor two-dimensional one-band Hubbard model in a variety of regimes, without adjustable parameter or phenomenological assumption. There is no need to assume the existence of quasiparticles, as is the case in Boltzmann equation approaches.

For illustrative purposes, we have presented the results of calculations for $U=6t$ and nearest neighbor hoppings only. For the DC resistivity, we find at high temperature that it saturates at the Ioffe-Regel value, namely when the mean-free path is of the order of the inter lattice spacing. The existence of this Ioffe-Regel limit is usually assumed on phenomenological grounds. Here we have demonstrated it. That limit may be exceeded in strong coupling\cite{Palsson:2001,Terletska:2010}, but that is beyond the TPSC approach. Vertex corrections have a dramatic influence for dopings smaller than the quantum critical doping. They can change the temperature dependence of the resistivity from insulating to metallic when the antiferromagnetic correlation length becomes larger than the thermal de Broglie wave length, namely in the pseudogap regime. At the quantum critical point, the resistivity is linear at low temperature, although vertex corrections tend to reduce the temperature range of the linearity compared to the calculation with the bubble only. At low temperature, the linear term persists at dopings larger than the quantum critical point, although we cannot exclude that it disappears at temperatures lower than those accessible to us. The coefficient of this linear term is also correlated with the vanishing of the superconducting $T_c$ obtained with the TPSC method, in qualitative agreement with experimental results on the cuprates, the pnictides and the organics.\cite{Doiron-Leyraud:2009_1,Doiron-Leyraud:2009} For dopings equal or greater than the critical doping, the resistivity with all vertex corrections is always smaller than the simple bubble result. In general, for most of the dopings and temperatures considered, the first vertex correction has the tendency to increase the resistivity, while the second has the opposite effect. Therefore the results of each type of term are always very different and one cannot neglect the Aslamasov-Larkin-like contribution in those regimes. The latter contribution vanishes only when there is exact particle-hole symmetry.

We observe in the optical conductivity that vertex corrections are important at all the dopings considered and that no term can be neglected. The effect is the strongest for dopings smaller than the critical doping near and below the pseudogap temperature. We also observe that the frequency at which the results with and without vertex corrections cease to differ decrease with increasing doping. The hump structure in the mid-infrared frequency range, related to the pseudogap, is observed both with and without vertex corrections, but with different amplitudes at a given temperature. At the quantum critical point and beyond, the effect is important both at high and low temperature though more important at low temperature. The low frequency part of the conductivity is higher with the vertex corrections and increases quite rapidly with doping.

This work is presently being extended in several directions. For example, one can study more realistic band parameters for high-temperature superconductors and investigate the connection between the single-particle scattering rate along the Fermi surface and the temperature dependence of the resistivity. The sensitivity of the resistivity to the details of the model in the pseudogap regime would also be interesting to investigate. Using a similar approach, one can also envisage calculating the thermopower and other transport properties.

\section*{ACKNOWLEDGMENTS}
We acknowledge Patrick S\'emon for useful discussions, especially about the Fourier transforms of splines. A.-M.S.T. also acknowledges Louis Taillefer for discussions about experiment and S. Okamoto and J. Carbotte about Appendix~\ref{sec:Ward}. Some of these discussions occured at the Aspen Center for Physics. We are grateful to L.-F. Arsenault for reading the manuscript. This work was partially supported by NSERC and by the Tier I Canada Research Chair Program (A.-M.S.T.). Computational resources were provided by CFI, MELS, the RQCHP, and Compute Canada.

\appendix

\section{\textit{f}-sum rule for the conductivity}\label{sec:Appendix_f_sum_rule}
In this appendix, we derive the general expression for the real part of the conductivity, Eq.~\eqref{eq:Re_sigma}, and we show that the zero-Matsubara frequency value of the current-current correlation function suffices to check numerically the validity of the \textit{f}-sum rule.
We begin from the continuity equation,
\begin{equation}
\frac{\partial \rho (\mathbf{r},t)}{\partial t}+\diverg \mathbf{j}(\mathbf{r},t)=0\,,
\end{equation}
which, in Fourier space, reads
\begin{equation}
-\omega \rho (\mathbf{q},\omega)+\mathbf{q}\cdot \mathbf{j}(\mathbf{q},\omega)=0\,.
\end{equation}
If $\mathbf{j}(\mathbf{q},\omega)=j_x(\mathbf{q},\omega)\hat{x}$, we have
\begin{equation}\label{eq:eq_contin_q_omega}
q_x j_x(\mathbf{q},\omega)=\omega \rho (\mathbf{q},\omega)\,.
\end{equation}
Using space translational invariance, the two-particle spectral function corresponding to an observable $A$ can be written formally as
\begin{equation}
\chi_{AA}^{\prime\prime}(\mathbf{q},\omega)=\frac{1}{N\top} \langle [A(\mathbf{q},\omega), A(-\mathbf{q},-\omega)]\rangle\,.
\end{equation}
with $\top\rightarrow\infty$, so that, from the continuity equation Eq.\eqref{eq:eq_contin_q_omega} we have the relation between current and charge correlation functions
\begin{equation}
\chi_{j_xj_x}^{\prime\prime}(\mathbf{q},\omega)=\frac{\omega^2}{q_x^2}\chi_{\rho\rho}^{\prime\prime}(\mathbf{q},\omega)\,,
\end{equation}
and thus,
\begin{equation}
\int \frac{d\omega}{\pi} \frac{\chi_{j_xj_x}^{\prime\prime}(\mathbf{q},\omega)}{\omega}=\frac{1}{q_x^2}\int \frac{d\omega}{\pi} \omega\chi_{\rho\rho}^{\prime\prime}(\mathbf{q},\omega)\,.
\end{equation}
The right-hand side can be obtained from equal-time commutators since
\begin{equation}
\int \frac{d\omega}{\pi} \omega\chi_{\rho\rho}^{\prime\prime}(\mathbf{q},\omega)=\left(i\frac{\partial }{\partial t}\int \frac{d\omega}{2\pi} e^{-i\omega t}2\chi_{\rho\rho}^{\prime\prime}(\mathbf{q},\omega) \right)\Big|_{t=0}\,
\end{equation}
and, by definition,
\begin{equation}
\int \frac{d\omega}{2\pi} e^{-i\omega t}2\chi_{\rho\rho}^{\prime\prime}(\mathbf{q},\omega)=\frac{1}{N} \langle[\rho(\mathbf{q},t),\rho(-\mathbf{q},0)]\rangle\,,
\end{equation}
so that
\begin{equation}
\begin{split}
\int \frac{d\omega}{\pi} \frac{\chi_{j_xj_x}^{\prime\prime}(\mathbf{q},\omega)}{\omega}&=\frac{1}{q_x^2} \frac{1}{N}\langle[i\frac{\partial }{\partial t}\rho(\mathbf{q},t),\rho(-\mathbf{q},0)]\rangle\Big|_{t=0}\\
&=\frac{1}{q_x^2} \frac{1}{N}\left\langle\big[ [\rho(\mathbf{q}),H](t),\rho(-\mathbf{q},0)\big]\right\rangle\Big|_{t=0}\,.
\end{split}
\end{equation}
Taking for $H$ the Hubbard hamiltonian and calculating the commutators, we get
\begin{equation}
\begin{split}
\int \frac{d\omega}{\pi} \frac{\chi_{j_xj_x}^{\prime\prime}(\mathbf{q},\omega)}{\omega}&=\frac{1}{q_x^2} \frac{1}{N}\sum_{\mathbf{k}\sigma}(\epsilon_{\mathbf{k}+\mathbf{q}}+\epsilon_{\mathbf{k}-\mathbf{q}}-2\epsilon_{\mathbf{k}})\mn{n_{\mathbf{k}\sigma}}\,.
\end{split}
\end{equation}
We are interested in the long wave length limit, namely $\mathbf{q}\rightarrow 0$, so that
\begin{multline}
\epsilon_{\mathbf{k}+\mathbf{q}}\approx \epsilon_{\mathbf{k}}+\frac{\partial \epsilon_{\mathbf{k}}}{\partial k_x}q_x+\frac{\partial \epsilon_{\mathbf{k}}}{\partial k_y}q_y\\
+\frac{1}{2}\frac{\partial^2 \epsilon_{\mathbf{k}}}{\partial k_x^2}q_x^2+\frac{1}{2}\frac{\partial^2 \epsilon_{\mathbf{k}}}{\partial k_y^2}q_y^2+\frac{\partial^2 \epsilon_{\mathbf{k}}}{\partial k_x\partial k_y}q_xq_y\,
\end{multline}
and, if we consider the longitudinal conductivity $\mathbf{q}=q_x\hat{x}$, we obtain
\begin{equation}\label{eq:sum_rule_chijj}
\begin{split}
\int \frac{d\omega}{\pi} \frac{\chi_{j_xj_x}^{\prime\prime}(q_x,\omega)}{\omega}&= \frac{1}{N}\sum_{\mathbf{k}\sigma}\frac{\partial^2 \epsilon_{\mathbf{k}}}{\partial k_x^2}\mn{n_{\mathbf{k}\sigma}}\\
&=-\langle k_x \rangle\,,
\end{split}
\end{equation}
using the earlier definition Eq.~\eqref{eq:def_kx0}.

To derive the general results for the real part of the conductivity Eq.~\eqref{eq:Re_sigma}, we begin with the spectral form for $\chi_{j_xj_x}(q_x,\omega)$ that reads
\begin{equation}
\chi_{j_xj_x}(q_x,\omega)=\int \frac{d\omega'}{\pi} \frac{\chi_{j_xj_x}^{\prime\prime}(q_x,\omega)}{\omega'-\omega-i\eta}\,,
\end{equation}
so that, from Eq.\eqref{eq:def_cond} and Eq.\eqref{eq:sum_rule_chijj}, we have
\begin{equation}\label{eq:derivation_Re_sigma_chijj}
\begin{split}
\sigma_{xx}(q_x,\omega)&=\frac{\left\langle k_x \right\rangle+\chi_{j_xj_x}(\omega)}{i(\omega+i\eta)}\\
&=\frac{1}{i(\omega+i\eta)}\left(-\int \frac{d\omega'}{\pi} \frac{\chi_{j_xj_x}^{\prime\prime}(q_x,\omega')}{\omega'}\right.\\
&\qquad\left.+\int \frac{d\omega'}{\pi} \frac{\chi_{j_xj_x}^{\prime\prime}(q_x,\omega')}{\omega'-\omega-i\eta}\right)\\
&=\frac{1}{i(\omega+i\eta)}\int \frac{d\omega'}{\pi} \frac{(\omega+i\eta)\chi_{j_xj_x}^{\prime\prime}(q_x,\omega')}{\omega'(\omega'-\omega-i\eta)}\\
&=\frac{1}{i}\int \frac{d\omega'}{\pi} \frac{\chi_{j_xj_x}^{\prime\prime}(q_x,\omega')}{\omega'(\omega'-\omega-i\eta)}
\end{split}
\end{equation}
and since
\begin{equation}
\frac{1}{\omega'-\omega-i\eta}=P\frac{1}{\omega'-\omega}+i\pi\delta(\omega'-\omega)\,,
\end{equation}
we obtain the desired result
\begin{equation}\label{eq:Re_sigma_chijj}
Re\, \sigma_{xx}(q_x,\omega)=\frac{\chi_{j_xj_x}^{\prime\prime}(q_x,\omega)}{\omega}\,.
\end{equation}
Substituting in the form found earlier Eq.\eqref{eq:sum_rule_chijj}, the \textit{f}-sum rule for the conductivity is
\begin{equation}\label{eq:sum_rule_sigma}
\begin{split}
\int_{-\infty}^{\infty} \frac{d\omega}{\pi} Re\, \sigma_{xx}(q_x,\omega)=\frac{1}{N}\sum_{\mathbf{k}\sigma}\frac{\partial^2 \epsilon_{\mathbf{k}}}{\partial k_x^2}\mn{n_{\mathbf{k}\sigma}}\,,
\end{split}
\end{equation}
for $q_x\rightarrow 0$. Since the spectral form for $\chi_{j_xj_x}(q_x,iq_n)$ is
\begin{equation}
\chi_{j_xj_x}(q_x,iq_n)=\int_{-\infty}^{\infty} \frac{d\omega}{\pi} \frac{\chi_{j_xj_x}^{\prime\prime}(q_x,\omega)}{\omega-iq_n}\,,
\end{equation}
we have the desired result,
\begin{equation}\label{eq:sum_rule_chijj_Matsubara}
\chi_{j_xj_x}(q_x,iq_n=0)=\frac{1}{N}\sum_{\mathbf{k}\sigma}\frac{\partial^2 \epsilon_{\mathbf{k}}}{\partial k_x^2}\mn{n_{\mathbf{k}\sigma}}\,.
\end{equation}
It is thus sufficient to look at the zero Matsubara frequency value of $\chi_{j_xj_x}$ to check whether the sum rule is satisfied (assuming that $\langle k_x \rangle$ has been calculated).

Since the results Eq.\eqref{eq:sum_rule_sigma}, or equivalently Eq.\eqref{eq:sum_rule_chijj_Matsubara}, are a consequence of the continuity equation, it means this sum rule must be respected when there is conservation of particle number.

\section{Ward Idendity}\label{sec:Ward}

In this appendix, we derive the general Ward identity that follows from charge conservation. It is far too complicated to be verified in full generality numerically within our approach. We also indicate that the Ward identity suffices to find the vertex correction simply in cases where frequency and momentum variations are smooth. These assumptions are not fulfilled in our case hence these simplifications cannot be used.

We begin from linear response theory in imaginary time. We omit the diamagnetic term, which is not relevant for the present discussion. We find,%
\begin{equation}
\left\langle \mathbf{j}_{\mathbf{q}}\left(  \tau\right)  \right\rangle
=\int_{0}^{\beta}d\tau^{\prime}\left\langle T_{\tau}\hat{\mathbf{j}}_{\mathbf{q}}\left(  \tau\right)  \hat{\mathbf{j}}_{-\mathbf{q}}\left(  \tau^{\prime}\right)  \right\rangle \cdot\mathbf{A}_{\mathbf{q}}\left(\tau^{\prime}\right)
\end{equation}
where $\hat{\mathbf{j}}$ is the paramagnetic current in the interaction representation. Take a single (bosonic) Matsubara frequency for the vector potential%
\begin{equation}
\mathbf{A}_{\mathbf{q}}\left(  \tau^{\prime}\right)  =T\mathbf{A}_{\mathbf{q}}\left(  q_{m}\right)  e^{-iq_{m}\tau^{\prime}}
\end{equation}
and extract the corresponding Matsubara frequency for the current, then the quantity to evaluate is
\begin{widetext}
\begin{equation}\label{Current in terms of four point function}
\begin{split}
\left\langle \mathbf{j}_{\mathbf{q}}\left(  q_{m}\right)  \right\rangle=&T\int_{0}^{\beta}d\tau\int_{0}^{\beta}d\tau^{\prime}e^{iq_{m}\left(  \tau-\tau^{\prime}\right)}\left\langle T_{\tau}\hat{\mathbf{j}}_{\mathbf{q}}\left(  \tau\right)\hat{\mathbf{j}}_{-\mathbf{q}}\left(  \tau^{\prime}\right)  \right\rangle\cdot\mathbf{A}_{\mathbf{q}}\left(  q_{m}\right)\\
=&T\int_{0}^{\beta}d\tau\int_{0}^{\beta}d\tau^{\prime}e^{iq_{m}\left(  \tau-\tau^{\prime}\right)}\frac{1}{N}\sum_{\mathbf{k,\sigma}}\sum_{\mathbf{k}^{\prime}}\mathbf{\nabla}_{\mathbf{k}}\varepsilon_{\mathbf{k}}\\
&\times\left\langle T_{\tau}c_{\mathbf{k\sigma}}^{\dagger}\left(  \tau\right)  c_{\mathbf{k+q\sigma}}\left(  \tau\right) c_{\mathbf{k}^{\prime}+\mathbf{q\sigma}}^{\dagger}\left(  \tau^{\prime}\right)  c_{\mathbf{k}^{\prime}\mathbf{\sigma}}\left(\tau^{\prime}\right)\right\rangle\mathbf{\nabla}_{\mathbf{k}^{\prime}}\varepsilon_{\mathbf{k}^{\prime}}\cdot\mathbf{A}_{\mathbf{q}}\left(  q_{m}\right)
\end{split}
\end{equation}
where, in the last equality, we have also used spin conservation with the fact that only the connected piece will contribute because the average current in equilibrium vanishes.

The Ward identity that we need can be obtained from the single spin component version of following equality that can be derived from current conservation,
\cite{Dare:1994}
\begin{equation}
\begin{split}
&\sum_{\mathbf{k}}\left[  \frac{\partial}{\partial\tau}+\left(\varepsilon_{\mathbf{k+q}}-\varepsilon_{\mathbf{k}}\right)  \right]\left\langle T_{\tau}c_{\mathbf{k\sigma}}^{\dagger}\left(  \tau\right)c_{\mathbf{k+q\sigma}}\left(  \tau\right) c_{\mathbf{k}^{\prime}+\mathbf{q\sigma}}^{\dagger}\left(  \tau_{1}\right)  c_{\mathbf{k}^{\prime}\mathbf{\sigma}}\left(\tau_{2}\right)  \right\rangle \\
&\qquad=\delta\left(  \tau-\tau_{1}\right)  G_{\mathbf{k}^{\prime}\sigma}\left(\tau_{2}-\tau\right)  -\delta\left(  \tau-\tau_{2}\right)  G_{\mathbf{k}^{\prime}+\mathbf{q}\sigma}\left(  \tau-\tau_{1}\right) \,.
\end{split}
\end{equation}
Current conservation, that we saw in the first two equations of appendix \ref{sec:Appendix_f_sum_rule}, would give a vanishing right-hand side were it not for the theta functions whose derivatives give delta functions and ultimately equal-time commutators that can be evaluated. Define the fermionic Matsubara frequency four-point correlation function
\begin{equation}\label{Four point function}
\begin{split}
&  \Lambda_{\sigma}\left(  \mathbf{k,k+q;}k_{m},\mathbf{k}^{\prime}+\mathbf{q;}k_{n}^{\prime},\mathbf{k}^{\prime}\right)  \\
&  \qquad\equiv\int_{0}^{\beta}d\tau_{1}\int_{0}^{\beta}d\tau_{2} e^{ik_{m}\left(  \tau-\tau_{1}\right)  }e^{ik_{n}^{\prime}\left(  \tau_{2}-\tau\right)  }\left\langle T_{\tau}c_{\mathbf{k\sigma}}^{\dagger}\left(\tau\right)  c_{\mathbf{k+q\sigma}}\left(  \tau\right)c_{\mathbf{k}^{\prime}+\mathbf{q\sigma}}^{\dagger}\left(  \tau_{1}\right)  c_{\mathbf{k}^{\prime}\mathbf{\sigma}}\left(\tau_{2}\right) \right\rangle .
\end{split}
\end{equation}
Then, taking the same fermionic components of the Ward identity, integrating by parts, it takes the form
\begin{equation}\label{Ward General}
\begin{split}
&  \sum_{\mathbf{k}}\left[  \left(  -ik_{m}+ik_{n}^{\prime}\right)  +\left(\varepsilon_{\mathbf{k+q}}-\varepsilon_{\mathbf{k}}\right)  \right] \Lambda_{\sigma}\left(  \mathbf{k,k+q;}k_{m},\mathbf{k}^{\prime}+\mathbf{q;}k_{n}^{\prime},\mathbf{k}^{\prime}\right)   =G_{\mathbf{k}^{\prime}\sigma}\left(  k_{n}^{\prime}\right)  -G_{\mathbf{k}^{\prime}+\mathbf{q}\sigma}\left(  k_{m}\right)  .
\end{split}
\end{equation}
We need the amputated function that is summed over all wave vectors to compute the current-current correlation function. So let us define the charge and current three point vertices (for a single spin component), valid in the small $\mathbf{q}$ limit
\begin{align}
&  \Gamma_{\rho}\left(  k_{m},\mathbf{k}^{\prime}+\mathbf{q;}k_{n}^{\prime},\mathbf{k}^{\prime}\right)   =-\sum_{\mathbf{k}}\Lambda_{\sigma}\left(  \mathbf{k,k+q;}k_{m}%
,\mathbf{k}^{\prime}+\mathbf{q;}k_{n}^{\prime},\mathbf{k}^{\prime}\right)G_{\mathbf{k}^{\prime}\sigma}^{-1}\left(  k_{n}^{\prime}\right)
G_{\mathbf{k}^{\prime}+\mathbf{q}\sigma}^{-1}\left(  k_{m}\right)  \\
&  \mathbf{q\cdot\Gamma}_{J}\left(  k_{m},\mathbf{k}^{\prime}+\mathbf{q;}k_{n}^{\prime},\mathbf{k}^{\prime}\right)   =-\mathbf{q\cdot}\sum_{\mathbf{k}}\mathbf{\nabla}_{\mathbf{k}}\varepsilon_{\mathbf{k}}\Lambda_{\sigma}\left(  \mathbf{k,k+q;}k_{m},\mathbf{k}^{\prime}+\mathbf{q;}k_{n}^{\prime},\mathbf{k}^{\prime}\right)
G_{\mathbf{k}^{\prime}\sigma}^{-1}\left(  k_{n}^{\prime}\right)G_{\mathbf{k}^{\prime}+\mathbf{q}\sigma}^{-1}\left(  k_{m}\right)\label{Current vertex}.
\end{align}
Then, in the long wave length limit, the general Ward identity Eq.(\ref{Ward General}) can be rewritten for the three-point functions as follows
\begin{equation}\label{Ward three point}
\begin{split}
\left(  ik_{m}-ik_{n}^{\prime}\right) \Gamma_{\rho}-\mathbf{q\cdot\Gamma}_{J}=&G_{\mathbf{k}^{\prime}+\mathbf{q}\sigma}^{-1}\left(  k_{m}\right)
-G_{\mathbf{k}^{\prime}\sigma}^{-1}\left(  k_{n}^{\prime}\right)\\
=&\left(  ik_{m}-ik_{n}^{\prime}\right)  -\mathbf{q\cdot\nabla}_{\mathbf{k}^{\prime}}\varepsilon_{\mathbf{k}^{\prime}}-\Sigma_{\mathbf{k}^{\prime}\mathbf{+q}\sigma}\left(  k_{m}\right)  +\Sigma_{\mathbf{k}^{\prime}\sigma}\left(  k_{n}^{\prime}\right)
\end{split}
\end{equation}

We can obtain the four-point function that we need in the expression for the current Eq.(\ref{Current in terms of four point function}) from the fermionic Matsubara expression Eq.(\ref{Four point function}) for the four point function at $\tau_{1}=\tau_{2}=\tau^{\prime}$ as follows
\begin{equation}
\begin{split}
\left\langle \mathbf{j}_{\mathbf{q}}\left(  q_{m}\right)  \right\rangle  &=T\int_{0}^{\beta}d\tau\int_{0}^{\beta}d\tau^{\prime}%
e^{iq_{m}\left(  \tau-\tau^{\prime}\right)  }\frac{1}{N}\sum_{\mathbf{k,\sigma}}\sum_{\mathbf{k}^{\prime}}T\sum_{k_{m}}T\sum_{k_{n}^{\prime}}e^{-ik_{m}\left(  \tau-\tau^{\prime}\right)  }e^{-ik_{n}^{\prime}\left(\tau^{\prime}-\tau\right)  }\\
&\qquad\qquad\times\mathbf{\nabla}_{\mathbf{k}}\varepsilon_{\mathbf{k}}\Lambda_{\sigma}\left(  \mathbf{k,k+q;}k_{m},\mathbf{k}^{\prime}+\mathbf{q;}k_{n}^{\prime},\mathbf{k}^{\prime}\right)  \mathbf{\nabla}_{\mathbf{k}^{\prime}}\varepsilon_{\mathbf{k}^{\prime}}\cdot\mathbf{A}_{\mathbf{q}}\left(  q_{m}\right)\\
&=\frac{1}{N}\sum_{\mathbf{k,\sigma}}\sum_{\mathbf{k}^{\prime}}T\sum_{k_{n}^{\prime}}\mathbf{\nabla}_{\mathbf{k}}\varepsilon_{\mathbf{k}}\Lambda_{\sigma}\left(\mathbf{k,k+q;}k_{n}^{\prime}+q_{m},\mathbf{k}^{\prime}+\mathbf{q;}k_{n}^{\prime},\mathbf{k}^{\prime}\right)  \mathbf{\nabla}_{\mathbf{k}^{\prime}}\varepsilon_{\mathbf{k}^{\prime}}\cdot\mathbf{A}_{\mathbf{q}}\left(q_{m}\right)\,.
\end{split}
\end{equation}
The sum over wave vectors $\mathbf{k}$\textbf{ }allows one to rewrite the latter in terms of the current vertex defined in Eq.(\ref{Current vertex})%
\begin{equation}
\left\langle \mathbf{j}_{\mathbf{q=0}}\left(  q_{m}\right)  \right\rangle =-\frac{2}{N}\sum_{\mathbf{k}^{\prime}}T\sum_{k_{n}^{\prime}}  \mathbf{\Gamma}_{J}\left(  k_{n}^{\prime}+q_{m},\mathbf{k}^{\prime}\mathbf{;}k_{n}^{\prime},\mathbf{k}^{\prime}\right)  G_{\mathbf{k}^{\prime}}\left(  k_{n}^{\prime}\right)  G_{\mathbf{k}^{\prime}}\left(  k_{n}^{\prime}+q_{m}\right)  \mathbf{\nabla}_{\mathbf{k}^{\prime}}\varepsilon_{\mathbf{k}^{\prime}}\cdot\mathbf{A}_{\mathbf{q=0}}\left(  q_{m}\right)  .
\end{equation}
We have performed the sum over spins, which explains the factor two, and taken the $\mathbf{q=0}$ limit first to represent a constant electric field.
\end{widetext}

There is a simple case where the Ward identity suffices to find the vertex correction. Consider the Ward identity for the three point function Eq.(\ref{Ward three point}). The rest of this Appendix is correct only if we can assume that in the $\mathbf{q\rightarrow0}$ finite frequency $ik_{m}-ik_{n}^{\prime}$ limit, the charge vertex $\Gamma_{\rho}\left(k_{m},\mathbf{k}^{\prime}+\mathbf{q;}k_{n}^{\prime},\mathbf{k}^{\prime}\right)  $ and current vertex $\Gamma_{J}\left(  k_{m},\mathbf{k}^{\prime}+\mathbf{q;}k_{n}^{\prime},\mathbf{k}^{\prime}\right)  $ are analytical in $\mathbf{q}$ and have a finite limit at $\mathbf{q=0}$. It can be checked that this is the case for the non-interacting system. In that case, all the non-analyticities are contained in the product of Green's functions appearing in our last expression for $\left\langle \mathbf{j}_{\mathbf{q=0}}\left(q_{m}\right)  \right\rangle .$ While these kinds of analytic properties can be assumed for Fermi liquids, this is \textit{not} appropriate in our case, that is more singular in the presence of strong antiferromagnetic fluctuations.

In the case where we can assume analyticity, the expansion for $\Gamma_{\rho}$ in powers of $\mathbf{q}$\textbf{ }must begin at order $q^{2}$ because it is a scalar. Then, the long wave length limit of the vertices can be found by identifying the coefficients of the $\left(  ik_{m}-ik_{n}^{\prime}\right)$ and of the $\mathbf{q}$ on the left and right-hand side of the Ward identity for the three point function Eq.(\ref{Ward three point}). For the charge three point function we thus find
\begin{equation}
\Gamma_{\rho}\left(  k_{m},\mathbf{k}^{\prime}\mathbf{;}k_{n}^{\prime
},\mathbf{k}^{\prime}\right)  =1-\frac{\Sigma_{\mathbf{k}^{\prime}\sigma
}\left(  k_{m}\right)  -\Sigma_{\mathbf{k}^{\prime}\sigma}\left(
k_{n}^{\prime}\right)  }{\left(  ik_{m}-ik_{n}^{\prime}\right)  }.
\end{equation}
while for the current vertex, taking the $\mathbf{q\rightarrow0}$ limit, one obtains
\begin{equation}\label{eq:Gamma_J_Ward}
\Gamma_{J}\left(  k_{m},\mathbf{k}^{\prime}\mathbf{;}k_{n}^{\prime}%
,\mathbf{k}^{\prime}\right)  =\mathbf{\nabla}_{\mathbf{k}^{\prime}}%
\varepsilon_{\mathbf{k}^{\prime}}+\mathbf{\nabla}_{\mathbf{k}^{\prime}}%
\Sigma_{\mathbf{k}^{\prime}\sigma}\left(  k_{m}\right)  \text{ }.
\end{equation}
The first term in the last two equations is the bare vertex and the last term the vertex correction. These results for the vertex corrections are valid, for example, for impurity scattering or electron-phonon interactions where the assumptions upon which they were derived are valid,\cite{Mahan:3rd} but not in our case where the gradient of the self-energy with respect to wave vector can be very large, for example near hot spots. The above two equations by themselves are often called the Ward identities.\cite{Mahan:3rd}

Note that, at zero external Matsubara frequency, as long as the charge vertex does not diverge, Eq. \eqref{eq:Gamma_J_Ward} becomes exact for $k_m=k_n'$. This form for the vertex generates ladder diagrams only, demonstrating that the Aslamasov-Larkin contribution vanishes, as found numerically in Sec. \ref{sec:f-sum-rule_res}.

\section{Fast-Fourier transforms, cubic splines and asymptotic expansions}\label{sec:FFT_splines_HF_exp}

The use of fast-Fourier transforms (FFTs) is absolutely essential for this calculation. We begin by describing their use for obtaining the Lindhard function, Eq.\eqref{eq:Lindhard}, and the self-energy, where the convolutions that render the use of FFTs possible are apparent. The convolutions are not so apparent in the case of the conductivity, particularly the Aslamasov-Larkin like terms, that require an elaborate discussion in the third subsection.

\subsection{The Lindhard function}\label{sec:Lindhard}

The Lindhard function Eq.\eqref{eq:Lindhard} is needed to compute the spin and charge susceptibilities Eq.\eqref{eq:chi_sp_q} and Eq.\eqref{eq:chi_ch_q}, which is defined in a more explicit way as
\begin{equation}\label{eq:Lindhard_2}
\chi_{0}(\mathbf{q},iq_n)=-2\frac{T}{N}\sum_{k,ik_n} G^{(1)}(\mathbf{k}+\mathbf{q}, ik_n+iq_n)G^{(1)}(\mathbf{k}, ik_n)\,,
\end{equation}
with
\begin{equation}
G^{(1)}(\mathbf{k}, ik_n)=\frac{1}{ik_n-\epsilon_\mathbf{k}+\mu_0}
\end{equation}
where
$\epsilon_\mathbf{k}$ is the bare particle dispersion relation, $k_n=(2n+1)\pi T$ is a Fermionic Matsubara frequency and $\mu_0$ is the bare chemical potential. Expression \eqref{eq:Lindhard_2} could be calculated by first performing analytically the sum over Matsubara frequencies to obtain
\begin{equation}\label{eq:chi0_FermiDirac}
\chi_0(\mathbf{q},iq_n)=-\frac{2}{N}\sum_{\mathbf{k}} \frac{f(\xi_\mathbf{k})-f(\xi_{\mathbf{k}+\mathbf{q}})}{iq_n+\xi_\mathbf{k}-\xi_{\mathbf{k}+\mathbf{q}}}\,,
\end{equation}
where $\xi_\mathbf{k}=\epsilon_{\mathbf{k}}-\mu_0$ and $f(\xi_\mathbf{k})$ is the Fermi-Dirac distribution, and then integrating numerically over $\mathbf{k}$ for each value of $(\mathbf{q},iq_n)$. While it may seem that we have saved some work by doing exactly the Matsubara sum, by doing so we do not use a property of Eq.\eqref{eq:Lindhard_2} that can make our calculation much easier. This is the fact that this expression is a convolution, and convolutions can be calculated in a very efficient way using fast-Fourier transforms (FFT). For instance, Eq.\eqref{eq:Lindhard_2} can be written as
\begin{multline}\label{eq:chi0_TF}
\chi_{0}(\mathbf{q},iq_n)=\\
-2 \int_0^\beta d\tau e^{iq_n\tau} \sum_{\mathbf{r}} e^{-i\mathbf{q}\cdot \mathbf{r}}G^{(1)}(\mathbf{r},\tau)G^{(1)}(-\mathbf{r},-\tau)\,,
\end{multline}
where $\beta=T^{-1}$.

To compute $\chi_0(\mathbf{q},iq_n)$ on a grid of size $N_\mathbf{q}N_{q_n}$ using the form where Matsubara sums have been done, Eq.\eqref{eq:chi0_FermiDirac}, one needs to do this number of integrals over the Brillouin zone, a number of operations that scales like $N_\mathbf{q}^2N_{q_n}$ if we consider that each integral scales like $N_\mathbf{q}$. On the other hand, using the convolution form Eq.\eqref{eq:chi0_TF}, we need to do a two-dimensional FFT on $G^{(1)}(\mathbf{k},\tau)$ to obtain $G^{(1)}(\mathbf{r},\tau)$ (here, $G^{(1)}(\mathbf{k},\tau)$ is known analytically), a task that scales like $N_\mathbf{q}N_{q_n}\log N_\mathbf{q}$, and then the three-dimensional FFT in Eq.\eqref{eq:chi0_TF} that scales like $N_\mathbf{q}N_{q_n}\log N_\mathbf{q}N_{q_n}$. We therefore have a gain proportional to $N_\mathbf{q}/\log N_\mathbf{q}N_{q_n}$.

There is one delicate point. Eq.\eqref{eq:chi0_TF} contains a continuous Fourier transform on the imaginary time $\tau$ while FFTs are discrete transforms. The simplest thing to do would be to dicretize $\tau$ and perform an ordinary FFT. This would give acceptable results for the low Matsubara frequencies but the high frequencies would be completely wrong since $\chi_{0}(\mathbf{q},iq_n)$ would be periodic in $q_n$, while it has to decrease as a series in even powers of $1/q_n$. To solve this problem we make use of cubic splines to approximate the integrand between the discrete imaginary time points and perform a continuous Fourier transform on this spline. In appendix \ref{sec:FT_spline}, we show how to compute the continuous Fourier transform of a cubic spline using in fact only a discrete Fourier transform. Let us consider first the imaginary time Fourier transform in Eq.\eqref{eq:chi0_TF}. Using the form \eqref{eq:TF_fx_spline} given in appendix \ref{sec:FT_spline}, we obtain
\begin{multline}\label{eq:chi0_TF_spline}
\chi_0(\mathbf{r},iq_n)=\frac{\chi_0'(\mathbf{r},\tau=\beta)-\chi_0'(\mathbf{r},\tau=0)}{q_n^2}\\
+\frac{1-e^{iq_n\Delta\tau}}{q_n^4}\sum_{j=0}^{N-1} S_{j+1}^{(3)}(\mathbf{r}) e^{iq_n\tau_j}\,,
\end{multline}
where $N$ is the number of intervals in the imaginary time grid, $\Delta\tau$ is the size of an interval, $S_j(\mathbf{r},\tau)$ is the cubic polynomial in the $j^{th}$ interval and $S_{j}^{(3)}(\mathbf{r})$ is the third derivative of $S_j(\mathbf{r},\tau)$.

Notice that Eq.\eqref{eq:chi0_TF_spline} contains the discrete Fourier transform of $S_{j}^{(3)}(\mathbf{r})$, so that the result of this transform itself will be periodic in $q_n$. But the factor $1/(q_n^4)$ in front makes this term important only for low frequencies.  Expression \eqref{eq:chi0_TF_spline} contains the derivatives of $\chi_0(\mathbf{r},\tau)$ with respect to $\tau$ at the boundaries. It is explained in appendix \ref{sec:FT_spline} how those derivatives are used to complete the linear system that must be solved to obtain the spline coefficients. They thus have to be calculated before the spline and must be known when Eq.\eqref{eq:chi0_TF_spline} is used.  Here since $\chi_0(\mathbf{r},\tau)=-2G^{(1)}(\mathbf{r},\tau)G^{(1)}(-\mathbf{r},-\tau)$, and we have analytical expressions for $G^{(1)}(\mathbf{k},\tau)$, it is straightforward to calculate this derivative. For $0<\tau<\beta$, we have
\begin{equation}
\begin{split}
G^{(1)}(\mathbf{k},\tau)&=-e^{-\xi_{\mathbf{k}}\tau} f(-\xi_{\mathbf{k}})\\
&=-e^{\xi_{\mathbf{k}}(\beta-\tau)} f(\xi_{\mathbf{k}})\,,
\end{split}
\end{equation}
so that
\begin{equation}\label{eq:dtau_Grtau}
\begin{split}
\frac{\partial G^{(1)}(\mathbf{r},\tau)}{\partial\tau}&=\frac{1}{N}\sum_{\mathbf{k}} e^{i\mathbf{k}\cdot\mathbf{r}}\xi_{\mathbf{k}}\,e^{-\xi_{\mathbf{k}}\tau}f(-\xi_{\mathbf{k}})\\
&=\frac{1}{N}\sum_{\mathbf{k}} e^{i\mathbf{k}\cdot\mathbf{r}}\xi_{\mathbf{k}}\,e^{\xi_{\mathbf{k}}(\beta-\tau)}f(\xi_{\mathbf{k}})\,.
\end{split}
\end{equation}
Then, still for $0<\tau<\beta$,
\begin{equation}
G^{(1)}(\mathbf{k},-\tau)=e^{\xi_{\mathbf{k}}\tau} f(\xi_{\mathbf{k}})
\end{equation}
and, therefore,
\begin{equation}
\frac{\partial G^{(1)}(-\mathbf{r},-\tau)}{\partial\tau}=\frac{1}{N}\sum_{\mathbf{k}} e^{-i\mathbf{k}\cdot\mathbf{r}}\xi_{\mathbf{k}} e^{\xi_{\mathbf{k}}\tau} f(\xi_{\mathbf{k}})\,.
\end{equation}
If we have inversion symmetry,
\begin{equation}
G^{(1)}(-\mathbf{r},-\tau)=G^{(1)}(\mathbf{r},-\tau)\,,
\end{equation}
then
\begin{equation}
\frac{\partial G^{(1)}(-\mathbf{r},-\tau)}{\partial\tau}=\frac{1}{N}\sum_{\mathbf{k}} e^{i\mathbf{k}\cdot\mathbf{r}}\xi_{\mathbf{k}} e^{\xi_{\mathbf{k}}\tau} f(\xi_{\mathbf{k}})\,,
\end{equation}
and we have
\begin{equation}
\frac{\partial G^{(1)}(-\mathbf{r},-\tau)}{\partial\tau}=-\frac{\partial G^{(1)}(\mathbf{r},\beta-\tau)}{\partial\tau}\,.
\end{equation}

\subsection{The self-energy}\label{sec:self-energy}

The next function we can calculate using FFTs is the self-energy Eq.\eqref{eq:self_2_1} that can be written as
\begin{multline}\label{eq:self_tpsc_TF_Vrtau}
\Sigma_\sigma^{(2)}(\mathbf{k}, ik_n)=Un_{-\sigma}\\
+\int d\tau e^{ik_n\tau} \sum_{\mathbf{r}} e^{-i\mathbf{k}\cdot \mathbf{r}}V(-\mathbf{r},-\tau) G_\sigma^{(1)}(\mathbf{r}, \tau)\,,
\end{multline}
where
\begin{equation}\label{eq:TF_Vrtau}
V(\mathbf{r},\tau)= \frac{U}{8}\left[3U_{sp}\chi_{sp}(\mathbf{r},\tau)+U_{ch}\chi_{ch}(\mathbf{r},\tau)\right]\,.
\end{equation}
To calculate $V(\mathbf{r},\tau)$ accurately, we can use the fact that $\chi_{sp}(\mathbf{q},iq_n)$ and $\chi_{ch}(\mathbf{q},iq_n)$ approach assymptotically  $\chi_0(\mathbf{q},iq_n)$ as $q_n$ increases and that $\chi_0(\mathbf{r},\tau)$ is known once $G_\sigma^{(1)}(\mathbf{r}, \tau)$ is. Then $V(\mathbf{r},\tau)$ is computed from
\begin{multline}\label{eq:TF_Vrtau_chi0}
V(\mathbf{r},\tau)=\\
 \frac{U}{8}\frac{T}{N}\sum_{\mathbf{q},iq_n} e^{i\mathbf{q}\cdot\mathbf{r}}e^{-iq_n\tau}\Big[ 3U_{sp}\chi_{sp}(\mathbf{q},iq_n)+U_{ch}\chi_{ch}(\mathbf{q},iq_n)\\
-\left(3U_{sp}+U_{ch}\right)\chi_{0}(\mathbf{q},iq_n)\Big]\\
+\frac{U}{8}\left(3U_{sp}+U_{ch}\right)\chi_0(\mathbf{r},\tau)\,,
\end{multline}
where $\chi_0(\mathbf{r},\tau)=-2G_\sigma^{(1)}(\mathbf{r}, \tau)G_\sigma^{(1)}(-\mathbf{r}, -\tau)$. Because the asymptotic part is removed in the Fourier transform, it converges as the transform of $1/(q_n^4)$ instead of $1/(q_n^2)$, so that a smaller cutoff can be used.

In Eq.\eqref{eq:self_tpsc_TF_Vrtau}, there is a continuous Fourier transform so that we have to use cubic splines to represent the $\tau$-dependence of the integrand. To compute the splines and then use formula \eqref{eq:TF_fx_spline} we need to compute the derivatives of $V(-\mathbf{r},-\tau) G_\sigma^{(1)}(\mathbf{r}, \tau)$ at $\tau=0$ and $\tau=\beta$, i.e. those derivatives of $V(-\mathbf{r},-\tau)$ and $G_\sigma^{(1)}(\mathbf{r}, \tau)$. The latter were already computed using Eq. \eqref{eq:dtau_Grtau} to calculate $\chi_0$. For $V(-\mathbf{r},-\tau)$, since this function is symetric with respect to $\tau=\beta/2$, we only need the derivative at $\tau=0$. If we differentiate expression \eqref{eq:TF_Vrtau_chi0} with respect to $\tau$ and set $\tau=0$, we notice that the sum disapears since $\chi_{sp}(\mathbf{q},iq_n)$ and $\chi_{ch}(\mathbf{q},iq_n)$ are even functions of $q_n$ while the derivative makes a factor $q_n$ appear in the sum. In other words, at $\tau=0$ (and thus $\tau=\beta$), the derivatives of $\chi_{sp}(\tau)$, $\chi_{ch}(\tau)$ and $\chi_0(\tau)$ are equal, which comes from the fact that they have the same asymptotic limit as can be seen from Eq.\eqref{eq:TF_fx_spline} (with $k=q_n$), considering that there are no odd terms for those functions. Therefore, we have
\begin{equation}\label{eq:dtau_Vrtau_chi0}
\frac{\partial V(\mathbf{r},\tau)}{\partial\tau}\Bigg|_{\tau=0}=\frac{U}{8}\left(3U_{sp}+U_{ch}\right)\frac{\partial \chi_0(\mathbf{r},\tau)}{\partial\tau}\Bigg|_{\tau=0}\,.
\end{equation}

Once the self-energy Eq.\eqref{eq:self_tpsc_TF_Vrtau} is obtained, we have to calculate the interacting chemical potential $\mu$ to define the Green's function $G^{(2)}$ as
\begin{equation}\label{eq:def_G2_k_ikn}
G^{(2)}(\mathbf{k},ik_n)=\frac{1}{ik_n-\epsilon_\mathbf{k}+\mu-\Sigma_\sigma^{(2)}(\mathbf{k}, ik_n)}\,.
\end{equation}
For a given filling $\mn{n}$, $\mu$ is defined implicitly by the equation
\begin{equation}
\begin{split}
\mn{n}&=2\frac{1}{N}\sum_{\mathbf{k}}  G^{(2)}(\mathbf{k},\tau=0^-)\\
&=2\frac{T}{N}\sum_{\mathbf{k},ik_n} e^{-ik_n0^-} G^{(2)}(\mathbf{k},ik_n).
\end{split}
\end{equation}
However, this expression is not very practical numerically because of the convergence factor $e^{-ik_n0^-}$ that is not well defined numerically. We use instead
\begin{equation}
2\frac{T}{N}\sum_{\mathbf{k},ik_n} \left[G^{(2)}(\mathbf{k},ik_n)-G^{(1)}(\mathbf{k},ik_n)\right]=0\,,
\end{equation}
where it is assumed that $G^{(1)}$ contains the chemical potential $\mu_0$ that gives the desired filling. No convergence factor is needed since the sum converges as the sum of $1/(k_n^2)$.

\subsection{The \textit{current-current} correlation function}\label{sec:chijj}

Finally we come to the calculation of $\chi_{j_xj_x}$, Eq. \eqref{eq:chi_jj_qn1}. The first term, called the ``bubble'' contribution because of its bubble shape in figure \ref{fig:chijj}, can be written as
\begin{equation}\label{eq:chijj_1}
\begin{split}
&\chi_{j_xj_x}^{b}(iq_n)=\\
&\frac{-2T}{N}\sum_\mathbf{k} \left(\frac{\partial \epsilon_\mathbf{k}}{\partial k_x}\right)^2 \sum_{ik_m} G^{(2)}(\mathbf{k},ik_m)G^{(2)}(\mathbf{k},ik_m+iq_n)\\
&=\frac{-2}{N}\sum_\mathbf{k} \left(\frac{\partial \epsilon_\mathbf{k}}{\partial k_x}\right)^2 \int_0^{\beta} d\tau\,e^{iq_n\tau} G^{(2)}(\mathbf{k},\tau) G^{(2)}(\mathbf{k},-\tau)\,.
\end{split}
\end{equation}
To use this formula we first have to compute $G^{(2)}(\mathbf{k},\tau)$ from $G^{(2)}(\mathbf{k},ik_n)$ defined in Eq.\eqref{eq:def_G2_k_ikn}. To do that we cannot simply perform a Fourier transform with respect to $ik_n$ on $G^{(2)}(\mathbf{k},ik_n)$ including only a finite number of frequencies. This is because $G^{(2)}(\mathbf{k},\tau)$ has a discontinuity at $\tau=0$ that will produce oscillations (the Gibbs phenomenon) close to $\tau=0$ and $\tau=\beta$ (remember that $G(\mathbf{k},\beta-\tau)=-G(\mathbf{k},-\tau)$). This problem can be solved by using the asymptotic form of $G^{(2)}(\mathbf{k},ik_n)$ to perform the transform on all frequencies. First, let us define $\bar{G}(\mathbf{k},\tau)$, the Fourier transform of some Green's function $G(\mathbf{k},ik_n)$ on a finite set of frequencies,
\begin{equation}\label{eq:finite_FT_G}
\begin{split}
\bar{G}(\mathbf{k},\tau_j)&=T\sideset{}{'}\sum_{ik_n} e^{-ik_n\tau_j} G(\mathbf{k},ik_n)\\
&=e^{i\pi (N_\tau-1)j/N_\tau}\,T\sum_{n=0}^{N_\tau-1} e^{-i2\pi nj/N_\tau} G(\mathbf{k},ik_{n-\frac{N_\tau}{2}})\,.
\end{split}
\end{equation}
where $\sideset{}{'}\sum$ means that the sum is finite, $N_\tau$ is the number of values of $\tau$ and we have used $k_n=(2n+1)\pi T$ and $\tau_j=j/(N_\tau T)$. The sum in second line has the form used in standard FFT routines. Now, to compute the self energy Eq. \eqref{eq:self_tpsc_TF_Vrtau}, we have used Fourier transforms of cubic splines so that $\Sigma^{(2)}(\mathbf{k},ik_n)$ has the form Eq.\eqref{eq:TF_fx_spline}. The asymptotic form of $G^{(2)}(\mathbf{k},ik_n)$ is therefore
\begin{widetext}
\begin{equation}\label{eq:G2_inf_k_ikn}
G^{(2)}_{inf}(\mathbf{k},ik_n)=\frac{1}{ik_m-\tilde{\epsilon}_{\mathbf{k}}-\Sigma_{inf}(\mathbf{k},ik_n)}
=\frac{1}{ik_n-\tilde{\epsilon}_{\mathbf{k}}-\left(\frac{s_1}{ik_n}+\frac{s_2(\mathbf{k})}{(ik_n)^2}+\frac{s_3(\mathbf{k})}{(ik_n)^3}\right)}
\end{equation}
where $\tilde{\epsilon}_{\mathbf{k}}=\epsilon_{\mathbf{k}}-\mu$. The Fourier transform over $ik_n$ of this function can be done analytically using the residue theorem. For a function $g(z)$ having only simple poles, we have
\begin{equation}\label{eq:sum_Res}
T\sum_{ik_m} e^{-ik_m\tau}g(ik_m)=
\begin{cases}
-\sum_j \underset{z=z_j}{Res}\left[g(z)\right]f(-z_j)e^{-z_j\tau}\,,\quad &0<\tau<\beta\,,\\
\sum_j \underset{z=z_j}{Res}\left[g(z)\right]f(z_j)e^{-z_j\tau}\,,& -\beta<\tau<0\,,
\end{cases}
\end{equation}
which, applied to \eqref{eq:G2_inf_k_ikn}, gives
\begin{multline}\label{eq:G2_inf_k_tau}
G^{(2)}_{inf}(\mathbf{k},\tau)=
\mp\sum_{j=1}^4 e^{-z_j(\mathbf{k})\tau}f(\mp z_j(\mathbf{k}))\frac{[z_j(\mathbf{k})]^3}{[z_j(\mathbf{k})-z_1(\mathbf{k})]\ldots[z_j(\mathbf{k})-z_{j-1}(\mathbf{k})][z_j(\mathbf{k})-z_{j+1}(\mathbf{k})]\ldots[z_j(\mathbf{k})-z_4(\mathbf{k})]}\,,
\end{multline}
\end{widetext}
where the minus sign is for $0<\tau<\beta$ and the plus sign, for $-\beta<\tau<0$. The $z_j(\mathbf{k})$ are the roots of the polynomial
\begin{equation}
z^4-\tilde{\epsilon}_{\mathbf{k}}z^3-s_1z^2-s_2(\mathbf{k})z-s_3(\mathbf{k})\label{eq:pol4_Ginf}\,.
\end{equation}
Those roots are given by quite imposing, but analytical formulas. Finally, assuming that $\bar{G}^{(2)}_{inf}(\mathbf{k},\tau)$ is the finite Fourier transform of $G^{(2)}_{inf}(\mathbf{k},ik_n)$ as defined in Eq.\eqref{eq:finite_FT_G}, we use
\begin{equation}\label{eq:G2_k_tau}
G^{(2)}(\mathbf{k},\tau)=\bar{G}^{(2)}(\mathbf{k},\tau)+\left[G^{(2)}_{inf}(\mathbf{k},\tau)-\bar{G}^{(2)}_{inf}(\mathbf{k},\tau)\right]
\end{equation}
where the term between the brackets is the contribution from frequencies beyond the cutoff used in Eq.\eqref{eq:finite_FT_G}. This expression will be very accurate if the asymptotic behavior is well obeyed beyond the cutoff. Using the somewhat complicated expression \eqref{eq:G2_inf_k_ikn} as the asymptotic form of $G^{(2)}(\mathbf{k},ik_n)$ may seem to complicate the calculations needlessly since the first term of the high frequency expansion of $G^{(1)}(\mathbf{k},ik_n)$ is identical to that of $G^{(2)}(\mathbf{k},ik_n)$. However, to compute the cubic spline of the integrand in Eq.\eqref{eq:chijj_1}, and then its Fourier transform, we need the derivatives of $G^{(2)}(\mathbf{k},\tau)$ at $\tau=0$ and $\tau=\beta$ and, for that purpose, we have to use a more accurate asymptotic form. The reason will be explained shortly.

As we have just mentioned, we need to compute derivatives of Eq.\eqref{eq:G2_k_tau} with respect to $\tau$ at the boundaries. Let us first rewrite expression \eqref{eq:G2_k_tau} using Eq.\eqref{eq:finite_FT_G},
\begin{multline}\label{eq:G2_k_tau_expl}
G^{(2)}(\mathbf{k},\tau)=T\sideset{}{'}\sum_{ik_n} e^{-ik_n\tau} \left[G^{(2)}(\mathbf{k},ik_n)-G^{(2)}_{inf}(\mathbf{k},ik_n)\right]\\
+G^{(2)}_{inf}(\mathbf{k},\tau)
\end{multline}
so that
\begin{multline}\label{eq:dG2_dtau0}
\frac{\partial G^{(2)}(\mathbf{k},\tau)}{\partial\tau}\Big|_{\tau=0}=\\
T\sideset{}{'}\sum_{ik_n} (-ik_n) \left[G^{(2)}(\mathbf{k},ik_n)-G^{(2)}_{inf}(\mathbf{k},ik_n)\right]\\
+\frac{\partial G^{(2)}_{inf}(\mathbf{k},\tau)}{\partial\tau}\Big|_{\tau=0}\,,
\end{multline}
where the last term is obtained from the derivative of Eq.\eqref{eq:G2_inf_k_tau}. If  $G^{(2)}_{inf}$ was taken to be $G^{(1)}$, the sum in \eqref{eq:G2_k_tau_expl} would converge like the sum of $1/(ik_n)^2$, but the sum in \eqref{eq:dG2_dtau0} would not converge. Thus, $G^{(2)}_{inf}$ must have at least the same first two terms in its high frequency expansion as $G^{(2)}$. If we use Eq.\eqref{eq:G2_inf_k_ikn}, the first five terms in the expansion are equal, the sum in \eqref{eq:G2_k_tau_expl} thus converges as the sum of $1/(ik_n)^6$, while the sum in \eqref{eq:dG2_dtau0}, as the sum of $1/(ik_n)^5$. This gives us a very precise evaluation of $G^{(2)}(\mathbf{k},\tau)$ and its derivatives. To obtain the derivatives at $\tau=\beta$ we use the relation
\begin{equation}\label{eq:dG_dtau_beta}
\frac{\partial G_{\sigma}(\mathbf{k},\tau)}{\partial \tau}\Bigg|_{\tau=\beta}=\tilde{\epsilon}_{\mathbf{k}}+Un_{-\sigma}-\frac{\partial G_{\sigma}(\mathbf{k},\tau)}{\partial \tau}\Bigg|_{\tau=0}
\end{equation}
that is derived  from the spectral representation of $G(\mathbf{k},\tau)$.

Once $G^{(2)}(\mathbf{k},\tau)$ and its derivatives are obtained, the integrals in expression \eqref{eq:chijj_1} are evaluated by computing the cubic splines for $G^{(2)}(\mathbf{k},\tau) G^{(2)}(\mathbf{k},-\tau)$ and using formula \eqref{eq:TF_fx_spline} for the Fourier transform. The rest of the calculation is simply a sum over the Brillouin zone, where, of course, it is preferable to use the symmetries of the system to save computational resources.

The second term of Eq.\eqref{eq:chi_jj_qn1} is
\begin{equation}\label{eq:chijj_2}
\begin{split}
\chi_{j_xj_x}^{v_1}(iq_n)=&-\frac{U}{4}\left(\frac{T}{N}\right)^2\sum_{k_1 k_2} G^{(2)}(k_1)G^{(2)}(k_1+iq_n)\\
&\times G^{(1)}(k_2)G^{(1)}(k_2+iq_n)\frac{\partial \epsilon_\mathbf{k}}{\partial k_x}(\mathbf{k}_1)\frac{\partial \epsilon_\mathbf{k}}{\partial k_x}(\mathbf{k}_2)\\
&\times  \left[3U_{sp}\chi_{sp}(k_2-k_1)+U_{ch}\chi_{ch}(k_2-k_1)\right]
\end{split}
\end{equation}
where $k_1+iq_n=(\mathbf{k}_1,ik_m+iq_n)$. If we define
\begin{equation}\label{eq:def_fn_V_k}
\begin{split}
f_n(k_2)&=\frac{\partial \epsilon_\mathbf{k}}{\partial k_x}(\mathbf{k}_2)G^{(1)}(k_2)G^{(1)}(k_2+iq_n)\,,\\
V(k_2-k_1)&=\frac{U}{8}\left[3U_{sp}\chi_{sp}(k_2-k_1)+U_{ch}\chi_{ch}(k_2-k_1)\right]\,,
\end{split}
\end{equation}
then, Eq.\eqref{eq:chijj_2} becomes
\begin{multline}
\chi_{j_xj_x}^{v_1}(iq_n)=-2\frac{T}{N}\sum_{k_1} \frac{\partial \epsilon_\mathbf{k}}{\partial k_x}(\mathbf{k}_1)G^{(2)}(k_1)G^{(2)}(k_1+iq_n)\\
\times\frac{T}{N}\sum_{k_2}f_n(k_2)V(k_2-k_1)\,.
\end{multline}
The sum over $k_2$ being a convolution, we have
\begin{multline}\label{eq:chijj_2_fV}
\chi_{j_xj_x}^{v_1}(iq_n)=-\frac{T}{N}\sum_{k} \frac{\partial \epsilon_\mathbf{k}}{\partial k_x}\,G^{(2)}(k)G^{(2)}(k+iq_n)\\
\times\sum_{\bar{1}}e^{-ik\cdot\bar{1}}f_n(\bar{1})V(\bar{1})\,.
\end{multline}
Note that the sum over $\bar{1}$ written in explicit form is
\begin{equation}
\sum_{j}\int_0^\beta d\tau\, e^{-i\mathbf{k}\cdot\mathbf{r}_j} e^{ik_m \tau} f_n(\mathbf{r}_j,\tau)V(\mathbf{r}_j,\tau)\,.
\end{equation}
where the function $f_n(\mathbf{r},\tau)$ is given by
\begin{equation}\label{eq:TF_fn_rt}
\begin{split}
f_n(\mathbf{r},\tau)&=\frac{1}{N}\sum_{\mathbf{k}}e^{i\mathbf{k}\cdot\mathbf{r}}\frac{\partial \epsilon_\mathbf{k}}{\partial k_x}\\
&\times T\sum_{ik_m} e^{-ik_m\tau}  G^{(1)}(\mathbf{k},ik_{m})G^{(1)}(\mathbf{k},ik_{m}+iq_n)\,.
\end{split}
\end{equation}
Since $G^{(1)}(\mathbf{k},ik_{m})$ is the non-interacting Green's function, the Fourier transform over $ik_m$ can be done analytically. Using \eqref{eq:sum_Res}, we obtain, for $q_n\neq 0$,
\begin{multline}\label{eq:TF_iqn_GG}
T\sum_{ik_m} e^{-ik_m\tau}G^{(1)}(\mathbf{k},ik_{m})G^{(1)}(\mathbf{k},ik_{m}+iq_n)\\
=\frac{e^{iq_n\tau}-1}{iq_n}e^{-\tilde{\epsilon}_\mathbf{k}\tau}\left[\theta(\tau)(1-f(\tilde{\epsilon}_\mathbf{k}))-\theta(-\tau)f(\tilde{\epsilon}_\mathbf{k})\right]\,.
\end{multline}
For $q_n=0$, there is a double pole so that the calculation with the residue theorem is slightly different. However, we can use the simple following trick,
\begin{equation}\label{eq:fqn0}
\begin{split}
&T\sum_{ik_m} e^{-ik_m\tau}G^{(1)}(\mathbf{k},ik_{m})^2= T\sum_{ik_m} e^{-ik_m\tau}\left(\frac{1}{ik_m-\tilde{\epsilon}_\mathbf{k}}\right)^2\\
&=\frac{\partial}{\partial\tilde{\epsilon}_\mathbf{k}}T\sum_{ik_m} e^{-ik_m\tau}\left(\frac{1}{ik_m-\tilde{\epsilon}_\mathbf{k}}\right)\\
&=\frac{\partial}{\partial\tilde{\epsilon}_k}\left(-e^{-\tilde{\epsilon}_k\tau}[(1-f(\tilde{\epsilon}_k))\theta(\tau)-f(\tilde{\epsilon}_k)\theta(-\tau)]\right)\\
&=e^{-\tilde{\epsilon}_\mathbf{k}\tau}\left(\tau [(1-f(\tilde{\epsilon}_\mathbf{k}))\theta(\tau)-f(\tilde{\epsilon}_\mathbf{k})\theta(-\tau)]+\frac{\partial f(\tilde{\epsilon}_\mathbf{k})}{\partial\tilde{\epsilon}_\mathbf{k}}\right)\,.
\end{split}
\end{equation}
For $\tau>0$ and $q_n\neq 0$, we have
\begin{equation}\label{eq:def_fn_r_tau}
\begin{split}
f_n(\mathbf{r},\tau)&=\frac{e^{iq_n\tau}-1}{iq_n}\frac{1}{N}\sum_{\mathbf{k}}e^{i\mathbf{k}\cdot\mathbf{r}}\frac{\partial \epsilon_\mathbf{k}}{\partial k_x}\,f(-\tilde{\epsilon}_\mathbf{k})e^{-\tilde{\epsilon}_\mathbf{k}\tau}\\
&=\frac{e^{iq_n\tau}-1}{iq_n}\frac{1}{N}\sum_{\mathbf{k}}e^{i\mathbf{k}\cdot\mathbf{r}}\frac{\partial \epsilon_\mathbf{k}}{\partial k_x}\,f(\tilde{\epsilon}_\mathbf{k})e^{(\beta-\tau)\tilde{\epsilon}_\mathbf{k}}\\
&=g_n(\tau)h(\mathbf{r},\tau)\,.
\end{split}
\end{equation}
The functions $f_n(\mathbf{r},\tau)$ for different values of $n$ are thus obtained by multiplying the $n$ independent function
\begin{equation}\label{eq:def_h_r_tau}
h(\mathbf{r},\tau)=\frac{1}{N}\sum_{\mathbf{k}}e^{i\mathbf{k}\cdot\mathbf{r}}\frac{\partial \epsilon_\mathbf{k}}{\partial k_x}\,f(\tilde{\epsilon}_\mathbf{k})e^{(\beta-\tau)\tilde{\epsilon}_\mathbf{k}}
\end{equation}
by $g_n(\tau)=(e^{iq_n\tau}-1)/(iq_n)$. Using this result, we find
\begin{equation}
\begin{split}
&\sum_{\bar{1}}e^{-ik_1\cdot\bar{1}}f_n(\bar{1})V(\bar{1})=\\
&=\int d\tau\, e^{ik_m \tau} g_n(\tau) \sum_{j} e^{-i\mathbf{k}_1\cdot\mathbf{r}_j} h(\mathbf{r}_j,\tau)V(\mathbf{r}_j,\tau)\,.
\end{split}
\end{equation}
Inserting this back into Eq.\eqref{eq:chijj_2_fV}, we obtain for the $q_n\neq 0$ terms,
\begin{equation}
\begin{split}
&\chi_{j_xj_x}^{v_1}(iq_n)=\\
&-2\frac{T}{N}\sum_{\mathbf{k}}\sum_{ik_m} \frac{\partial \epsilon_\mathbf{k}}{\partial k_x}\,G^{(2)}(\mathbf{k},ik_m)G^{(2)}(\mathbf{k},ik_m+iq_n)\\
&\qquad\times\int d\tau\, e^{ik_m \tau} g_n(\tau) \sum_{j} e^{-i\mathbf{k}\cdot\mathbf{r}_j} h(\mathbf{r}_j,\tau)V(\mathbf{r}_j,\tau)\\
&=-\frac{2}{iq_n}\frac{1}{N}\sum_{\mathbf{k}}\frac{\partial \epsilon_\mathbf{k}}{\partial k_x}\,T\sum_{ik_m} G^{(2)}(\mathbf{k},ik_m)G^{(2)}(\mathbf{k},ik_m+iq_n)\\
&\qquad\times\Big(\int d\tau\, e^{i(k_m+q_n) \tau} \sum_{j} e^{-i\mathbf{k}\cdot\mathbf{r}_j} h(\mathbf{r}_j,\tau)V(\mathbf{r}_j,\tau)\\
&\qquad-\int d\tau\, e^{ik_m \tau} \sum_{j} e^{-i\mathbf{k}\cdot\mathbf{r}_j} h(\mathbf{r}_j,\tau)V(\mathbf{r}_j,\tau)\Big)\,.
\end{split}
\end{equation}
To make the convolutions more apparent, we define
\begin{multline}\label{eq:def_J_k_ikn}
J(\mathbf{k},ik_m)=\\
G^{(2)}(\mathbf{k},ik_m)\int d\tau\, e^{ik_m \tau} \sum_{j} e^{-i\mathbf{k}\cdot\mathbf{r}_j} h(\mathbf{r}_j,\tau)V(\mathbf{r}_j,\tau)\,
\end{multline}
so that
\begin{multline}
\chi_{j_xj_x}^{v_1}(iq_n)=\\
-\frac{2}{iq_n}\frac{1}{N}\sum_{\mathbf{k}}\frac{\partial \epsilon_\mathbf{k}}{\partial k_x}\Big(T\sum_{ik_m} G^{(2)}(\mathbf{k},ik_m)J(\mathbf{k},ik_m+iq_n)\\
-T\sum_{ik_m} G^{(2)}(\mathbf{k},ik_m+iq_n)J(\mathbf{k},ik_m)\Big)
\end{multline}
and we can use the convolution theorem to obtain
\begin{multline}\label{eq:chijj_2_fin}
\chi_{j_xj_x}^{v_1}(iq_n)=\\
-\frac{2}{iq_n}\frac{1}{N}\sum_{\mathbf{k}}\frac{\partial \epsilon_\mathbf{k}}{\partial k_x}\,\Big(\int_0^{\beta}d\tau\, e^{-iq_n\tau}G^{(2)}(\mathbf{k},\tau)J(\mathbf{k},-\tau)\\
-\int_0^{\beta}d\tau\, e^{iq_n\tau}G^{(2)}(\mathbf{k},\tau)J(\mathbf{k},-\tau)\Big)\,.
\end{multline}

Note that, since all the Matsubara sums have been transformed into Fourier transforms that can be done with FFTs and that an FFT gives all $N$ values of a $N$ point transform at the same time, we obtain the values of $\chi_{j_xj_x}^{v_1}(iq_n)$ for all $iq_n$ except $iq_n=0$ at the same time when the last FFT is done.

We still need to do the Fourier transform over $\tau$ in Eq.\eqref{eq:def_J_k_ikn} using cubic splines. For that we need the derivatives of $h(\mathbf{r}_j,\tau)V(\mathbf{r}_j,\tau)$ at $\tau=0$ and $\tau=\beta$. The derivatives of $V(\mathbf{r}_j,\tau)$ have already been used to compute the self-energy and are obtained using Eq.\eqref{eq:dtau_Vrtau_chi0}. For $h(\mathbf{r}_j,\tau)$, the derivative is obtained from the definition \eqref{eq:def_h_r_tau}.

Next, we have to compute $J(\mathbf{k},-\tau)$. First, we define the function
\begin{equation}
Q(\mathbf{k},ik_n)=\int d\tau\, e^{ik_n \tau} \sum_{j} e^{-i\mathbf{k}\cdot\mathbf{r}_j} h(\mathbf{r}_j,\tau)V(\mathbf{r}_j,\tau)
\end{equation}
so that the definition \eqref{eq:def_J_k_ikn} reads $J(\mathbf{k},ik_m)=G^{(2)}(\mathbf{k},ik_m)Q(\mathbf{k},ik_n)$. The Fourier transform over $\tau$ is done using the method described in appendix  \ref{sec:FT_spline}. The asymptotic form of $Q(\mathbf{k},ik_n)$ is
\begin{equation}
Q_{inf}(\mathbf{k},ik_n)=\frac{q_1(\mathbf{k})}{ik_n}+\frac{q_2(\mathbf{k})}{(ik_n)^2}+\frac{q_3(\mathbf{k})}{(ik_n)^3}\,,
\end{equation}
so that, using Eq.\eqref{eq:G2_inf_k_ikn} for $G^{(2)}_{inf}(\mathbf{k},ik_m)$, the asymptotic form for $J(\mathbf{k},ik_m)$ is,
\begin{multline}
J_{inf}(\mathbf{k},ik_n)=\frac{1}{ik_n-\tilde{\epsilon}_{\mathbf{k}}-\left(\frac{s_1}{ik_n}+\frac{s_2(\mathbf{k})}{(ik_n)^2}+\frac{s_3(\mathbf{k})}{(ik_n)^3}\right)}\\
\times\left(\frac{q_1(\mathbf{k})}{ik_n}+\frac{q_2(\mathbf{k})}{(ik_n)^2}+\frac{q_3(\mathbf{k})}{(ik_n)^3}\right)
\end{multline}
and its Fourier transform is
\begin{multline}
J_{inf}(\mathbf{k},-\tau)=\sum_{j=1}^4e^{z_j(\mathbf{k})\tau}f(z_j(\mathbf{k}))\\
\times\frac{q_1(\mathbf{k})[z_j(\mathbf{k})]^2+q_2(\mathbf{k})z_j(\mathbf{k})+q_3(\mathbf{k})}{\prod_{i\neq j} [z_j(\mathbf{k})-z_i(\mathbf{k})]}\,,
\end{multline}
where the $z_j(\mathbf{k})$ are the roots of the polynomial \eqref{eq:pol4_Ginf}. By analogy with our previous calculations, the function $J(\mathbf{k},-\tau)$ is then obtained from
\begin{multline}\label{eq:J_k_tau}
J(\mathbf{k},-\tau)=T\sideset{}{'}\sum_{ik_n} e^{ik_n\tau} [J(\mathbf{k},ik_n)-J_{inf}(\mathbf{k},ik_n)]\\
+J_{inf}(\mathbf{k},-\tau)
\end{multline}
where $\sideset{}{'}\sum$ means that the sum is done up to a cutoff frequency.

Once these results are substituted in the expression for the Maki-Thompson term, Eq.\eqref{eq:chijj_2_fin}, we need the derivatives of $G^{(2)}(\mathbf{k},\tau)J(\mathbf{k},-\tau)$ at $\tau=0$ and $\tau=\beta$.  For $G^{(2)}(\mathbf{k},\tau)$, they are given by Eq.\eqref{eq:dG2_dtau0} and Eq.\eqref{eq:dG_dtau_beta}. As for the derivatives of $J(\mathbf{k},-\tau)$, they are obtained by differentiating Eq.\eqref{eq:J_k_tau}.

Finally, we need to evaluate separately the $q_n=0$ term for $\tau>0$. From Eq.\eqref{eq:fqn0}, we have
\begin{equation}
\begin{split}
&f_0(\mathbf{r},\tau)=\\
&\qquad\frac{1}{N}\sum_{\mathbf{k}}e^{i\mathbf{k}\cdot\mathbf{r}}\frac{\partial \epsilon_\mathbf{k}}{\partial k_x}(\mathbf{k})\;T\sum_{ik_m} e^{-ik_m\tau}  G^{(1)}(\mathbf{k},ik_{m})^2\\
&=\frac{1}{N}\sum_{\mathbf{k}}e^{i\mathbf{k}\cdot\mathbf{r}}\frac{\partial \epsilon_\mathbf{k}}{\partial k_x}(\mathbf{k})\;e^{-\tilde{\epsilon}_\mathbf{k}\tau}\left(\tau f(-\tilde{\epsilon}_\mathbf{k}))+\frac{\partial f(\tilde{\epsilon}_\mathbf{k})}{\partial\tilde{\epsilon}_\mathbf{k}}\right)\,.
\end{split}\label{eq:f0_r_t}
\end{equation}
so that
\begin{multline}
\chi_{j_xj_x}^{v_1}(0)=\\
-2\frac{1}{N}\sum_{\mathbf{k}} \frac{\partial \epsilon_\mathbf{k}}{\partial k_x}(\mathbf{k})\;T\sum_{ik_m}G^{(2)}(\mathbf{k},ik_m)G^{(2)}(\mathbf{k},ik_m)\\
\times\int_0^{\beta}d\tau\, \sum_{j} e^{-i\mathbf{k}_1\cdot\mathbf{r}_j} e^{ik_m \tau} f_0(\mathbf{r}_j,\tau)V(\mathbf{r}_j,\tau)\,.
\end{multline}

Using our previous definition, Eq.\eqref{eq:def_fn_V_k}, for $f_n(k_2)$, the third term in Eq.\eqref{eq:chi_jj_qn1} can be rewritten as
\begin{widetext}
\begin{multline}\label{eq:chijj_3_1}
\chi_{j_xj_x}^{v_2}(iq_n)=\frac{U}{2}\left(\frac{T}{N}\right)^2\sum_{k_1,q_1}\frac{\partial \epsilon_\mathbf{k}}{\partial k_x}(\mathbf{k}_1)G^{(2)}(k_1)G^{(2)}(k_1+iq_n)G^{(1)}(k_1+q_1+iq_n) \\
\qquad\times \Bigg(3 U_{sp}\frac{1}{1 - \frac{U_{sp}}{2} \chi_0(q_1)} \,\frac{1}{1 - \frac{U_{sp}}{2} \chi_0(q_1+iq_n)}+U_{ch}\frac{1}{1 + \frac{U_{ch}}{2} \chi_0(q_1)} \,\frac{1}{1 + \frac{U_{ch}}{2} \chi_0(q_1+iq_n)}\Bigg)\\
\times\frac{T}{N}\sum_{k_2} f_n(k_2)\left[G^{(1)}(k_2+q_1+iq_n)+G^{(1)}(k_2-q_1)\right]\,.
\end{multline}

The sum over $k_2$ is the sum of two convolutions and can be written as
\begin{equation}\label{eq:TF_fn_G}
\begin{split}
&\frac{T}{N}\sum_{k_2} f_n(k_2)\left[G^{(1)}(k_2+q_1+iq_n)+G^{(1)}(k_2-q_1)\right]=\sum_{\bar{1}} \left(e^{i(q_1+iq_n)\cdot\bar{1}}+e^{-iq_1\cdot\bar{1}}\right)f_n(\bar{1})G^{(1)}(-\bar{1})\\
&\qquad=\sum_{j}\int_{0}^{\beta} d\tau\, \left(e^{i\mathbf{q}_1\cdot\mathbf{r}_j}e^{-i(q_{m}+q_n)\tau}+e^{-i\mathbf{q}_1\cdot\mathbf{r}_j}e^{iq_m\tau}\right) f_n(\mathbf{r}_j,\tau)G^{(1)}(-\mathbf{r}_j,-\tau)\\
&\qquad=\int_{0}^{\beta} d\tau\, \left(-e^{-i(q_{m}+q_n)\tau}+e^{iq_m\tau}\right)\sum_{j} e^{-i\mathbf{q}_1\cdot\mathbf{r}_j} f_n(\mathbf{r}_j,\tau)G^{(1)}(\mathbf{r}_j,-\tau)\,,
\end{split}
\end{equation}
where we have used $f_n(-\mathbf{r}_j,\tau)=-f_n(\mathbf{r}_j,\tau)$ and $G^{(1)}(-\mathbf{r}_j,-\tau)=G^{(1)}(\mathbf{r}_j,-\tau)$. Using $f_n(\mathbf{r},\tau)=g_n(\tau)h(\mathbf{r},\tau)$, Eq.\eqref{eq:def_fn_r_tau}, we have
\begin{equation}
\begin{split}
&\frac{T}{N}\sum_{k_2} f_n(k_2)\left[G^{(1)}(k_2+q_1+iq_n)+G^{(1)}(k_2-q_1)\right]=\\
&\qquad\qquad \int d\tau\, \left(-e^{-i(q_{m}+q_n)\tau}+e^{iq_m\tau}\right) g_n(\tau)\sum_{j}e^{-i\mathbf{q}_1\cdot\mathbf{r}_j}h(\mathbf{r}_j,\tau)G^{(1)}(\mathbf{r}_j,-\tau)\\
&\qquad=\frac{1}{iq_n}\int d\tau\, \left(-e^{-iq_{m}\tau}+e^{-i(q_{m}+q_n)\tau}+e^{i(q_m+q_n)\tau}-e^{iq_m\tau}\right)\sum_{j}e^{-i\mathbf{q}_1\cdot\mathbf{r}_j} h(\mathbf{r}_j,\tau)G^{(1)}(\mathbf{r}_j,-\tau)\,.
\end{split}
\end{equation}
assuming  $q_n\neq 0$. Inserting this expression into Eq.\eqref{eq:chijj_3_1}, we get
\begin{multline}\label{eq:chijj_3_2}
\chi_{j_xj_x}^{v_2}(iq_n)=\frac{1}{iq_n}\frac{U}{2}\left(\frac{T}{N}\right)^2\sum_{k_1,q_1}\frac{\partial \epsilon_\mathbf{k}}{\partial k_x}(\mathbf{k}_1)G^{(2)}(k_1)G^{(2)}(k_1+iq_n)G^{(1)}(k_1+q_1+iq_n)\\
\times  \Bigg(3 U_{sp}\frac{1}{1 - \frac{U_{sp}}{2} \chi_0(q_1)} \,\frac{1}{1 - \frac{U_{sp}}{2} \chi_0(q_1+iq_n)}+U_{ch}\frac{1}{1 + \frac{U_{ch}}{2} \chi_0(q_1)} \,\frac{1}{1 + \frac{U_{ch}}{2} \chi_0(q_1+iq_n)}\Bigg)\\
\times\int d\tau\, \left(-e^{-iq_{m}\tau}+e^{-i(q_{m}+q_n)\tau}+e^{i(q_m+q_n)\tau}-e^{iq_m\tau}\right)\sum_{j}e^{-i\mathbf{q}_1\cdot\mathbf{r}_j} h(\mathbf{r}_j,\tau)G^{(1)}(\mathbf{r}_j,-\tau)\,.
\end{multline}
Using the definitions
\begin{equation}
\bar{G}^{(1)}_n(k_1+q_1)=G^{(1)}(k_1+q_1+iq_n)
\end{equation}
and
\begin{multline}\label{eq:def_fn2_Hn}
H_n(q_1)=\Bigg(3 U_{sp}\frac{1}{1 - \frac{U_{sp}}{2} \chi_0(q_1)} \,\frac{1}{1 - \frac{U_{sp}}{2} \chi_0(q_1+iq_n)}+U_{ch}\frac{1}{1 + \frac{U_{ch}}{2} \chi_0(q_1)} \,\frac{1}{1 + \frac{U_{ch}}{2} \chi_0(q_1+iq_n)}\Bigg)\\
\times\int d\tau\, \left(e^{i(q_m+q_n)\tau}+e^{-i(q_{m}+q_n)\tau}-e^{iq_m\tau}-e^{-iq_m\tau}\right)\sum_{j}e^{-i\mathbf{q}_1\cdot\mathbf{r}_j} h(\mathbf{r}_j,\tau)G^{(1)}(\mathbf{r}_j,-\tau)\,,
\end{multline}
Eq.\eqref{eq:chijj_3_2} reads
\begin{equation}\label{eq:chijj_3_3}
\begin{split}
\chi_{j_xj_x}^{v_2}(iq_n)&=\frac{1}{iq_n}\frac{U}{2}\frac{T}{N}\sum_{k_1}\frac{\partial \epsilon_\mathbf{k}}{\partial k_x}(\mathbf{k}_1)G^{(2)}(k_1)G^{(2)}(k_1+iq_n)\frac{T}{N}\sum_{q_1} \bar{G}^{(1)}_n(k_1+q_1)H_n(q_1)\\
&=\frac{1}{iq_n}\frac{U}{2}\frac{T}{N}\sum_{k_1}\frac{\partial \epsilon_\mathbf{k}}{\partial k_x}(\mathbf{k}_1)G^{(2)}(k_1)G^{(2)}(k_1+iq_n)\sum_{\bar{1}} e^{ik_1\cdot\bar{1}} \bar{G}^{(1)}_n(-\bar{1})H_n(\bar{1})\,.
\end{split}
\end{equation}
\end{widetext}
where the sum over $q_1$ was written as a Fourier transform. Unfortunately, since this transform does not give a function of $k_1$ or $k_1+iq_n$, or a sum of either, because the dependence on $iq_n$ in $H_n(q_1)$ does not factor out as a sum of exponentials, the sum over $k_1$ does not have the form of a convolution or a sum of convolutions and has to be done in the form Eq.\eqref{eq:chijj_3_3} for each different frequency $iq_n$.

Now, some work remains to be done before one can do the Fourier transform over $\bar{1}$ in Eq.\eqref{eq:chijj_3_3}. First, $G^{(1)}_n(\mathbf{r},\tau)$ is explicitly given by
\begin{equation}
G^{(1)}_n(\mathbf{r},\tau)=\frac{1}{N}\sum_{\mathbf{k}}e^{i\mathbf{k}\cdot\mathbf{r}}\; T\sum_m  e^{-ik_m\tau}\frac{1}{ik_m+iq_n-\tilde{\epsilon}_\mathbf{k}}
\end{equation}
so that changing summation variable above or using Eq.\eqref{eq:sum_Res}, we obtain
\begin{equation}\label{eq:Gn_vs_G}
G^{(1)}_n(\mathbf{r},\tau)=e^{iq_n\tau}G^{(1)}(\mathbf{r},\tau)\,.
\end{equation}
Eq.\eqref{eq:chijj_3_3} thus becomes
\begin{multline}\label{eq:chijj_3_4}
\chi_{j_xj_x}^{v_2}(iq_n)=\\
\frac{1}{iq_n}\frac{U}{2}\frac{T}{N}\sum_{\mathbf{k},ik_m}\frac{\partial \epsilon_\mathbf{k}}{\partial k_x}(\mathbf{k})G^{(2)}(\mathbf{k},ik_m)G^{(2)}(\mathbf{k},ik_m+iq_n)\\
\times\sum_{\mathbf{r}} \int d\tau \, e^{i\mathbf{k}\cdot\mathbf{r}} e^{-i(k_m+q_n)\tau} G^{(1)}(\mathbf{r},-\tau)H_n(\mathbf{r},\tau)\,.
\end{multline}

There is however a problem if we directly use $H_n(\mathbf{r},\tau)$ in Eq.\eqref{eq:chijj_3_4}.  From the definition \eqref{eq:def_fn2_Hn}, one notices that $H_n(\mathbf{q},iq_m)$ is peaked at the frequency $q_m=-q_n$ because of the factor
\begin{equation}
\frac{1}{1-\frac{U_{sp}}{2}\chi_0(\mathbf{q},iq_m+iq_n)}
\end{equation}
in the spin part. Therefore, $H_n(\mathbf{r},\tau)$ has oscillations in $\tau$ at the frequency $q_n$ and it is necessary to refine the grid in $\tau$ when $q_n$ increases. $H_n(\mathbf{q},iq_m)$ is also peaked at $q_m=0$ because of the factor
\begin{equation}
\frac{1}{1-\frac{U_{sp}}{2}\chi_0(\mathbf{q},iq_m)}\,.
\end{equation}
It is therefore not sufficient to express $H_n(\mathbf{r},\tau)$ using a function shifted in frequency multiplied by an oscillating function. By doing that we would reduce at most the oscillation frequency by half. Instead, we express $H_n(\mathbf{r},\tau)$ as a sum of two functions, one peaked at $q_m=0$ and the other, at $q_m=-q_n$. Then we will be able to apply a translation in frequency to the latter and factor out the oscillating part. First, we write
\begin{multline}
\frac{1}{1-\frac{U_{sp}}{2}\chi_0(\mathbf{q},iq_m+iq_n)}\frac{1}{1-\frac{U_{sp}}{2}\chi_0(\mathbf{q},iq_m)}=\\
\frac{1}{\chi_0(\mathbf{q},iq_m)-\chi_0(\mathbf{q},iq_m+iq_n)}\frac{\chi_0(\mathbf{q},iq_m)}{1-\frac{U_{sp}}{2}\chi_0(\mathbf{q},iq_m)}\\
+\frac{1}{\chi_0(\mathbf{q},iq_m+iq_n)-\chi_0(\mathbf{q},iq_m)}\frac{\chi_0(\mathbf{q},iq_m+iq_n)}{1-\frac{U_{sp}}{2}\chi_0(\mathbf{q},iq_m+iq_n)}\,.
\end{multline}
If we define
\begin{multline}
D(\mathbf{q},iq_{m})=\\
\int d\tau\, \left(e^{iq_m\tau}+e^{-iq_m\tau}\right)\sum_{j}e^{-i\mathbf{q}\cdot\mathbf{r}_j} h(\mathbf{r}_j,\tau)G^{(1)}(\mathbf{r}_j,-\tau)\,,
\end{multline}
the spin part of $H_n(\mathbf{q},iq_m)$ reads
\begin{multline}\label{eq:Hn_part_frac}
-\frac{D(\mathbf{q},iq_{m})-D(\mathbf{q},iq_{m}+iq_n)}{\chi_0(\mathbf{q},iq_m)-\chi_0(\mathbf{q},iq_m+iq_n)}\chi_{sp}(\mathbf{q},iq_m)\\
+\frac{D(\mathbf{q},iq_{m}+iq_n)-D(\mathbf{q},iq_{m})}{\chi_0(\mathbf{q},iq_m+iq_n)-\chi_0(\mathbf{q},iq_m)}\chi_{sp}(\mathbf{q},iq_m+iq_n)\,.
\end{multline}
Note that, since $D(\mathbf{q},iq_{m})$ and $\chi_0(\mathbf{q},iq_m)$ are even functions, the function
\begin{equation}\label{eq:diffD_sur_diff_chi0}
\frac{D(\mathbf{q},iq_{m})-D(\mathbf{q},iq_{m}+iq_n)}{\chi_0(\mathbf{q},iq_m)-\chi_0(\mathbf{q},iq_m+iq_n)}
\end{equation}
is undetermined when $q_m=-q_n/2$, which happens when $n$ is even. However, we know  that $H_n(\mathbf{q},-\frac{iq_n}{2})$ vanishes so that one can assume an arbitrary value for Eq.\eqref{eq:diffD_sur_diff_chi0} at that point. Numerically, it is better for that factor to be smooth. This is achieved by using a simple interpolation to fix that value.

Now, if we define the function
\begin{multline}\label{eq:def_In_q_qm}
I_n(\mathbf{q},iq_m)=\\
3U_{sp}\frac{D(\mathbf{q},iq_{m})-D(\mathbf{q},iq_{m}+iq_n)}{\chi_0(\mathbf{q},iq_m)-\chi_0(\mathbf{q},iq_m+iq_n)}\frac{\chi_0(\mathbf{q},iq_m)}{1-\frac{U_{sp}}{2}\chi_0(\mathbf{q},iq_m)}\\
+U_{ch}\frac{D(\mathbf{q},iq_{m})-D(\mathbf{q},iq_{m}+iq_n)}{\chi_0(\mathbf{q},iq_m)-\chi_0(\mathbf{q},iq_m+iq_n)}\frac{\chi_0(\mathbf{q},iq_m)}{1+\frac{U_{ch}}{2}\chi_0(\mathbf{q},iq_m)}\,,
\end{multline}
which is peaked at $q_m=0$, we make explicit that $H_n(\mathbf{q},iq_m)$ is peaked at two values of frequency
\begin{equation}
H_n(\mathbf{q},iq_m)=-I_n(\mathbf{q},iq_m)+I_n(\mathbf{q},-iq_m-iq_n)\,.
\end{equation}

Using the bosonic Matsubara frequency representation for $I_n(\mathbf{q},\tau)$ we obtain
\begin{equation}\label{eq:Hn_tau_vs_In_tau}
\begin{split}
H_n(\mathbf{q},\tau)&=T\sum_{iq_m}e^{-iq_m\tau}H_n(\mathbf{q},iq_m)\\
&=-I_n(\mathbf{q},\tau)+e^{iq_n\tau}I_n(\mathbf{q},-\tau)\,.
\end{split}
\end{equation}

To compute $I_n(\mathbf{q},\tau)$ from $I_n(\mathbf{q},iq_m)$ we use the same procedure as for the previous Fourier transforms over Matsubara frequencies. Using
\begin{equation}
\begin{split}
D_{inf}(\mathbf{q},iq_{m})&=\frac{d(\mathbf{q})}{(iq_m)^2}\\
\chi_{0,inf}(\mathbf{q},iq_m)&=\frac{c(\mathbf{q})}{(iq_m)^2}
\end{split}
\end{equation}
the asymptotic form for $I_n(\mathbf{q},iq_m)$ is
\begin{multline}\label{eq:In_inf_qm}
I_n^{inf}(\mathbf{q},iq_m)=3U_{sp}\frac{d(\mathbf{q})}{(iq_m)^2-\frac{U_{sp}}{2}c(\mathbf{q})}\\
+U_{ch}\frac{d(\mathbf{q})}{(iq_m)^2+\frac{U_{ch}}{2}c(\mathbf{q})}\,.
\end{multline}
Using a complex plane integration and the residue theorem, we get
\begin{multline}
I_n^{inf}(\mathbf{q},\tau)=3U_{sp}\frac{d(\mathbf{q})}{2z_{sp}(\mathbf{q})}\left(n_B(-z_{sp}(\mathbf{q}))e^{-z_{sp}(\mathbf{q})\tau}\right.\\
\qquad\left.-n_B(z_{sp}(\mathbf{q}))e^{z_{sp}(\mathbf{q})\tau}\right)\\
+U_{ch}\frac{d(\mathbf{q})}{2iz_{ch}(\mathbf{q})}\left(n_B(-iz_{ch}(\mathbf{q}))e^{-iz_{ch}(\mathbf{q})\tau}\right.\\
\qquad\left.-n_B(iz_{ch}(\mathbf{q}))e^{iz_{ch}(\mathbf{q})\tau}\right)\,,
\end{multline}
where $n_B(z)$ is the Bose-Einstein distribution and
\begin{equation}
\begin{split}
z_{sp}(\mathbf{q})&=\sqrt{\frac{U_{sp}}{2}c(\mathbf{q})}\,,\\
z_{ch}(\mathbf{q})&=\sqrt{\frac{U_{ch}}{2}c(\mathbf{q})}\,,
\end{split}
\end{equation}
or, using $n_B(-z)=-e^{\beta z}n_B(z)$,
\begin{multline}
I_n^{inf}(\mathbf{q},\tau)=\\
-3U_{sp}\frac{d(\mathbf{q})}{z_{sp}(\mathbf{q})}n_B(z_{sp}(\mathbf{q}))e^{\frac{\beta}{2}z_{sp}(\mathbf{q})}\cosh\left[\left(\frac{\beta}{2}-\tau\right)z_{sp}(\mathbf{q})\right]\\
-U_{ch}\frac{d(\mathbf{q})}{iz_{ch}(\mathbf{q})}n_B(iz_{ch}(\mathbf{q}))e^{i\frac{\beta}{2}z_{ch}(\mathbf{q})}\cos\left[\left(\frac{\beta}{2}-\tau\right)z_{ch}(\mathbf{q})\right]\,.
\end{multline}
Using that result we write, as before, the transform in such a way that it converges quickly
\begin{multline}\label{eq:def_In_tau_In_qm_inf}
I_n(\mathbf{q},\tau)=T\sideset{}{'}\sum_{iq_m}e^{-iq_m\tau}\left[I_n(\mathbf{q},iq_m)-I_n^{inf}(\mathbf{q},iq_m)\right]\\
+I_n^{inf}(\mathbf{q},\tau)\,.
\end{multline}

There is however one last difficulty. The asymptotic form Eq.\eqref{eq:In_inf_qm} approaches Eq.\eqref{eq:def_In_q_qm} only when both $q_m$ and $q_m+q_n$ are large with respect to the bandwidth. Thus, the frequency range of the finite sum in Eq.\eqref{eq:def_In_tau_In_qm_inf} has to be chosen such that this condition is satisfied. If $q_n$ is positive, the range of negative $q_m$ must therefore be extended to make sure that, at the cutoff, $q_m+q_n$ is large.

Coming back to expression \eqref{eq:chijj_3_3}, using the results \eqref{eq:Gn_vs_G} and \eqref{eq:Hn_tau_vs_In_tau}, the sum over $\bar{1}$ becomes
\begin{equation}
\begin{split}
&\sum_{\bar{1}} e^{ik_1\cdot\bar{1}} G^{(1)}_n(-\bar{1})H_n(\bar{1})=\\
&\qquad\sum_{\mathbf{r}} \int d\tau \, e^{i\mathbf{k}\cdot\mathbf{r}} e^{-i(k_m+q_n)\tau} G^{(1)}(\mathbf{r},-\tau)\\
&\qquad\qquad\times\left[-I_n(\mathbf{r},\tau)+e^{iq_n\tau}I_n(\mathbf{r},-\tau)\right]\\
&=\sum_{\mathbf{r}} \int d\tau \, e^{i\mathbf{k}\cdot\mathbf{r}} e^{-ik_m\tau} \left[-e^{-iq_n\tau}G^{(1)}(\mathbf{r},-\tau)I_n(\mathbf{r},\tau)\right.\\
&\qquad\qquad\qquad\qquad\qquad\qquad\left.+G^{(1)}(\mathbf{r},-\tau)I_n(\mathbf{r},-\tau)\right]\,.
\end{split}
\end{equation}

For $q_n=0$, expression \eqref{eq:chijj_3_1} vanishes, as explained at the end of appendix B.

To reach reasonably low temperatures without seeing any finite size effect, we use a system of $512\times 512$ sites and $8192$ frequencies. If we were using a brute force approach to compute Eq.\eqref{eq:chijj_3_1} we would have to sum $(8192(512^2))^3=9.9\times 10^{27}$ terms. Assuming we could sum $10^9$ terms per second, it would take about 300 billion years to calculate $\chi_{j_xj_x}^{v_2}(iq_n)$ for one value of $q_n$ and thus about 30 000 billion years for a hundred values of $q_n$. Using the approach described above, this calculation is done in less than 2 days.

\section{Choice of Matsubara frequencies}\label{eq:grid_Matsubara}

While all external frequencies are obtained at the same time from the last FFT for the bubble and Maki-Thompson like term in $\chi_{j_xj_x}(iq_n)$, this is not the case for the Aslamasov-Larkin like term, Eq.\eqref{eq:chijj_3_1}, even in the form of Eq.\eqref{eq:chijj_3_3} that makes maximal use of fast-Fourier transforms. Hence, this term cannot be calculated for thousands of values of $q_n$, or for the same number as that used in internal Matsubara frequency sums. Therefore we have to compute Eq.\eqref{eq:chijj_3_3} for a reasonable number of carefully chosen frequencies. Assuming that most of the information is in the low frequencies and that, as their magnitude increases, it becomes less important to include all the high frequencies, we use the following non-uniform Matsubara frequency index grid. That grid consists of subintervals within which the Matsubara frequencies are equally spaced, with larger space in between Matsubara frequencies at large frequency. The spacing between frequencies in different subintervals increases by powers of 2.

First we define

$N_0=2^r$ : the last frequency index (cutoff), taken as a power of 2 ($r$ integer),

$m$: integer that determines how dense the grid is. A large $m$ gives a low density. $0 \leq m < r$,

$N_1=\frac{N_0}{2^m}$ : $N_1+1$ is the number of adjacent frequencies close to $n=0$,

$N_2=\frac{N_1}{2}$ : number of frequencies in each subinterval with a fixed spacing between Matsubara frequencies,

$N=N_1+mN_2+1$ :  total number of frequencies.

Then the indices of Matsubara frequencies on the grid are given by the following algorithm
\begin{equation}
n(j)=
\begin{cases}
j\,,\qquad j=0,\ldots, N_1-1,\\
\,\\
N_1 + 2^{l_j+1}mod(j-N_1,N_2) + N_1(2^{l_j}-1)\,,\\
\qquad l_j=floor\left(\frac{j-N_1}{N_2}\right)\,,\quad j=N_1,\ldots,  N-1\,.
\end{cases}
\end{equation}

For example, taking $N_0=256$, $m=5$, so that $N_1=8$, $N_2=4$ and a total number of frequencies $N=29$, we obtain the following indices:
$n=$ 0, 1, 2, 3, 4, 5, 6, 7, 8, 10, 12, 14, 16, 20, 24, 28, 32, 40, 48, 56, 64, 80, 96, 112, 128, 160, 192, 224, 256.

With this kind of grid we greatly reduce the number of frequencies for which we have to calculate $\chi_{j_xj_x}^{v_2}$ with Eq.\eqref{eq:chijj_3_3} while retaining the essential information. Note that this also speeds up the analytical continuation with our maximum entropy method described in appendix \ref{sec:analytical_cont_app}.

\section{Fourier transform of a cubic spline}\label{sec:FT_spline}

Assume we have the following integral to do
\begin{equation}\label{eq:TF_g_x}
f(k)=\int_{x_0}^{x_N} dx \,g(x) e^{-ikx}
\end{equation}
 but that we know only $N+1$ discrete values of $g(x_i)$. Let us approximate $g(x)$ in the interval using a cubic spline $S(x)$ defined as
\begin{equation}\label{eq:def_S_x}
S(x)=
\begin{cases}
S_1(x) & x_0<x<x_1\\
S_2(x) & x_1<x<x_2\\
\qquad\vdots &\\
S_{N}(x) & x_{N-1}<x<x_N\,,
\end{cases}
\end{equation}
where the $S_n(x)$ are cubic polynomials, with the conditions
\begin{equation}\label{eq:eqns_Sn_x}
\begin{split}
S_n(x_{n-1})&=g(x_{n-1})\\
S_n(x_{n})&=g(x_{n})\\
S_{n}'(x_{n-1})&=S_{n-1}'(x_{n-1})\qquad n>1\\
S_{n}''(x_{n-1})&=S_{n-1}''(x_{n-1})\qquad n>1\\
S_{1}'(x_{0})&=g'(x_{0})\\
S_{N}'(x_{N})&=g'(x_{N})\,,
\end{split}
\end{equation}
defining the $4N$ equations necessary to determine the $4N$ coefficients of the spline. The integral \eqref{eq:TF_g_x} becomes
\begin{equation}\label{eq:integ_spline}
f(k)\approx \sum_{n=1}^N \int_{x_{n-1}}^{x_n} dx \,S_n(x) e^{-ikx}\,.
\end{equation}
Integrating by parts we obtain,
\begin{equation}\label{eq:integ_part_I}
\begin{split}
f(k)&=-\frac{1}{ik} \sum_{n=1}^N \left(  e^{-ikx}S_n(x)\Big|_{x_{n-1}}^{x_n}\right)\\
&\qquad+ \frac{1}{ik}\sum_{n=1}^N \int_{x_{n-1}}^{x_n} dx \,S_n'(x) e^{-ikx} \\
&= \frac{e^{-ikx_{0}}S_1(x_{0})-e^{-ikx_N}S_N(x_N)}{ik}\\
&\qquad+\frac{1}{ik}\sum_{n=1}^N\int_{x_{n-1}}^{x_n} dx \,S_n'(x) e^{-ikx}
\end{split}
\end{equation}
where we have used the continuity of the spline at the points $x_n$ to eliminate all the intermediate terms in the first sum. Now, if we integrate by parts in the second term and use the continuity of the derivatives $S_n'(x)$ at the points $x=x_n$, we obtain
\begin{multline}
f(k)=\frac{e^{-ikx_{0}}S_1(x_{0})-e^{-ikx_N}S_N(x_N)}{ik}\\
+\frac{e^{-ikx_{0}}S_1'(x_{0})-e^{-ikx_N}S_N'(x_N)}{(ik)^2}\\
+\frac{1}{(ik)^2}\sum_{n=1}^N\int_{x_{n-1}}^{x_n} dx \,S_n''(x) e^{-ikx}\,.
\end{multline}
Doing it one last time, using the fact that we also have the continuity of the second derivatives $S_n''(x)$ at $x_n$, we finally obtain
\begin{multline}\label{eq:TF_fx_spline}
f(k)=\frac{e^{-ikx_{0}}S_1(x_{0})-e^{-ikx_N}S_N(x_N)}{ik}\\
+\frac{e^{-ikx_{0}}S_1'(x_{0})-e^{-ikx_N}S_N'(x_N)}{(ik)^2}\\
+\frac{e^{-ikx_{0}}S_1''(x_{0})-e^{-ikx_N}S_N''(x_N)}{(ik)^3}\\
+\frac{1-e^{-ik\Delta x}}{(ik)^4}\sum_{n=0}^{N-1} S_{n+1}^{(3)} e^{-ikx_{n}}\,,
\end{multline}
where $\Delta x=x_{n+1}-x_n$. The remaining discrete transform can be done using an FFT. The result of this transform will be periodic since it is discrete, but because of the factor $1/(ik)^4$ in front, it will only have a relevant contribution for small $k$. This periodicity in the sum therefore produces only a very small noise at high $k$, the remaining discretization noise, that decreases with increasing $k$. Note that expression \eqref{eq:TF_fx_spline} is valid only for $k\neq 0$. For $k=0$, the result is simply the sum, over all subintervals, of integrals of cubic polynomials with their respective coefficients.

Finally, note that we have chosen to fix the value of the derivatives at the boundaries to complete the system of equations \eqref{eq:eqns_Sn_x} defining the spline. But other interesting choices are also possible. For example, if we know the coefficients of the terms in $1/(k^2)$ and $1/(k^3)$, i.e. the numerator in the second and third terms of Eq.\eqref{eq:TF_fx_spline}, fixing those coefficients is a good alternative. In the case where $k$ is a frequency, this choice is convenient because we often know the high frequency expansion from sum rules.

\section{Analytical continuation for the conductivity}\label{sec:analytical_cont_app}

Let us start by rewriting the spectral representation for the Matsubara \textit{current-current} correlation function $\chi_{j_xj_x}(iq_n)$,
\begin{equation}
\begin{split}
\chi_{j_xj_x}(iq_n)&=\int \frac{d\omega}{\pi}\, \frac{\chi_{j_xj_x}''(\omega)}{\omega-iq_n}\\
&=iq_n\int \frac{d\omega}{\pi}\, \frac{\chi_{j_xj_x}''(\omega)}{\omega^2+q_n^2}+\int \frac{d\omega}{\pi}\, \frac{\omega\, \chi_{j_xj_x}''(\omega)}{\omega^2+q_n^2}\,.
\end{split}
\end{equation}
Since $\chi_{j_xj_x}''(\omega)$ is odd,
\begin{equation}
\chi_{j_xj_x}(iq_n)=\int \frac{d\omega}{\pi}\, \frac{\omega\, \chi_{j_xj_x}''(\omega)}{\omega^2+q_n^2}\,.
\end{equation}
The real part of the optical conductivity is
\begin{equation}
\sigma^r(\omega)=\frac{\chi_{j_xj_x}''(\omega)}{\omega}\,,
\end{equation}
so that we obtain
\begin{equation}\label{eq:chijj_iqn_vs_sigma}
\chi_{j_xj_x}(iq_n)=2\int_{0}^{\infty} \frac{d\omega}{\pi}\, \frac{\omega^2 }{\omega^2+q_n^2}\sigma^r(\omega)\,,
\end{equation}
using the fact that the integrand is even. This is our starting point. The objective is to obtain the real part of the conductivity, on the right-hand side, from the Matsubara expression for the susceptibility on the left-hand side. Most analytical continuation is done for imaginary-time data, but not in our case.~\cite{Jarrell1996133}

Suppose we could be satisfied with $\sigma^r(\omega)$ on a discrete set of points $\omega_j$. We can use a numerical integration method to approximate the integral \eqref{eq:chijj_iqn_vs_sigma}, which would then have the form
\begin{equation}\label{eq:chijj_K_sigma}
\chi_{j_xj_x}(iq_n)\approx \sum_j K_{nj}\sigma^r_j\,,
\end{equation}
where $\sigma^r_j=\sigma^r(\omega_j)$ and $K_{nj}$ is a $N_{q_n}\times N_\omega$ matrix, $N_{q_n}$ being the size of the vector $\chi_{j_xj_x}(iq_n)$ and $N_\omega$, the size of the vector $\sigma^r_j$. Now, $\sigma^r_j$ is the quantity we want to determine. If $N_{q_n}=N_\omega$ then $\sigma^r_j$ is completely determined by the linear system \eqref{eq:chijj_K_sigma}. However, the matrix $K_{nj}$ is ill-conditioned so that a small noise in $\chi_{j_xj_x}(iq_n)$ would result in a very noisy solution $\sigma^r_j$. Also, $N_{q_n}$ is generally smaller than $N_\omega$, the number of real frequencies for which we want to determine $\sigma^r(\omega)$. We therefore need to include more information in the problem to find a unique $\sigma^r_j$. The way to do this is to use a maximum entropy approach. In this approach, we minimize the function
\begin{equation}\label{eq:chi2_S_maxent}
\chi^2-\alpha S\,,
\end{equation}
where
\begin{equation}
\begin{split}
\chi^2=&\sum_{\{iq_n\}} \left(\frac{\chi_{j_xj_x}(iq_n)-\sum_j K_{nj}\sigma^r_j}{\epsilon_n}\right)^2\\
&=\sum_{\{iq_n\}} \left(\frac{\chi_{j_xj_x}(iq_n)-\bar{\chi}_{j_xj_x}(iq_n)}{\epsilon_n}\right)^2
\end{split}
\end{equation}
measures the deviation of $\bar{\chi}_{j_xj_x}(iq_n)=\sum_j K_{nj}\sigma^r_j$ with respect to $\chi_{j_xj_x}(iq_n)$, $\epsilon_n$ being an estimate of the error of $\chi_{j_xj_x}(iq_n)$ with respect to the ``exact'' function. $S$ is the differential entropy defined as
\begin{equation}\label{eq:diff_entropy}
S=-\int_0^{\infty} d\omega\, \sigma^r(\omega) \ln \frac{\sigma^r(\omega)}{m(\omega)}
\end{equation}
where $m(\omega)$ is called the default model. The value of $\alpha$ can be chosen according to different criteria. As is often done, we choose it such that $\chi^2\approx N_{q_n}$, so that $|\chi_{j_xj_x}(iq_n)-\bar{\chi}_{j_xj_x}(iq_n)|$ be equal to $\epsilon_n$ on average.

Errors in the numerical evaluation of the integral \eqref{eq:chijj_iqn_vs_sigma}, i.e. in the definition of $K_{nj}$ in Eq.\eqref{eq:chijj_K_sigma}, are equivalent to having larger errors in the data $\chi_{j_xj_x}(iq_n)$.  Therefore, when Eq.\eqref{eq:chi2_S_maxent} is minimized to find a solution $\sigma^r_j$, this could lead to large errors in $\sigma^r_j$ with respect to the optimal solution because the inversion of expression \eqref{eq:chijj_iqn_vs_sigma} is an ill-conditioned problem. It is thus clear that the error we make by replacing Eq.\eqref{eq:chijj_iqn_vs_sigma} by Eq.\eqref{eq:chijj_K_sigma} must be smaller than the estimated error $\epsilon_n$ on the original data $\chi_{j_xj_x}(iq_n)$. That is why we need a very accurate numerical integration technique to define $K_{nj}$. Because we use the spectral representation in Matsubara frequencies, the weight function $\omega^2/(\omega^2+q_n^2)$ in the integrand of Eq.\eqref{eq:chijj_iqn_vs_sigma} is simple and can be integrated analytically. Hence, if we use, for example, a polynomial approximation for $\sigma^r(\omega)$ in a given interval $[\omega_{j-1},\omega_{j}]$, the integral can also be done analytically in the interval. If we use a good piecewise polynomial approximation for $\sigma^r(\omega)$, then we can evaluate the integral \eqref{eq:chijj_iqn_vs_sigma} precisely.

Maybe the most efficient approach to integrate Eq.\eqref{eq:chijj_iqn_vs_sigma} in one dimension is Gaussian quadratures. However, in the latter, both the weights and the grid points depend on the weight function, which in our case depends on $q_n$. We would therefore need a different grid in $\omega$ for each frequency $iq_n$. This is not possible because we can only search for a unique vector $\sigma^r_j$ defined on a unique grid $\omega_j$. Another very efficient way of doing the integrals \eqref{eq:chijj_iqn_vs_sigma} is to model $\sigma^r(\omega)$ using a cubic spline. This approach allows to use a fixed grid and is very precise because cubic splines are very good to approximate smooth function, which is the case of $\sigma^r(\omega)$. It seems therefore the best approach for our problem. However, to be able to perform the integral \eqref{eq:chijj_iqn_vs_sigma} over the whole frequency range $[0,\infty$, and at the same time reduce the number of frequencies in the grid, which helps speed up the minimization of Eq.\eqref{eq:chi2_S_maxent}, the spline we use is divided into two parts, a low frequency part that is a cubic spline in $\omega$ and a high frequency part cubic in $u=1/\omega$. Then, to make the spline linear system well-conditioned, we use a grid that is uniform in $\omega$ in the low frequency part and uniform in $u$ in the hight frequency part. Finally, integrating analytically in a piecewise manner, keeping the weight function $\omega^2/(\omega^2+q_n^2)$ intact in the integrand is a great advantage as the temperature decreases since this function then becomes sharper and sharper, and is thus increasingly difficult to integrate numerically. Of course the conductivity $\sigma^r(\omega)$ itself becomes also sharper as $T$ decreases and we have to adjust the grid to resolve its structure. But the numerical integration is still much easier and precise with this approach. We describe below how the $K_{nj}$ is defined and also our choice of grid in $\omega$.

We start with the following representation for $\sigma^r(\omega)$,
\begin{multline}\label{eq:def_spline_sigma}
\sigma^r(\omega)=\\
\begin{cases}
s_j(\omega)\,, & \omega_{j-1}\leq\omega\leq\omega_j\,,\qquad 1\leq j\leq N\\
s_j(\frac{1}{\omega})\,, & \omega_{j-1}\leq\omega\leq\omega_j\,,\qquad N<j\leq N+M\,,
\end{cases}
\end{multline}
where $s_j(x)=a_jx^3+b_jx^2+c_jx+d_j$, with the conditions
\begin{equation}\label{eq:conditions_spline_sigma}
\begin{split}
s_j(\omega_{j-1})&=\sigma^r_{j-1}\,,\\
s_j(\omega_{j})&=\sigma^r_{j}\,,\\
s_j'(\omega_{j-1})&=s_{j-1}'(\omega_{j-1})\,,\\
s_j''(\omega_{j-1})&=s_{j-1}''(\omega_{j-1})\,,\\
s_1'(0)&=0\,,
\end{split}
\end{equation}
for $j\leq N$, while
\begin{equation}\label{eq:conditions_spline_sigma_w_u}
\begin{split}
s_{N+1}\left(\frac{1}{\omega_{N}}\right)&=\sigma^r_{N}\,,\\
s_{N+1}\left(\frac{1}{\omega_{N+1}}\right)&=\sigma^r_{N+1}\,,\\
s_{N+1}'\left(\frac{1}{\omega_{N}}\right)&=s_{N}'\left(\omega_{N}\right)\,,\\
s_{N+1}''\left(\frac{1}{\omega_{N}}\right)&=s_{N}''\left(\omega_{N}\right)\,,
\end{split}
\end{equation}
and
\begin{equation}\label{eq:conditions_spline_sigma_u}
\begin{split}
s_j\left(\frac{1}{\omega_{j-1}}\right)&=\sigma^r_{j-1}\,,\\
s_j\left(\frac{1}{\omega_{j}}\right)&=\sigma^r_{j}\,,\\
s_j'\left(\frac{1}{\omega_{j-1}}\right)&=s_{j-1}'\left(\frac{1}{\omega_{j-1}}\right)\,,\\
s_j''\left(\frac{1}{\omega_{j-1}}\right)&=s_{j-1}''\left(\frac{1}{\omega_{j-1}}\right)\,,\\
\frac{\partial s_{N+M}\left(\frac{1}{\omega}\right)}{\partial \frac{1}{\omega}}\Big |_{\omega=\infty}&=0\,,
\end{split}
\end{equation}
for $N+1<j\leq N+M$. All the derivatives noted with $'$ and $''$ are taken with respect to $\omega$. The second condition for $j=N+M$ and the last condition in Eq.\eqref{eq:conditions_spline_sigma_u} make sure that there is no constant and no $1/\omega$ terms in $s_{N+M}(1/\omega)$ so that the integral of the spline converges. Note that this spline is not physically correct for $\omega\rightarrow\infty$ since the moments of $\sigma^r(\omega)$ are not defined. But this does not have any significant importance in the numerical solution if $\omega_{N+M-1}$ is chosen large enough.  The advantage of this spline is that it is very easy to implement.

Let us now define the kernel $K_{nj}$ in Eq. \eqref{eq:chijj_K_sigma}. For the low-frequency part of Eq.\eqref{eq:chijj_iqn_vs_sigma} we have
\begin{equation}
\begin{split}
\chi_{j_xj_x}^{lf}(iq_n)&=2\int_{0}^{\omega_N} \frac{d\omega}{\pi}\, \frac{\omega^2 }{\omega^2+q_n^2}\sigma^r(\omega)\,,\\
&=\frac{2}{\pi}\sum_{j=1}^N\int_{\omega_j-1}^{\omega_j} d\omega\, \frac{\omega^2 }{\omega^2+q_n^2}s_j(\omega)\,,\\
&=\frac{2}{\pi}\sum_{j=1}^N\int_{\omega_j-1}^{\omega_j} d\omega\, \frac{a_j\omega^5+b_j\omega^4+c_j\omega^3+d_j\omega^2 }{\omega^2+q_n^2}\,,
\end{split}
\end{equation}
where $\omega_0=0$. For $q_n\neq 0$, we obtain
\begin{multline}\label{eq:chijj_lf_qn}
\chi_{j_xj_x}^{lf}(iq_n)=\\
\frac{2}{\pi}\sum_{j=1}^N \Bigg(\left[\frac{q_n^4}{2}\ln\left(q_n^2+\omega^2\right)-\frac{q_n^2}{2}\omega^2+\frac{\omega^4}{4}\right]_{\omega_j-1}^{\omega_j}a_j\\
+\left[q_n^3\arctan\left(\frac{\omega}{q_n}\right)-q_n^2\omega+\frac{\omega^3}{3}\right]_{\omega_j-1}^{\omega_j}b_j\\
+\left[\frac{\omega^2}{2}-\frac{q_n^2}{2}\ln\left(q_n^2+\omega^2\right)\right]_{\omega_j-1}^{\omega_j}c_j\\
+\left[\omega-q_n\arctan\left(\frac{\omega}{q_n}\right)\right]_{\omega_j-1}^{\omega_j}d_j\Bigg)\,.
\end{multline}
and, for $q_n=0$,
\begin{equation}\label{eq:chijj_lf_qn0}
\begin{split}
\chi_{j_xj_x}^{lf}(0)&=\frac{2}{\pi}\sum_{j=1}^N\int_{\omega_j-1}^{\omega_j} d\omega\, \left(a_j\omega^3+b_j\omega^2+c_j\omega+d_j\right)\\
&=\frac{2}{\pi}\sum_{j=1}^N\left(\frac{\omega_j^4-\omega_{j-1}^4}{4}\,a_j+\frac{\omega_j^3-\omega_{j-1}^3}{3}\,b_j\right.\\
&\qquad\qquad\left.+\frac{\omega_j^2-\omega_{j-1}^2}{2}\,c_j+(\omega_j-\omega_{j-1})\,d_j\right)
\end{split}
\end{equation}

The high frequency part of Eq.\eqref{eq:chijj_iqn_vs_sigma} is
\begin{equation}
\chi_{j_xj_x}^{hf}(iq_n)=2\int_{\omega_N}^{\infty} \frac{d\omega}{\pi}\, \frac{\omega^2 }{\omega^2+q_n^2}\sigma^r(\omega)\,.
\end{equation}
Using
\begin{equation}
\omega =\frac{1}{u}\,,\quad\Rightarrow\quad d\omega=-\frac{1}{u^2}du\,,
\end{equation}
it becomes
\begin{equation}
\chi_{j_xj_x}^{hf}(iq_n)=\frac{2}{\pi}\int_{0}^{u_M} du \, \frac{1 }{ u^2+u^4q_n^2}\sigma^r\left(\frac{1}{u}\right)\,,
\end{equation}
where $u_M=1/\omega_N$. Now, with Eq.\eqref{eq:def_spline_sigma}, we have
\begin{equation}
\begin{split}
\chi_{j_xj_x}^{hf}(iq_n)&=\frac{2}{\pi}\sum_{j=1}^M \int_{u_j-1}^{u_j} du \, \frac{1 }{ u^2+u^4q_n^2}s_j(u)\\
&=\frac{2}{\pi}\sum_{j=1}^M \int_{u_{j-1}}^{u_j} du \, \frac{\alpha_j u^3+\beta_ju^2+\gamma_ju+\delta_j}{ u^2+u^4q_n^2}\,,
\end{split}
\end{equation}
where $1\leq j\leq M$, $u_0=0$, and we have used a different notation for the coefficients to match their indices with those of grid points in $u$, which are indexed in order of decreasing $\omega$. For  $q_n\neq 0$, we obtain
\begin{equation}\label{eq:chijj_qn_hf}
\begin{split}
\chi_{j_xj_x}^{hf}(iq_n)=&\frac{2}{\pi}\sum_{j=1}^M\Bigg(\left[\frac{1}{2q_n^2}\ln(1+q_n^2u^2)\right]_{u_j-1}^{u_j} \alpha_j\\
&+\left[\frac{1}{q_n}\arctan(q_nu)\right]_{u_j-1}^{u_j}\beta_j\\
&+\left[\ln(u) -\frac{1}{2}\ln(1+q_n^2u^2)\right]_{u_j-1}^{u_j}\gamma_j\\
&+\left[-\frac{1}{u}-q_n\arctan(q_nu)\right]_{u_j-1}^{u_j}\delta_j\Bigg)\,,
\end{split}
\end{equation}
and, for $q_n=0$,
\begin{equation}\label{eq:chijj_qn0_hf}
\begin{split}
\chi_{j_xj_x}^{hf}(0)&=\frac{2}{\pi}\sum_{j=1}^M \int_{u_j-1}^{u_j} du \, \left(\alpha_ju+\beta_j+\frac{\gamma_j}{ u}+\frac{\delta_j}{ u^2}\right)\\
&=\frac{2}{\pi}\sum_{j=1}^M \Bigg[\frac{u_j^2-u_{j-1}^2}{2}\alpha_j+(u_j-u_{j-1})\beta_j\\
&\qquad\qquad+\left(\ln\frac{u_j}{u_{j-1}}\right)\gamma_j+\left(\frac{1}{u_{j-1}}-\frac{1}{u_j}\right)\delta_j\Bigg] \,.
\end{split}
\end{equation}
Note that $\gamma_1=0$ and $\delta_1=0$, so that this expression has no problem with $u=0$.

The spline coefficients for the complete spline are obtained at the same time for both the low frequency and high frequency parts using the conditions \eqref{eq:conditions_spline_sigma}, \eqref{eq:conditions_spline_sigma_w_u} and \eqref{eq:conditions_spline_sigma_u}. Because the spline has two parts made of polynomials in $\omega$ and $u=1/\omega$, one must define the spline linear system for that particular case, but this is a rather straightforward task. Note that, for the high frequency part, the condition of continuity of the first derivatives with respect to $u$ or $\omega$ are the same, but that is not true for the continuity of the second derivatives. However, we do not observe any loss of accuracy if the continuity of the second derivatives with respect to $u$ instead of $\omega$ is used.

The conditions \eqref{eq:conditions_spline_sigma}, \eqref{eq:conditions_spline_sigma_w_u} and \eqref{eq:conditions_spline_sigma_u}, put in matrix form are written as $Av=\bar{\sigma}$, where $v$ is a vector containing the spline coefficients and $\bar{\sigma}$ is a vector containing the values $\sigma^r_j$ (that are repeated) and zeros otherwise. We want to obtain the coefficients as linear forms of the vector $\sigma^r$. For that we have to invert the matrix $A$ and first obtain a linear form for $v$. The first two lines of the condition \eqref{eq:conditions_spline_sigma} tell us that $\bar{\sigma}$ contains the elements  $\sigma^r_j$ repeated twice each, except for $\sigma^r_0$, and the other elements of $\bar{\sigma}$ are zeros. So if we define the matrix $P_{A\sigma}$ that sum up the pairs of columns in $A^{-1}$ corresponding to the same $\sigma^r_j$ and removes all columns that correspond to zeros in $\bar{\sigma}$ we get
\begin{equation}\label{eq:v_vs_sigma}
\begin{split}
v&=A^{-1}P_{A\sigma}\sigma^r\\
&=T_{v\sigma}\sigma^r
\end{split}
\end{equation}
Then the vectors formed with the coefficients are given by expressions like
\begin{equation}\label{eq:coeffs_vs_v}
a=P_av\,,\quad b=P_bv\,,\quad\ldots
\end{equation}
where $P_a$ extracts a column formed of all the coefficients of the cubic terms, $P_b$ the coefficients of the quadratic terms, etc. Now, expressions \eqref{eq:chijj_lf_qn} and \eqref{eq:chijj_lf_qn0} have the matrix form
\begin{equation}
\chi_{j_xj_x}^{lf}(iq_n)=\frac{2}{\pi} \Bigg(\bar{K}_n^{a}a+\bar{K}_n^{b}b+\bar{K}_n^{c}c+\bar{K}_n^{d}d\Bigg)\,,
\end{equation}
which becomes, using Eq.\eqref{eq:coeffs_vs_v} and Eq.\eqref{eq:v_vs_sigma},
\begin{equation}\label{eq:chijj_lf_K_sigma}
\chi_{j_xj_x}^{lf}(iq_n)=\frac{2}{\pi} \Bigg(\bar{K}_n^{a}P_a+\bar{K}_n^{b}P_b+\bar{K}_n^{c}P_c+\bar{K}_n^{d}P_d\Bigg)T_{v\sigma}\sigma^r\,.
\end{equation}

Similarly, if we define the projectors for the vectors $\alpha$, $\beta$, $\gamma$ and $\delta$, then Eq.\eqref{eq:chijj_qn_hf} and Eq.\eqref{eq:chijj_qn0_hf} take the form
\begin{equation}\label{eq:chijj_hf_K_sigma}
\chi_{j_xj_x}^{hf}(iq_n)=\frac{2}{\pi} \Bigg(\bar{K}_n^{\alpha}P_\alpha+\bar{K}_n^{\beta}P_\beta+\bar{K}_n^{\gamma}P_\gamma+\bar{K}_n^{\delta}P_\delta\Bigg)T_{v\sigma}\sigma^r\,.
\end{equation}
Summing Eq.\eqref{eq:chijj_lf_K_sigma} and Eq.\eqref{eq:chijj_hf_K_sigma} we obtain
\begin{equation}
\chi_{j_xj_x}(iq_n)=K_n\sigma^r\,,
\end{equation}
with
\begin{multline}
K_n=\frac{2}{\pi} \Bigg(\bar{K}_n^{a}P_a+\bar{K}_n^{b}P_b+\bar{K}_n^{c}P_c+\bar{K}_n^{d}P_d\\
\qquad\qquad  +\bar{K}_n^{\alpha}P_\alpha+\bar{K}_n^{\beta}P_\beta+\bar{K}_n^{\gamma}P_\gamma+\bar{K}_n^{\delta}P_\delta\Bigg)T_{v\sigma}\,.
\end{multline}

To conclude this section, a few other points must be addressed. First, the form of expression \eqref{eq:chijj_lf_qn} becomes numerically unstable when $q_n/\omega$ are large. For example, the first two terms in the high $q_n/\omega$ expansion of $\frac{q_n^4}{2}\ln\left(q_n^2+\omega^2\right)$ in $\bar{K}_n^{a}$ cancel out the terms $-\frac{q_n^2}{2}\omega^2$ and $\frac{\omega^4}{4}$. When $q_n/\omega$  increases, the magnitude of those terms becomes much larger than $\bar{K}_n^{a}$ itself so that if one computes numerically the three terms of $\bar{K}_n^{a}$ separately and then adds them, the finite precision error becomes larger than the true result. To overcome this problem one simply has to use the large $q_n/\omega$ expansion of $\bar{K}_n^{a}$ starting at a certain cutoff, instead of directly the form appearing in Eq.\eqref{eq:chijj_lf_qn}. The same kind of cancelation appears in the other terms of Eq.\eqref{eq:chijj_lf_qn} so that the large $q_n/\omega$ expansions must be used for those terms as well. In the case of Eq.\eqref{eq:chijj_qn_hf} it is when $q_nu$ is small that some simplifications occur when replacing the expressions by their expansions, which will also improve the accuracy of the numerical result.

Second, since we invert the spline matrix $A$ to define $T_{v\sigma}$ in Eq.\eqref{eq:v_vs_sigma}, it is preferable that this matrix be well-conditioned. For that purpose we define our grid to be uniform in $\omega$ for the low frequency part of the spline and uniform in $u$ for the high frequency part. The grid in $u$ is
\begin{equation}
u=0, \frac{1}{\omega_{N+M-1}}\,,\quad \frac{1}{\omega_{N+M-2}}\,,\quad\ldots , \frac{1}{\omega_{N}}\,.
\end{equation}
If $u_j-u_{j-1}$ is constant we have
\begin{multline}
\omega_{N+M-2}=\frac{\omega_{N+M-1}}{2}\,,\quad \omega_{N+M-3}=\frac{\omega_{N+M-1}}{3}\,,\\
\ldots\,, \qquad\omega_{N}=\frac{\omega_{N+M-1}}{M}\,,
\end{multline}
or
\begin{multline}
\omega_{N+1}=\frac{M\omega_{N}}{M-1}\,,\quad \omega_{N+2}=\frac{M\omega_{N}}{M-2}\,,\\
\ldots\,, \qquad\omega_{N+M-1}=M\omega_{N}\,.
\end{multline}
Also, to have a density of points $\omega_j$ that varies continuously when we change from high to low frequency, we assume that $\omega_{N+1}-\omega_{N}=\omega_{N}-\omega_{N-1}$. Defining $\Delta \omega_{lf}=\omega_{N}-\omega_{N-1}$, we have
\begin{equation}
\frac{M\omega_{N}}{M-1}-\omega_{N}=\Delta \omega_{lf}\,,
\end{equation}
so that
\begin{equation}
M=\frac{\omega_{N}}{\Delta \omega_{lf}}+1
\end{equation}
and, if $\omega_j-\omega_{j-1}$ is constant for $j\leq N$,
\begin{equation}
\Delta \omega_{lf}=\frac{\omega_{N}}{N}\,,
\end{equation}
so that $M=N+1$. Note that when choosing $\omega_N$ and $N$, we have to check that the last frequency $\omega_{N+M-1}=(N+1)\omega_N$ is large enough while not so large that it would uselessly make the calculation heavy.
To further improve the conditioning of the matrix $A$, we use normalized frequencies $\omega_j'=\omega_j/\omega_N$ so that $\omega_N'=1$ and $u_M'=1$. In the $j^{th}$ interval, if  $a_j'$, $b_j'$, $c_j'$ et $d_j'$ are the coefficients in the normalized grid, we have
\begin{equation}
\begin{split}
\sigma^r(\omega_j)&=a_j\omega_j^3+b_j\omega_j^2+c_j\omega_j+d_j\,,\\
&=a_j'\frac{\omega_j^3}{\omega_N^3}+b_j'\frac{\omega_j^2}{\omega_N^2}+c_j'\frac{\omega_j}{\omega_N}+d_j'\,,
\end{split}
\end{equation}
and therefore,
\begin{equation}
a_j=\frac{a_j'}{\omega_N^3}\,,\quad b_j=\frac{b_j'}{\omega_N^2}\,,\quad c_j=\frac{c_j'}{\omega_N}\,,\quad d_j=d_j'\,.
\end{equation}
Finally, for the high frequency part of the spline, we have
\begin{equation}
\alpha_j=\omega_N^3\alpha_j'\,,\quad\beta_j=\omega_N^2\beta_j'\,,\quad\gamma_j=\omega_N\gamma_j'\,,\quad\delta_j=\delta_j'\,.
\end{equation}

To end this section, we comment on the differential entropy and the minimization procedure. For the differential entropy \eqref{eq:diff_entropy}, we use a default model that is almost flat in the region where $\sigma^r(\omega)$ is expected to have its main structure, and that decreases gradually to very small values for frequencies much larger than the bandwidth. This ensures that the solution that we find is as unbiased as possible. The minimization of Eq.\eqref{eq:chi2_S_maxent} is performed using a Matlab routine called \textit{fmincon}, which uses a Trust-Region-Reflective algorithm that has been proven quite efficient with not as much tendency to get trapped into local minima as other optimization routines. Our procedure is to start with a very large value of $\alpha$ such that the minimization process gives a solution very close to the default model $m(\omega)$ (the minimization of Eq.\eqref{eq:diff_entropy} alone has in fact a solution proportional to $m(\omega)$). Then, $\alpha$ is decreased and a new optimal solution is found, using the previous solution as a starting point in the optimization routine. This step is then repeated until $\chi^2\approx N_{q_n}$ or $\chi^2$ does not decrease anymore when $\alpha$ is reduced. Using an augmented lagrangian method, we also include inequality constraints to restrict the roughness of the solution $\sigma^r(\omega_j)$. This roughness appears at some point in the procedure when we try to make $K_n\sigma^r$ closer to $\chi_{j_xj_x}(iq_n)$. It is related to oscillations present in $K_{nj}$ as a function of the frequency index $j$ for a given $n$. Those oscillations are in fact the price to pay to work with an accurate numerical integration method since they are present in all approximations more sophisticated than a piecewise linear function for $\sigma^r(\omega)$ in the numerical integration (think about the unequal weights in a Simpson $1/3$ or in gaussian quadratures). In fact, the oscillations in our $K_{nj}$ have an extremely small relative amplitude, but they
appear greatly amplified in the solution $\sigma^r(\omega_j)$ when the relative distance $|\chi_{j_xj_x}(iq_n)-K_n\sigma^r|$ becomes very small. The link between oscillations in $\sigma^r(\omega_j)$ and oscillations in
$K_{nj}$ for $n$ fixed is clear because they are correlated in the two functions. So they are not related to noise in the data $\chi_{j_xj_x}(iq_n)$. However, those inequality constraints are not absolutely necessary to  obtain good, quantitative, results with our approach. They just ensure that $\sigma^r(\omega_j)$ is a smooth function.
 
 %


\begin{thebibliography}{10}%
\makeatletter
\providecommand \@ifxundefined [1]{%
 \ifx #1\undefined \expandafter \@firstoftwo
 \else \expandafter \@secondoftwo
\fi
}%
\providecommand \@ifnum [1]{%
 \ifnum #1\expandafter \@firstoftwo
 \else \expandafter \@secondoftwo
\fi
}%
\providecommand \enquote [1]{``#1''}%
\providecommand \bibnamefont  [1]{#1}%
\providecommand \bibfnamefont [1]{#1}%
\providecommand \citenamefont [1]{#1}%
\providecommand\href[0]{\@sanitize\@href}%
\providecommand\@href[1]{\endgroup\@@startlink{#1}\endgroup\@@href}%
\providecommand\@@href[1]{#1\@@endlink}%
\providecommand \@sanitize [0]{\begingroup\catcode`\&12\catcode`\#12\relax}%
\@ifxundefined \pdfoutput {\@firstoftwo}{%
 \@ifnum{\z@=\pdfoutput}{\@firstoftwo}{\@secondoftwo}%
}{%
 \providecommand\@@startlink[1]{\leavevmode}%
 \providecommand\@@endlink[0]{}%
}{%
 \providecommand\@@startlink[1]{%
  \leavevmode
  \pdfstartlink
   attr{/Border[0 0 1 ]/H/I/C[0 1 1]}%
   user{/Subtype/Link/A<</Type/Action/S/URI/URI(#1)>>}%
  \relax
 }%
 \providecommand\@@endlink[0]{\pdfendlink}%
}%
\providecommand \url  [0]{\begingroup\@sanitize \@url }%
\providecommand \@url [1]{\endgroup\@href {#1}{\urlprefix}}%
\providecommand \urlprefix [0]{URL }%
\providecommand \Eprint[0]{\href }%
\@ifxundefined \urlstyle {%
  \providecommand \doi [1]{doi:\discretionary{}{}{}#1}%
}{%
  \providecommand \doi [0]{doi:\discretionary{}{}{}\begingroup
  \urlstyle{rm}\Url }%
}%
\providecommand \doibase [0]{http://dx.doi.org/}%
\providecommand \Doi[1]{\href{\doibase#1}}%
\providecommand \bibAnnote [3]{%
  \BibitemShut{#1}%
  \begin{quotation}\noindent
    \textsc{Key:}\ #2\\\textsc{Annotation:}\ #3%
  \end{quotation}%
}%
\providecommand \bibAnnoteFile [2]{%
  \IfFileExists{#2}{\bibAnnote {#1} {#2} {\input{#2}}}{}%
}%
\providecommand \typeout [0]{\immediate \write \m@ne }%
\providecommand \selectlanguage [0]{\@gobble}%
\providecommand \bibinfo [0]{\@secondoftwo}%
\providecommand \bibfield [0]{\@secondoftwo}%
\providecommand \translation [1]{[#1]}%
\providecommand \BibitemOpen[0]{}%
\providecommand \bibitemStop [0]{}%
\providecommand \bibitemNoStop [0]{.\EOS\space}%
\providecommand \EOS [0]{\spacefactor3000\relax}%
\providecommand \BibitemShut [1]{\csname bibitem#1\endcsname}%
\bibitem{Anderson:1987}%
  \BibitemOpen
  \bibfield{author}{%
  \bibinfo {author} {\bibfnamefont{P.}~\bibnamefont{Anderson}},\ }%
  \bibfield{journal}{%
  \bibinfo {journal} {Science}\ }%
  \textbf{\bibinfo {volume} {235}},\ \bibinfo {pages} {1196 } (\bibinfo {year}
  {1987})%
  \bibAnnoteFile{NoStop}{Anderson:1987}%
\bibitem{Powell:2006}%
  \BibitemOpen
  \bibfield{author}{%
  \bibinfo {author} {\bibfnamefont{B.~J.}\ \bibnamefont{Powell}}\ and\ \bibinfo
  {author} {\bibfnamefont{R.~H.}\ \bibnamefont{McKenzie}},\ }%
  \bibfield{journal}{%
  \bibinfo {journal} {J. Phys. Cond. Mat.}\ }%
  \textbf{\bibinfo {volume} {18}},\ \bibinfo {pages} {R827} (\bibinfo {year}
  {2006})%
  \bibAnnoteFile{NoStop}{Powell:2006}%
\bibitem{Nakano:1999}%
  \BibitemOpen
  \bibfield{author}{%
  \bibinfo {author} {\bibfnamefont{H.}~\bibnamefont{Nakano}}\ and\ \bibinfo
  {author} {\bibfnamefont{M.}~\bibnamefont{Imada}},\ }%
  \bibfield{journal}{%
  \bibinfo {journal} {Journal of the Physical Society of Japan}\ }%
  \textbf{\bibinfo {volume} {68}},\ \bibinfo {pages} {1458 } (\bibinfo {year}
  {1999/04/})%
  \bibAnnoteFile{NoStop}{Nakano:1999}%
\bibitem{Tohyama:2005}%
  \BibitemOpen
  \bibfield{author}{%
  \bibinfo {author} {\bibfnamefont{T.}~\bibnamefont{Tohyama}}, \bibinfo
  {author} {\bibfnamefont{Y.}~\bibnamefont{Inoue}}, \bibinfo {author}
  {\bibfnamefont{K.}~\bibnamefont{Tsutsui}},\ and\ \bibinfo {author}
  {\bibfnamefont{S.}~\bibnamefont{Maekawa}},\ }%
  \bibfield{journal}{%
  \bibinfo {journal} {Physical Review B (Condensed Matter and Materials
  Physics)}\ }%
  \textbf{\bibinfo {volume} {72}},\ \bibinfo {pages} {45113 } (\bibinfo {year}
  {2005/07/15})%
  \bibAnnoteFile{NoStop}{Tohyama:2005}%
\bibitem{Dagotto:1992}%
  \BibitemOpen
  \bibfield{author}{%
  \bibinfo {author} {\bibfnamefont{E.}~\bibnamefont{Dagotto}}, \bibinfo
  {author} {\bibfnamefont{A.}~\bibnamefont{Moreo}}, \bibinfo {author}
  {\bibfnamefont{F.}~\bibnamefont{Ortolani}}, \bibinfo {author}
  {\bibfnamefont{J.}~\bibnamefont{Riera}},\ and\ \bibinfo {author}
  {\bibfnamefont{D.}~\bibnamefont{Scalapino}},\ }%
  \bibfield{journal}{%
  \bibinfo {journal} {Physical Review B (Condensed Matter)}\ }%
  \textbf{\bibinfo {volume} {45}},\ \bibinfo {pages} {10107 } (\bibinfo {year}
  {1992/05/01})%
  \bibAnnoteFile{NoStop}{Dagotto:1992}%
\bibitem{Riera:1994}%
  \BibitemOpen
  \bibfield{author}{%
  \bibinfo {author} {\bibfnamefont{J.}~\bibnamefont{Riera}}\ and\ \bibinfo
  {author} {\bibfnamefont{E.}~\bibnamefont{Dagotto}},\ }%
  \bibfield{journal}{%
  \bibinfo {journal} {Physical Review B (Condensed Matter)}\ }%
  \textbf{\bibinfo {volume} {50}},\ \bibinfo {pages} {452 } (\bibinfo {year}
  {1994/07/01})%
  \bibAnnoteFile{NoStop}{Riera:1994}%
\bibitem{Li:1994}%
  \BibitemOpen
  \bibfield{author}{%
  \bibinfo {author} {\bibfnamefont{W.-Z.}\ \bibnamefont{Li}}, \bibinfo {author}
  {\bibfnamefont{F.}~\bibnamefont{Chen}}, \bibinfo {author}
  {\bibfnamefont{T.-F.}\ \bibnamefont{Xu}},\ and\ \bibinfo {author}
  {\bibfnamefont{H.-P.}\ \bibnamefont{Ying}},\ }%
  \bibfield{journal}{%
  \bibinfo {journal} {Communications in Theoretical Physics}\ }%
  \textbf{\bibinfo {volume} {22}},\ \bibinfo {pages} {273 } (\bibinfo {year}
  {1994/10/30})%
  \bibAnnoteFile{NoStop}{Li:1994}%
\bibitem{Scalapino92_1}%
  \BibitemOpen
  \bibfield{author}{%
  \bibinfo {author} {\bibfnamefont{D.~J.}\ \bibnamefont{Scalapino}}, \bibinfo
  {author} {\bibfnamefont{S.~R.}\ \bibnamefont{White}},\ and\ \bibinfo {author}
  {\bibfnamefont{S.~C.}\ \bibnamefont{Zhang}},\ }%
  \bibfield{journal}{%
  \Doi{10.1103/PhysRevLett.68.2830}{\bibinfo {journal} {Phys. Rev. Lett.}}\ }%
  \textbf{\bibinfo {volume} {68}},\ \bibinfo {pages} {2830} (\bibinfo {month}
  {May}\ \bibinfo {year} {1992})%
  \bibAnnoteFile{NoStop}{Scalapino92_1}%
\bibitem{comanac:2008}%
  \BibitemOpen
  \bibfield{author}{%
  \bibinfo {author} {\bibfnamefont{A.}~\bibnamefont{Comanac}}, \bibinfo
  {author} {\bibfnamefont{L.}~\bibnamefont{De'~Medici}}, \bibinfo {author}
  {\bibfnamefont{M.}~\bibnamefont{Capone}},\ and\ \bibinfo {author}
  {\bibfnamefont{A.~J.}\ \bibnamefont{Millis}},\ }%
  \bibfield{journal}{%
  \Doi{10.1038/nphys883}{\bibinfo {journal} {Nat Phys}}\ }%
  \textbf{\bibinfo {volume} {4}},\ \bibinfo {pages} {287} (\bibinfo {month}
  {Apr.}\ \bibinfo {year} {2008})%
  \bibAnnoteFile{NoStop}{comanac:2008}%
\bibitem{Chakraborty:2008}%
  \BibitemOpen
  \bibfield{author}{%
  \bibinfo {author} {\bibfnamefont{S.}~\bibnamefont{Chakraborty}}, \bibinfo
  {author} {\bibfnamefont{D.}~\bibnamefont{Galanakis}},\ and\ \bibinfo {author}
  {\bibfnamefont{P.}~\bibnamefont{Phillips}},\ }%
  \bibfield{journal}{%
  \Doi{10.1103/PhysRevB.78.212504}{\bibinfo {journal} {Phys. Rev. B}}\ }%
  \textbf{\bibinfo {volume} {78}},\ \bibinfo {pages} {212504} (\bibinfo {month}
  {Dec}\ \bibinfo {year} {2008})%
  \bibAnnoteFile{NoStop}{Chakraborty:2008}%
\bibitem{Mancini:2004}%
  \BibitemOpen
  \bibfield{author}{%
  \bibinfo {author} {\bibfnamefont{F.}~\bibnamefont{Mancini}}\ and\ \bibinfo
  {author} {\bibfnamefont{A.}~\bibnamefont{Avella}},\ }%
  \bibfield{journal}{%
  \bibinfo {journal} {Advances in Physics}\ }%
  \textbf{\bibinfo {volume} {53}},\ \bibinfo {pages} {537 } (\bibinfo {year}
  {2004})%
  \bibAnnoteFile{NoStop}{Mancini:2004}%
\bibitem{Maier:2003}%
  \BibitemOpen
  \bibfield{author}{%
  \bibinfo {author} {\bibfnamefont{T.~A.}\ \bibnamefont{{Maier}}},\ }%
  \bibfield{journal}{%
  \bibinfo {journal} {ArXiv Condensed Matter e-prints}}%
   (\bibinfo {month} {Dec.}\ \bibinfo {year} {2003}),\
  \Eprint{http://arxiv.org/abs/arXiv:cond-mat/0312447}{arXiv:cond-mat/0312447}%
  \bibAnnoteFile{NoStop}{Maier:2003}%
\bibitem{Hettler:2000}%
  \BibitemOpen
  \bibfield{author}{%
  \bibinfo {author} {\bibfnamefont{M.~H.}\ \bibnamefont{Hettler}}, \bibinfo
  {author} {\bibfnamefont{M.}~\bibnamefont{Mukherjee}}, \bibinfo {author}
  {\bibfnamefont{M.}~\bibnamefont{Jarrell}},\ and\ \bibinfo {author}
  {\bibfnamefont{H.~R.}\ \bibnamefont{Krishnamurthy}},\ }%
  \bibfield{journal}{%
  \bibinfo {journal} {Phys. Rev. B}\ }%
  \textbf{\bibinfo {volume} {61}},\ \bibinfo {pages} {12739 } (\bibinfo {year}
  {2000})%
  \bibAnnoteFile{NoStop}{Hettler:2000}%
\bibitem{Haule:2007}%
  \BibitemOpen
  \bibfield{author}{%
  \bibinfo {author} {\bibfnamefont{K.}~\bibnamefont{Haule}}\ and\ \bibinfo
  {author} {\bibfnamefont{G.}~\bibnamefont{Kotliar}},\ }%
  \bibfield{journal}{%
  \Doi{10.1103/PhysRevB.76.104509}{\bibinfo {journal} {Physical Review B
  (Condensed Matter and Materials Physics)}}\ }%
  \textbf{\bibinfo {volume} {76}},\ \bibinfo {eid} {104509} (\bibinfo {year}
  {2007})%
  \bibAnnoteFile{NoStop}{Haule:2007}%
\bibitem{Haule:2007_2}%
  \BibitemOpen
  \bibfield{author}{%
  \bibinfo {author} {\bibfnamefont{K.}~\bibnamefont{Haule}}\ and\ \bibinfo
  {author} {\bibfnamefont{G.}~\bibnamefont{Kotliar}},\ }%
  \bibfield{journal}{%
  \bibinfo {journal} {Europhysics Letters}\ }%
  \textbf{\bibinfo {volume} {77}},\ \bibinfo {pages} {6 pp. } (\bibinfo {year}
  {2007/01/}),\ ISSN \bibinfo {issn} {0295-5075}%
  \bibAnnoteFile{NoStop}{Haule:2007_2}%
\bibitem{Lin:2009_1}%
  \BibitemOpen
  \bibfield{author}{%
  \bibinfo {author} {\bibfnamefont{N.}~\bibnamefont{{Lin}}}, \bibinfo {author}
  {\bibfnamefont{E.}~\bibnamefont{{Gull}}},\ and\ \bibinfo {author}
  {\bibfnamefont{A.~J.}\ \bibnamefont{{Millis}}},\ }%
  \bibfield{journal}{%
  \Doi{10.1103/PhysRevB.80.161105}{\bibinfo {journal} {\prb}}\ }%
  \textbf{\bibinfo {volume} {80}},\ \bibinfo {pages} {161105} (\bibinfo {month}
  {Oct.}\ \bibinfo {year} {2009}),\
  \Eprint{http://arxiv.org/abs/0909.1625}{arXiv:0909.1625 [cond-mat.str-el]}%
  \bibAnnoteFile{NoStop}{Lin:2009_1}%
\bibitem{Okamoto:2010}%
  \BibitemOpen
  \bibfield{author}{%
  \bibinfo {author} {\bibfnamefont{S.}~\bibnamefont{Okamoto}}, \bibinfo
  {author} {\bibfnamefont{D.}~\bibnamefont{S\'en\'echal}}, \bibinfo {author}
  {\bibfnamefont{M.}~\bibnamefont{Civelli}},\ and\ \bibinfo {author}
  {\bibfnamefont{A.-M.~S.}\ \bibnamefont{Tremblay}},\ }%
  \bibfield{journal}{%
  \Doi{10.1103/PhysRevB.82.180511}{\bibinfo {journal} {Phys. Rev. B}}\ }%
  \textbf{\bibinfo {volume} {82}},\ \bibinfo {pages} {180511} (\bibinfo {month}
  {Nov}\ \bibinfo {year} {2010})%
  \bibAnnoteFile{NoStop}{Okamoto:2010}%
\bibitem{Kotliar:2001}%
  \BibitemOpen
  \bibfield{author}{%
  \bibinfo {author} {\bibfnamefont{G.}~\bibnamefont{Kotliar}}, \bibinfo
  {author} {\bibfnamefont{S.~Y.}\ \bibnamefont{Savrasov}}, \bibinfo {author}
  {\bibfnamefont{G.}~\bibnamefont{P{\'a}lsson}},\ and\ \bibinfo {author}
  {\bibfnamefont{G.}~\bibnamefont{Biroli}},\ }%
  \bibfield{journal}{%
  \bibinfo {journal} {Phys. Rev. Lett.}\ }%
  \textbf{\bibinfo {volume} {87}},\ \bibinfo {pages} {186401} (\bibinfo {year}
  {2001})%
  \bibAnnoteFile{NoStop}{Kotliar:2001}%
\bibitem{Maier:2005}%
  \BibitemOpen
  \bibfield{author}{%
  \bibinfo {author} {\bibfnamefont{T.}~\bibnamefont{Maier}}, \bibinfo {author}
  {\bibfnamefont{M.}~\bibnamefont{Jarrell}}, \bibinfo {author}
  {\bibfnamefont{T.}~\bibnamefont{Pruschke}},\ and\ \bibinfo {author}
  {\bibfnamefont{M.~H.}\ \bibnamefont{Hettler}},\ }%
  \bibfield{journal}{%
  \bibinfo {journal} {Reviews of Modern Physics}\ }%
  \textbf{\bibinfo {volume} {77}},\ \bibinfo {pages} {1027} (\bibinfo {year}
  {2005})%
  \bibAnnoteFile{NoStop}{Maier:2005}%
\bibitem{KotliarRMP:2006}%
  \BibitemOpen
  \bibfield{author}{%
  \bibinfo {author} {\bibfnamefont{G.}~\bibnamefont{Kotliar}}, \bibinfo
  {author} {\bibfnamefont{S.~Y.}\ \bibnamefont{Savrasov}}, \bibinfo {author}
  {\bibfnamefont{K.}~\bibnamefont{Haule}}, \bibinfo {author}
  {\bibfnamefont{V.~S.}\ \bibnamefont{Oudovenko}}, \bibinfo {author}
  {\bibfnamefont{O.}~\bibnamefont{Parcollet}},\ and\ \bibinfo {author}
  {\bibfnamefont{C.~A.}\ \bibnamefont{Marianetti}},\ }%
  \bibfield{journal}{%
  \Doi{10.1103/RevModPhys.78.865}{\bibinfo {journal} {Reviews of Modern
  Physics}}\ }%
  \textbf{\bibinfo {volume} {78}},\ \bibinfo {eid} {865} (\bibinfo {year}
  {2006})%
  \bibAnnoteFile{NoStop}{KotliarRMP:2006}%
\bibitem{Moriya:1990}%
  \BibitemOpen
  \bibfield{author}{%
  \bibinfo {author} {\bibfnamefont{T.}~\bibnamefont{Moriya}}, \bibinfo {author}
  {\bibfnamefont{Y.}~\bibnamefont{Takahashi}},\ and\ \bibinfo {author}
  {\bibfnamefont{K.}~\bibnamefont{Ueda}},\ }%
  \bibfield{journal}{%
  \bibinfo {journal} {Journal of the Physical Society of Japan}\ }%
  \textbf{\bibinfo {volume} {59}},\ \bibinfo {pages} {2905 } (\bibinfo {year}
  {1990/08})%
  \bibAnnoteFile{NoStop}{Moriya:1990}%
\bibitem{Hlubina:1995}%
  \BibitemOpen
  \bibfield{author}{%
  \bibinfo {author} {\bibfnamefont{R.}~\bibnamefont{Hlubina}}\ and\ \bibinfo
  {author} {\bibfnamefont{T.}~\bibnamefont{Rice}},\ }%
  \bibfield{journal}{%
  \bibinfo {journal} {Physical Review B (Condensed Matter)}\ }%
  \textbf{\bibinfo {volume} {51}},\ \bibinfo {pages} {9253 } (\bibinfo {year}
  {1995/04/01})%
  \bibAnnoteFile{NoStop}{Hlubina:1995}%
\bibitem{Rosch:1999}%
  \BibitemOpen
  \bibfield{author}{%
  \bibinfo {author} {\bibfnamefont{A.}~\bibnamefont{Rosch}},\ }%
  \bibfield{journal}{%
  \Doi{10.1103/PhysRevLett.82.4280}{\bibinfo {journal} {Phys. Rev. Lett.}}\ }%
  \textbf{\bibinfo {volume} {82}},\ \bibinfo {pages} {4280} (\bibinfo {month}
  {May}\ \bibinfo {year} {1999})%
  \bibAnnoteFile{NoStop}{Rosch:1999}%
\bibitem{RoschRMP:2007}%
  \BibitemOpen
  \bibfield{author}{%
  \bibinfo {author} {\bibfnamefont{H.~v.}\ \bibnamefont{L\"ohneysen}}, \bibinfo
  {author} {\bibfnamefont{A.}~\bibnamefont{Rosch}}, \bibinfo {author}
  {\bibfnamefont{M.}~\bibnamefont{Vojta}},\ and\ \bibinfo {author}
  {\bibfnamefont{P.}~\bibnamefont{W\"olfle}},\ }%
  \bibfield{journal}{%
  \Doi{10.1103/RevModPhys.79.1015}{\bibinfo {journal} {Rev. Mod. Phys.}}\ }%
  \textbf{\bibinfo {volume} {79}},\ \bibinfo {pages} {1015} (\bibinfo {month}
  {Aug}\ \bibinfo {year} {2007})%
  \bibAnnoteFile{NoStop}{RoschRMP:2007}%
\bibitem{Wermbter:1993}%
  \BibitemOpen
  \bibfield{author}{%
  \bibinfo {author} {\bibfnamefont{S.}~\bibnamefont{Wermbter}}\ and\ \bibinfo
  {author} {\bibfnamefont{L.}~\bibnamefont{Tewordt}},\ }%
  \bibfield{journal}{%
  \Doi{10.1103/PhysRevB.48.10514}{\bibinfo {journal} {Phys. Rev. B}}\ }%
  \textbf{\bibinfo {volume} {48}},\ \bibinfo {pages} {10514} (\bibinfo {month}
  {Oct}\ \bibinfo {year} {1993})%
  \bibAnnoteFile{NoStop}{Wermbter:1993}%
\bibitem{Dahm:1994}%
  \BibitemOpen
  \bibfield{author}{%
  \bibinfo {author} {\bibfnamefont{T.}~\bibnamefont{Dahm}}, \bibinfo {author}
  {\bibfnamefont{L.}~\bibnamefont{Tewordt}},\ and\ \bibinfo {author}
  {\bibfnamefont{S.}~\bibnamefont{Wermbter}},\ }%
  \bibfield{journal}{%
  \bibinfo {journal} {Physical Review B (Condensed Matter)}\ }%
  \textbf{\bibinfo {volume} {49}},\ \bibinfo {pages} {748 } (\bibinfo {year}
  {1994/01/01})%
  \bibAnnoteFile{NoStop}{Dahm:1994}%
\bibitem{Kontani:2007}%
  \BibitemOpen
  \bibfield{author}{%
  \bibinfo {author} {\bibfnamefont{H.}~\bibnamefont{Kontani}},\ }%
  \bibfield{journal}{%
  \bibinfo {journal} {Journal of the Physical Society of Japan}\ }%
  \textbf{\bibinfo {volume} {76}},\ \bibinfo {pages} {074707 } (\bibinfo {year}
  {2007/07})%
  \bibAnnoteFile{NoStop}{Kontani:2007}%
\bibitem{Kontani:1999}%
  \BibitemOpen
  \bibfield{author}{%
  \bibinfo {author} {\bibfnamefont{H.}~\bibnamefont{Kontani}}, \bibinfo
  {author} {\bibfnamefont{K.}~\bibnamefont{Kanki}},\ and\ \bibinfo {author}
  {\bibfnamefont{K.}~\bibnamefont{Ueda}},\ }%
  \bibfield{journal}{%
  \bibinfo {journal} {Physical Review B (Condensed Matter)}\ }%
  \textbf{\bibinfo {volume} {59}},\ \bibinfo {pages} {14723 } (\bibinfo {year}
  {1999/06/01})%
  \bibAnnoteFile{NoStop}{Kontani:1999}%
\bibitem{Kontani:2008}%
  \BibitemOpen
  \bibfield{author}{%
  \bibinfo {author} {\bibfnamefont{H.}~\bibnamefont{Kontani}},\ }%
  \bibfield{journal}{%
  \bibinfo {journal} {Reports on Progress in Physics}\ }%
  \textbf{\bibinfo {volume} {71}},\ \bibinfo {pages} {026501 } (\bibinfo {year}
  {2008/02/})%
  \bibAnnoteFile{NoStop}{Kontani:2008}%
\bibitem{Yanase:2002}%
  \BibitemOpen
  \bibfield{author}{%
  \bibinfo {author} {\bibfnamefont{Y.}~\bibnamefont{Yanase}},\ }%
  \bibfield{journal}{%
  \bibinfo {journal} {Journal of the Physical Society of Japan}\ }%
  \textbf{\bibinfo {volume} {71}},\ \bibinfo {pages} {278 } (\bibinfo {year}
  {2002/01/})%
  \bibAnnoteFile{NoStop}{Yanase:2002}%
\bibitem{Hartnoll:2011}%
  \BibitemOpen
  \bibfield{author}{%
  \bibinfo {author} {\bibfnamefont{S.~A.}\ \bibnamefont{Hartnoll}}, \bibinfo
  {author} {\bibfnamefont{D.~M.}\ \bibnamefont{Hofman}}, \bibinfo {author}
  {\bibfnamefont{M.~A.}\ \bibnamefont{Metlitski}},\ and\ \bibinfo {author}
  {\bibfnamefont{S.}~\bibnamefont{Sachdev}},\ }%
  \bibinfo {journal} {Private communication}%
  \bibAnnoteFile{NoStop}{Hartnoll:2011}%
\bibitem{Sadovskii:2002}%
  \BibitemOpen
\bibfield{journal}{%
    }%
  \bibfield{author}{%
  \bibinfo {author} {\bibfnamefont{M.~V.}\ \bibnamefont{Sadovskii}}\ and\
  \bibinfo {author} {\bibfnamefont{N.~A.}\ \bibnamefont{Strigina}},\ }%
  \bibfield{journal}{%
  \bibinfo {journal} {J. Exp. Theor. Phys. Lett.}\ }%
  \textbf{\bibinfo {volume} {95}},\ \bibinfo {pages} {526} (\bibinfo {year}
  {2002})%
  \bibAnnoteFile{NoStop}{Sadovskii:2002}%
\bibitem{LinMillis:2011}%
  \BibitemOpen
  \bibfield{author}{%
  \bibinfo {author} {\bibfnamefont{J.}~\bibnamefont{Lin}}\ and\ \bibinfo
  {author} {\bibfnamefont{A.~J.}\ \bibnamefont{Millis}},\ }%
  \bibfield{journal}{%
  \Doi{10.1103/PhysRevB.83.125108}{\bibinfo {journal} {Phys. Rev. B}}\ }%
  \textbf{\bibinfo {volume} {83}},\ \bibinfo {pages} {125108} (\bibinfo {month}
  {Mar}\ \bibinfo {year} {2011})%
  \bibAnnoteFile{NoStop}{LinMillis:2011}%
\bibitem{Maebashi:1997}%
  \BibitemOpen
  \bibfield{author}{%
  \bibinfo {author} {\bibfnamefont{H.}~\bibnamefont{Maebashi}}\ and\ \bibinfo
  {author} {\bibfnamefont{H.}~\bibnamefont{Fukuyama}},\ }%
  \bibfield{journal}{%
  \bibinfo {journal} {Journal of the Physical Society of Japan}\ }%
  \textbf{\bibinfo {volume} {66}},\ \bibinfo {pages} {3577 } (\bibinfo {year}
  {1997/11/})%
  \bibAnnoteFile{NoStop}{Maebashi:1997}%
\bibitem{Maebashi:1998}%
  \BibitemOpen
  \bibfield{author}{%
  \bibinfo {author} {\bibfnamefont{H.}~\bibnamefont{Maebashi}}\ and\ \bibinfo
  {author} {\bibfnamefont{H.}~\bibnamefont{Fukuyama}},\ }%
  \bibfield{journal}{%
  \bibinfo {journal} {Journal of the Physical Society of Japan}\ }%
  \textbf{\bibinfo {volume} {67}},\ \bibinfo {pages} {242 } (\bibinfo {year}
  {1998/01/})%
  \bibAnnoteFile{NoStop}{Maebashi:1998}%
\bibitem{Vilk:1994}%
  \BibitemOpen
  \bibfield{author}{%
  \bibinfo {author} {\bibfnamefont{Y.~M.}\ \bibnamefont{Vilk}}, \bibinfo
  {author} {\bibfnamefont{L.}~\bibnamefont{Chen}},\ and\ \bibinfo {author}
  {\bibfnamefont{A.-M.~S.}\ \bibnamefont{Tremblay}},\ }%
  \bibfield{journal}{%
  \bibinfo {journal} {Phys. Rev. B}\ }%
  \textbf{\bibinfo {volume} {49}},\ \bibinfo {pages} {13267 } (\bibinfo {year}
  {1994})%
  \bibAnnoteFile{NoStop}{Vilk:1994}%
\bibitem{Vilk:1997}%
  \BibitemOpen
  \bibfield{author}{%
  \bibinfo {author} {\bibfnamefont{Y.}~\bibnamefont{Vilk}}\ and\ \bibinfo
  {author} {\bibfnamefont{A.-M.}\ \bibnamefont{Tremblay}},\ }%
  \bibfield{journal}{%
  \bibinfo {journal} {J. Phys I (France)}\ }%
  \textbf{\bibinfo {volume} {7}},\ \bibinfo {pages} {1309 } (\bibinfo {year}
  {1997})%
  \bibAnnoteFile{NoStop}{Vilk:1997}%
\bibitem{Allen:2003}%
  \BibitemOpen
  \bibfield{author}{%
  \bibinfo {author} {\bibfnamefont{S.}~\bibnamefont{Allen}}, \bibinfo {author}
  {\bibfnamefont{A.-M.}\ \bibnamefont{Tremblay}},\ and\ \bibinfo {author}
  {\bibfnamefont{Y.~M.}\ \bibnamefont{Vilk}},\ }%
  in\ \emph{\bibinfo {booktitle} {Theoretical Methods for Strongly Correlated
  Electrons}},\ \bibinfo {editor} {edited by\ \bibinfo {editor}
  {\bibfnamefont{D.}~\bibnamefont{S{\'e}n{\'e}chal}}, \bibinfo {editor}
  {\bibfnamefont{C.}~\bibnamefont{Bourbonnais}},\ and\ \bibinfo {editor}
  {\bibfnamefont{A.-M.}\ \bibnamefont{Tremblay}}}\ (\bibinfo {year} {2003})\
  \Eprint{http://arxiv.org/abs/arXiv:cond-mat/0110130}{arXiv:cond-mat/0110130}%
  \bibAnnoteFile{NoStop}{Allen:2003}%
\bibitem{TremblayMancini:2011}%
  \BibitemOpen
  \bibfield{author}{%
  \bibinfo {author} {\bibfnamefont{A.-M.~S.}\ \bibnamefont{Tremblay}},\ }%
  \enquote{\bibinfo {title} {Theoretical methods for strongly correlated
  systems},}\ \ (\bibinfo {publisher} {Springer, New York},\ \bibinfo {year}
  {2011})\ Chap.~\bibinfo {chapter} {13}, p.\ \bibinfo {pages} {409},\
  \Eprint{http://arxiv.org/abs/arXiv:cond-mat/1107.1534}{arXiv:cond-mat/1107.1%
534}%
  \bibAnnoteFile{NoStop}{TremblayMancini:2011}%
\bibitem{Mermin:1966}%
  \BibitemOpen
  \bibfield{author}{%
  \bibinfo {author} {\bibfnamefont{N.~D.}\ \bibnamefont{Mermin}}\ and\ \bibinfo
  {author} {\bibfnamefont{H.}~\bibnamefont{Wagner}},\ }%
  \bibfield{journal}{%
  \Doi{10.1103/PhysRevLett.17.1133}{\bibinfo {journal} {Phys. Rev. Lett.}}\ }%
  \textbf{\bibinfo {volume} {17}},\ \bibinfo {pages} {1133} (\bibinfo {month}
  {Nov}\ \bibinfo {year} {1966})%
  \bibAnnoteFile{NoStop}{Mermin:1966}%
\bibitem{Hohenberg:1966}%
  \BibitemOpen
  \bibfield{author}{%
  \bibinfo {author} {\bibfnamefont{P.~C.}\ \bibnamefont{Hohenberg}},\ }%
  \bibfield{journal}{%
  \Doi{10.1103/PhysRev.158.383}{\bibinfo {journal} {Phys. Rev.}}\ }%
  \textbf{\bibinfo {volume} {158}},\ \bibinfo {pages} {383} (\bibinfo {month}
  {Jun}\ \bibinfo {year} {1967})%
  \bibAnnoteFile{NoStop}{Hohenberg:1966}%
\bibitem{Vilk:1995}%
  \BibitemOpen
  \bibfield{author}{%
  \bibinfo {author} {\bibfnamefont{Y.}~\bibnamefont{Vilk}}\ and\ \bibinfo
  {author} {\bibfnamefont{A.-M.}\ \bibnamefont{Tremblay}},\ }%
  \bibfield{journal}{%
  \bibinfo {journal} {J. Phys. Chem. Solids (UK)}\ }%
  \textbf{\bibinfo {volume} {56}},\ \bibinfo {pages} {1769 } (\bibinfo {year}
  {1995})%
  \bibAnnoteFile{NoStop}{Vilk:1995}%
\bibitem{Moukouri:2000}%
  \BibitemOpen
  \bibfield{author}{%
  \bibinfo {author} {\bibfnamefont{S.}~\bibnamefont{Moukouri}}, \bibinfo
  {author} {\bibfnamefont{S.}~\bibnamefont{Allen}}, \bibinfo {author}
  {\bibfnamefont{F.}~\bibnamefont{Lemay}}, \bibinfo {author}
  {\bibfnamefont{B.}~\bibnamefont{Kyung}}, \bibinfo {author}
  {\bibfnamefont{D.}~\bibnamefont{Poulin}}, \bibinfo {author}
  {\bibfnamefont{Y.~M.}\ \bibnamefont{Vilk}},\ and\ \bibinfo {author}
  {\bibfnamefont{A.-M.~S.}\ \bibnamefont{Tremblay}},\ }%
  \bibfield{journal}{%
  \bibinfo {journal} {Phys. Rev. B}\ }%
  \textbf{\bibinfo {volume} {61}},\ \bibinfo {pages} {7887 } (\bibinfo {year}
  {2000})%
  \bibAnnoteFile{NoStop}{Moukouri:2000}%
\bibitem{Kyung:2004}%
  \BibitemOpen
  \bibfield{author}{%
  \bibinfo {author} {\bibfnamefont{B.}~\bibnamefont{Kyung}}, \bibinfo {author}
  {\bibfnamefont{V.}~\bibnamefont{Hankevych}}, \bibinfo {author}
  {\bibfnamefont{A.-M.}\ \bibnamefont{Dar\'e}},\ and\ \bibinfo {author}
  {\bibfnamefont{A.-M.~S.}\ \bibnamefont{Tremblay}},\ }%
  \bibfield{journal}{%
  \bibinfo {journal} {Phys. Rev. Lett.}\ }%
  \textbf{\bibinfo {volume} {93}},\ \bibinfo {pages} {147004} (\bibinfo {year}
  {2004})%
  \bibAnnoteFile{NoStop}{Kyung:2004}%
\bibitem{Bickers:1989}%
  \BibitemOpen
  \bibfield{author}{%
  \bibinfo {author} {\bibfnamefont{N.}~\bibnamefont{Bickers}}\ and\ \bibinfo
  {author} {\bibfnamefont{D.}~\bibnamefont{Scalapino}},\ }%
  \bibfield{journal}{%
  \bibinfo {journal} {Ann. Phys. (USA)}\ }%
  \textbf{\bibinfo {volume} {193}},\ \bibinfo {pages} {206 } (\bibinfo {year}
  {1989})%
  \bibAnnoteFile{NoStop}{Bickers:1989}%
\bibitem{Bickers_dwave:1989}%
  \BibitemOpen
  \bibfield{author}{%
  \bibinfo {author} {\bibfnamefont{N.~E.}\ \bibnamefont{Bickers}}, \bibinfo
  {author} {\bibfnamefont{D.~J.}\ \bibnamefont{Scalapino}},\ and\ \bibinfo
  {author} {\bibfnamefont{S.~R.}\ \bibnamefont{White}},\ }%
  \bibfield{journal}{%
  \bibinfo {journal} {Phys. Rev. Lett.}\ }%
  \textbf{\bibinfo {volume} {62}},\ \bibinfo {pages} {961} (\bibinfo {year}
  {1989})%
  \bibAnnoteFile{NoStop}{Bickers_dwave:1989}%
\bibitem{Moriya:2003}%
  \BibitemOpen
  \bibfield{author}{%
  \bibinfo {author} {\bibfnamefont{T.}~\bibnamefont{Moriya}}\ and\ \bibinfo
  {author} {\bibfnamefont{K.}~\bibnamefont{Ueda}},\ }%
  \bibfield{journal}{%
  \bibinfo {journal} {Rep. Prog. Phys.}\ }%
  \textbf{\bibinfo {volume} {66}},\ \bibinfo {pages} {1299} (\bibinfo {year}
  {2003})%
  \bibAnnoteFile{NoStop}{Moriya:2003}%
\bibitem{Lonzarich:1985}%
  \BibitemOpen
  \bibfield{author}{%
  \bibinfo {author} {\bibfnamefont{G.~G.}\ \bibnamefont{Lonzarich}}\ and\
  \bibinfo {author} {\bibfnamefont{L.}~\bibnamefont{Taillefer}},\ }%
  \bibfield{journal}{%
  \bibinfo {journal} {J. Phys. C}\ }%
  \textbf{\bibinfo {volume} {18}},\ \bibinfo {pages} {4339} (\bibinfo {year}
  {1985})%
  \bibAnnoteFile{NoStop}{Lonzarich:1985}%
\bibitem{Moriya:1985}%
  \BibitemOpen
  \bibfield{author}{%
  \bibinfo {author} {\bibfnamefont{T.}~\bibnamefont{Moriya}},\ }%
  \emph{\bibinfo {title} {Spin Fluctuations in Itinerant Electron Magnetism}}\
  (\bibinfo {publisher} {Springer-Verlag, Berlin},\ \bibinfo {year} {1985})%
  \bibAnnoteFile{NoStop}{Moriya:1985}%
\bibitem{Dzyaloshinskii:1987}%
  \BibitemOpen
  \bibfield{author}{%
  \bibinfo {author} {\bibfnamefont{I.}~\bibnamefont{Dzyaloshinskii}},\ }%
  \bibfield{journal}{%
  \bibinfo {journal} {Sov. Phys. JETP}\ }%
  \textbf{\bibinfo {volume} {66}},\ \bibinfo {pages} {848} (\bibinfo {year}
  {1987})%
  \bibAnnoteFile{NoStop}{Dzyaloshinskii:1987}%
\bibitem{Schulz:1987}%
  \BibitemOpen
  \bibfield{author}{%
  \bibinfo {author} {\bibfnamefont{H.~J.}\ \bibnamefont{Schulz}},\ }%
  \bibfield{journal}{%
  \bibinfo {journal} {Europhys. Lett.}\ }%
  \textbf{\bibinfo {volume} {4}},\ \bibinfo {pages} {609} (\bibinfo {year}
  {1987})%
  \bibAnnoteFile{NoStop}{Schulz:1987}%
\bibitem{Lederer:1987}%
  \BibitemOpen
  \bibfield{author}{%
  \bibinfo {author} {\bibfnamefont{P.}~\bibnamefont{Lederer}}, \bibinfo
  {author} {\bibfnamefont{G.}~\bibnamefont{Montambaux}},\ and\ \bibinfo
  {author} {\bibfnamefont{D.}~\bibnamefont{Poilblanc}},\ }%
  \bibfield{journal}{%
  \bibinfo {journal} {J. Phys.}\ }%
  \textbf{\bibinfo {volume} {48}},\ \bibinfo {pages} {1613} (\bibinfo {year}
  {1987})%
  \bibAnnoteFile{NoStop}{Lederer:1987}%
\bibitem{Honerkamp:2001}%
  \BibitemOpen
  \bibfield{author}{%
  \bibinfo {author} {\bibfnamefont{C.}~\bibnamefont{Honerkamp}}\ and\ \bibinfo
  {author} {\bibfnamefont{M.}~\bibnamefont{Salmhofer}},\ }%
  \bibfield{journal}{%
  \bibinfo {journal} {Phys. Rev. Lett.}\ }%
  \textbf{\bibinfo {volume} {87}},\ \bibinfo {pages} {187004} (\bibinfo {year}
  {2001})%
  \bibAnnoteFile{NoStop}{Honerkamp:2001}%
\bibitem{Logan:1996}%
  \BibitemOpen
  \bibfield{author}{%
  \bibinfo {author} {\bibfnamefont{M.~A.}\ \bibnamefont{Tusch}}, \bibinfo
  {author} {\bibfnamefont{Y.~H.}\ \bibnamefont{Szczech}},\ and\ \bibinfo
  {author} {\bibfnamefont{D.~E.}\ \bibnamefont{Logan}},\ }%
  \bibfield{journal}{%
  \Doi{10.1103/PhysRevB.53.5505}{\bibinfo {journal} {Phys. Rev. B}}\ }%
  \textbf{\bibinfo {volume} {53}},\ \bibinfo {pages} {5505} (\bibinfo {month}
  {Mar}\ \bibinfo {year} {1996})%
  \bibAnnoteFile{NoStop}{Logan:1996}%
\bibitem{Motoyama:2007}%
  \BibitemOpen
  \bibfield{author}{%
  \bibinfo {author} {\bibfnamefont{E.}~\bibnamefont{Motoyama}}, \bibinfo
  {author} {\bibfnamefont{G.}~\bibnamefont{Yu}}, \bibinfo {author}
  {\bibfnamefont{I.}~\bibnamefont{Vishik}}, \bibinfo {author}
  {\bibfnamefont{O.}~\bibnamefont{Vajk}}, \bibinfo {author}
  {\bibfnamefont{P.}~\bibnamefont{Mang}},\ and\ \bibinfo {author}
  {\bibfnamefont{M.}~\bibnamefont{Greven}},\ }%
  \bibfield{journal}{%
  \bibinfo {journal} {Nature}\ }%
  \textbf{\bibinfo {volume} {445}},\ \bibinfo {pages} {186} (\bibinfo {year}
  {2007})%
  \bibAnnoteFile{NoStop}{Motoyama:2007}%
\bibitem{Veilleux:1995}%
  \BibitemOpen
  \bibfield{author}{%
  \bibinfo {author} {\bibfnamefont{A.~F.}\ \bibnamefont{Veilleux}}, \bibinfo
  {author} {\bibfnamefont{A.-M.}\ \bibnamefont{Dar\'e}}, \bibinfo {author}
  {\bibfnamefont{L.}~\bibnamefont{Chen}}, \bibinfo {author}
  {\bibfnamefont{Y.~M.}\ \bibnamefont{Vilk}},\ and\ \bibinfo {author}
  {\bibfnamefont{A.-M.~S.}\ \bibnamefont{Tremblay}},\ }%
  \bibfield{journal}{%
  \bibinfo {journal} {Phys. Rev. B}\ }%
  \textbf{\bibinfo {volume} {52}},\ \bibinfo {pages} {16255 } (\bibinfo {year}
  {1995})%
  \bibAnnoteFile{NoStop}{Veilleux:1995}%
\bibitem{Kyung:2003a}%
  \BibitemOpen
  \bibfield{author}{%
  \bibinfo {author} {\bibfnamefont{B.}~\bibnamefont{Kyung}}, \bibinfo {author}
  {\bibfnamefont{J.~S.}\ \bibnamefont{Landry}}, \bibinfo {author}
  {\bibfnamefont{D.}~\bibnamefont{Poulin}},\ and\ \bibinfo {author}
  {\bibfnamefont{A.-M.~S.}\ \bibnamefont{Tremblay}},\ }%
  \bibfield{journal}{%
  \bibinfo {journal} {Phys. Rev. Lett.}\ }%
  \textbf{\bibinfo {volume} {90}},\ \bibinfo {pages} {099702 } (\bibinfo {year}
  {2003})%
  \bibAnnoteFile{NoStop}{Kyung:2003a}%
\bibitem{LTP:2006}%
  \BibitemOpen
  \bibfield{author}{%
  \bibinfo {author} {\bibfnamefont{A.~M.~S.}\ \bibnamefont{Tremblay}}, \bibinfo
  {author} {\bibfnamefont{B.}~\bibnamefont{Kyung}},\ and\ \bibinfo {author}
  {\bibfnamefont{D.}~\bibnamefont{Senechal}},\ }%
  \bibfield{journal}{%
  \bibinfo {journal} {Low Temp. Phys.}\ }%
  \textbf{\bibinfo {volume} {32}},\ \bibinfo {pages} {424} (\bibinfo {year}
  {2006})%
  \bibAnnoteFile{NoStop}{LTP:2006}%
\bibitem{Baym:1962}%
  \BibitemOpen
  \bibfield{author}{%
  \bibinfo {author} {\bibfnamefont{G.}~\bibnamefont{Baym}},\ }%
  \bibfield{journal}{%
  \bibinfo {journal} {Physical review}\ }%
  \textbf{\bibinfo {volume} {127}},\ \bibinfo {pages} {1391} (\bibinfo {year}
  {1962})%
  \bibAnnoteFile{NoStop}{Baym:1962}%
\bibitem{Kadanoff_Baym:1962}%
  \BibitemOpen
  \bibfield{author}{%
  \bibinfo {author} {\bibfnamefont{L.}~\bibnamefont{Kadanoff}}\ and\ \bibinfo
  {author} {\bibfnamefont{G.}~\bibnamefont{Baym}},\ }%
  \emph{\bibinfo {title} {{Quantum Statistical Mechanics}}}\ (\bibinfo
  {publisher} {Benjamin},\ \bibinfo {year} {1962})%
  \bibAnnoteFile{NoStop}{Kadanoff_Baym:1962}%
\bibitem{Paul:2003}%
  \BibitemOpen
  \bibfield{author}{%
  \bibinfo {author} {\bibfnamefont{I.}~\bibnamefont{Paul}}\ and\ \bibinfo
  {author} {\bibfnamefont{G.}~\bibnamefont{Kotliar}},\ }%
  \bibfield{journal}{%
  \Doi{10.1103/PhysRevB.67.115131}{\bibinfo {journal} {Phys. Rev. B}}\ }%
  \textbf{\bibinfo {volume} {67}},\ \bibinfo {pages} {115131} (\bibinfo {month}
  {Mar.}\ \bibinfo {year} {2003}),\
  \Eprint{http://arxiv.org/abs/arXiv:cond-mat/0211538}{arXiv:cond-mat/0211538}%
  \bibAnnoteFile{NoStop}{Paul:2003}%
\bibitem{Vilk:1996}%
  \BibitemOpen
  \bibfield{author}{%
  \bibinfo {author} {\bibfnamefont{Y.}~\bibnamefont{Vilk}}\ and\ \bibinfo
  {author} {\bibfnamefont{A.-M.}\ \bibnamefont{Tremblay}},\ }%
  \bibfield{journal}{%
  \bibinfo {journal} {Europhys. Lett.}\ }%
  \textbf{\bibinfo {volume} {33}},\ \bibinfo {pages} {159 } (\bibinfo {year}
  {1996})%
  \bibAnnoteFile{NoStop}{Vilk:1996}%
\bibitem{MartinSchwinger:1959}%
  \BibitemOpen
  \bibfield{author}{%
  \bibinfo {author} {\bibfnamefont{P.~C.}\ \bibnamefont{Martin}}\ and\ \bibinfo
  {author} {\bibfnamefont{J.}~\bibnamefont{Schwinger}},\ }%
  \bibfield{journal}{%
  \Doi{10.1103/PhysRev.115.1342}{\bibinfo {journal} {Phys. Rev.}}\ }%
  \textbf{\bibinfo {volume} {115}},\ \bibinfo {pages} {1342} (\bibinfo {month}
  {Sep}\ \bibinfo {year} {1959})%
  \bibAnnoteFile{NoStop}{MartinSchwinger:1959}%
\bibitem{Luttinger_Ward:1960}%
  \BibitemOpen
  \bibfield{author}{%
  \bibinfo {author} {\bibfnamefont{J.~M.}\ \bibnamefont{Luttinger}}\ and\
  \bibinfo {author} {\bibfnamefont{J.~C.}\ \bibnamefont{Ward}},\ }%
  \bibfield{journal}{%
  \Doi{10.1103/PhysRev.118.1417}{\bibinfo {journal} {Phys. Rev.}}\ }%
  \textbf{\bibinfo {volume} {118}},\ \bibinfo {pages} {1417} (\bibinfo {month}
  {Jun}\ \bibinfo {year} {1960})%
  \bibAnnoteFile{NoStop}{Luttinger_Ward:1960}%
\bibitem{Hedeyati:1989}%
  \BibitemOpen
  \bibfield{author}{%
  \bibinfo {author} {\bibfnamefont{M.~R.}\ \bibnamefont{Hedayati}}\ and\
  \bibinfo {author} {\bibfnamefont{G.}~\bibnamefont{Vignale}},\ }%
  \bibfield{journal}{%
  \bibinfo {journal} {Phys. Rev. B}\ }%
  \textbf{\bibinfo {volume} {40}},\ \bibinfo {pages} {9044} (\bibinfo {year}
  {1989})%
  \bibAnnoteFile{NoStop}{Hedeyati:1989}%
\bibitem{Mahan:2000}%
  \BibitemOpen
  \bibfield{author}{%
  \bibinfo {author} {\bibfnamefont{G.}~\bibnamefont{Mahan}},\ }%
  \emph{\bibinfo {title} {Many-Particle Physics, 3rd edition, Section 6.4.4}}\
  (\bibinfo {publisher} {Kluwer/Plenum},\ \bibinfo {year} {2000})%
  \bibAnnoteFile{NoStop}{Mahan:2000}%
\bibitem{Berk:1966}%
  \BibitemOpen
  \bibfield{author}{%
  \bibinfo {author} {\bibfnamefont{N.~F.}\ \bibnamefont{Berk}}\ and\ \bibinfo
  {author} {\bibfnamefont{J.~R.}\ \bibnamefont{Schrieffer}},\ }%
  \bibfield{journal}{%
  \Doi{10.1103/PhysRevLett.17.433}{\bibinfo {journal} {Phys. Rev. Lett.}}\ }%
  \textbf{\bibinfo {volume} {17}},\ \bibinfo {pages} {433} (\bibinfo {month}
  {Aug}\ \bibinfo {year} {1966})%
  \bibAnnoteFile{NoStop}{Berk:1966}%
\bibitem{Larkin:2001}%
  \BibitemOpen
  \bibfield{author}{%
  \bibinfo {author} {\bibfnamefont{A.}~\bibnamefont{Larkin}}\ and\ \bibinfo
  {author} {\bibfnamefont{A.}~\bibnamefont{Varlamov}},\ }%
  \bibfield{journal}{%
  \bibinfo {journal} {ArXiv Condensed Matter e-prints}}%
   (\bibinfo {month} {Sep.}\ \bibinfo {year} {2001}),\
  \Eprint{http://arxiv.org/abs/arXiv:cond-mat/0109177}{arXiv:cond-mat/0109177}%
  \bibAnnoteFile{NoStop}{Larkin:2001}%
\bibitem{Maki:1968:1}%
  \BibitemOpen
  \bibfield{author}{%
  \bibinfo {author} {\bibfnamefont{K.}~\bibnamefont{Maki}},\ }%
  \bibfield{journal}{%
  \Doi{10.1143/PTP.39.897}{\bibinfo {journal} {Progress of Theoretical
  Physics}}\ }%
  \textbf{\bibinfo {volume} {39}},\ \bibinfo {pages} {897} (\bibinfo {year}
  {1968})%
  \bibAnnoteFile{NoStop}{Maki:1968:1}%
\bibitem{Maki:1968:2}%
  \BibitemOpen
  \bibfield{author}{%
  \bibinfo {author} {\bibfnamefont{K.}~\bibnamefont{Maki}},\ }%
  \bibfield{journal}{%
  \Doi{10.1143/PTP.40.193}{\bibinfo {journal} {Progress of Theoretical
  Physics}}\ }%
  \textbf{\bibinfo {volume} {40}},\ \bibinfo {pages} {193} (\bibinfo {year}
  {1968})%
  \bibAnnoteFile{NoStop}{Maki:1968:2}%
\bibitem{Maki:1969}%
  \BibitemOpen
  \bibfield{author}{%
  \bibinfo {author} {\bibfnamefont{K.}~\bibnamefont{Maki}},\ }%
  \bibfield{journal}{%
  \bibinfo {journal} {Journal of Low Temperature Physics}\ }%
  \textbf{\bibinfo {volume} {1}},\ \bibinfo {pages} {513} (\bibinfo {year}
  {1969})%
  \bibAnnoteFile{NoStop}{Maki:1969}%
\bibitem{Thompson:1970}%
  \BibitemOpen
  \bibfield{author}{%
  \bibinfo {author} {\bibfnamefont{R.~S.}\ \bibnamefont{Thompson}},\ }%
  \bibfield{journal}{%
  \Doi{10.1103/PhysRevB.1.327}{\bibinfo {journal} {Phys. Rev. B}}\ }%
  \textbf{\bibinfo {volume} {1}},\ \bibinfo {pages} {327} (\bibinfo {month}
  {Jan}\ \bibinfo {year} {1970})%
  \bibAnnoteFile{NoStop}{Thompson:1970}%
\bibitem{Aslamasov:1968}%
  \BibitemOpen
  \bibfield{author}{%
  \bibinfo {author} {\bibfnamefont{L.}~\bibnamefont{Aslamasov}}\ and\ \bibinfo
  {author} {\bibfnamefont{A.}~\bibnamefont{Larkin}},\ }%
  \bibfield{journal}{%
  \Doi{DOI: 10.1016/0375-9601(68)90623-3}{\bibinfo {journal} {Physics Letters
  A}}\ }%
  \textbf{\bibinfo {volume} {26}},\ \bibinfo {pages} {238 } (\bibinfo {year}
  {1968})%
  \bibAnnoteFile{NoStop}{Aslamasov:1968}%
\bibitem{Oudovenko:2002}%
  \BibitemOpen
  \bibfield{author}{%
  \bibinfo {author} {\bibfnamefont{V.~S.}\ \bibnamefont{Oudovenko}}\ and\
  \bibinfo {author} {\bibfnamefont{G.}~\bibnamefont{Kotliar}},\ }%
  \bibfield{journal}{%
  \bibinfo {journal} {Phys. Rev. B}\ }%
  \textbf{\bibinfo {volume} {65}},\ \bibinfo {pages} {075102 } (\bibinfo {year}
  {2002/02/15})%
  \bibAnnoteFile{NoStop}{Oudovenko:2002}%
\bibitem{Jarrell1996133}%
  \BibitemOpen
  \bibfield{author}{%
  \bibinfo {author} {\bibfnamefont{M.}~\bibnamefont{Jarrell}}\ and\ \bibinfo
  {author} {\bibfnamefont{J.~E.}\ \bibnamefont{Gubernatis}},\ }%
  \bibfield{journal}{%
  \Doi{DOI: 10.1016/0370-1573(95)00074-7}{\bibinfo {journal} {Physics
  Reports}}\ }%
  \textbf{\bibinfo {volume} {269}},\ \bibinfo {pages} {133 } (\bibinfo {year}
  {1996})%
  \bibAnnoteFile{NoStop}{Jarrell1996133}%
\bibitem{Beach_Gooding_2000}%
  \BibitemOpen
  \bibfield{author}{%
  \bibinfo {author} {\bibfnamefont{K.~S.~D.}\ \bibnamefont{Beach}}, \bibinfo
  {author} {\bibfnamefont{R.~J.}\ \bibnamefont{Gooding}},\ and\ \bibinfo
  {author} {\bibfnamefont{F.}~\bibnamefont{Marsiglio}},\ }%
  \bibfield{journal}{%
  \Doi{10.1103/PhysRevB.61.5147}{\bibinfo {journal} {Phys. Rev. B}}\ }%
  \textbf{\bibinfo {volume} {61}},\ \bibinfo {pages} {5147} (\bibinfo {month}
  {Feb}\ \bibinfo {year} {2000})%
  \bibAnnoteFile{NoStop}{Beach_Gooding_2000}%
\bibitem{Roy:2008}%
  \BibitemOpen
  \bibfield{author}{%
  \bibinfo {author} {\bibfnamefont{S.}~\bibnamefont{Roy}}\ and\ \bibinfo
  {author} {\bibfnamefont{A.-M.~S.}\ \bibnamefont{Tremblay}},\ }%
  \bibfield{journal}{%
  \bibinfo {journal} {Europhys. Lett.}\ }%
  \textbf{\bibinfo {volume} {84}},\ \bibinfo {pages} {37013} (\bibinfo {year}
  {2008})%
  \bibAnnoteFile{NoStop}{Roy:2008}%
\bibitem{Dare:1996}%
  \BibitemOpen
  \bibfield{author}{%
  \bibinfo {author} {\bibfnamefont{A.-M.}\ \bibnamefont{Dar{\'e}}}, \bibinfo
  {author} {\bibfnamefont{Y.~M.}\ \bibnamefont{Vilk}},\ and\ \bibinfo {author}
  {\bibfnamefont{A.-M.~S.}\ \bibnamefont{Tremblay}},\ }%
  \bibfield{journal}{%
  \bibinfo {journal} {Phys. Rev. B}\ }%
  \textbf{\bibinfo {volume} {53}},\ \bibinfo {pages} {14236 } (\bibinfo {year}
  {1996})%
  \bibAnnoteFile{NoStop}{Dare:1996}%
\bibitem{Onose:2004}%
  \BibitemOpen
  \bibfield{author}{%
  \bibinfo {author} {\bibfnamefont{Y.}~\bibnamefont{Onose}}, \bibinfo {author}
  {\bibfnamefont{Y.}~\bibnamefont{Taguchi}}, \bibinfo {author}
  {\bibfnamefont{K.}~\bibnamefont{Ishizaka}},\ and\ \bibinfo {author}
  {\bibfnamefont{Y.}~\bibnamefont{Tokura}},\ }%
  \bibfield{journal}{%
  \bibinfo {journal} {Physical Review B (Condensed Matter and Materials
  Physics)}\ }%
  \textbf{\bibinfo {volume} {69}},\ \bibinfo {pages} {24504 } (\bibinfo {year}
  {2004/01/01})%
  \bibAnnoteFile{NoStop}{Onose:2004}%
\bibitem{Zimmers:2005}%
  \BibitemOpen
  \bibfield{author}{%
  \bibinfo {author} {\bibfnamefont{A.}~\bibnamefont{Zimmers}}, \bibinfo
  {author} {\bibfnamefont{J.}~\bibnamefont{Tomczak}}, \bibinfo {author}
  {\bibfnamefont{R.}~\bibnamefont{Lobo}}, \bibinfo {author}
  {\bibfnamefont{N.}~\bibnamefont{Bontemps}}, \bibinfo {author}
  {\bibfnamefont{C.}~\bibnamefont{Hill}}, \bibinfo {author}
  {\bibfnamefont{M.}~\bibnamefont{Barr}}, \bibinfo {author}
  {\bibfnamefont{Y.}~\bibnamefont{Dagan}}, \bibinfo {author}
  {\bibfnamefont{R.}~\bibnamefont{Greene}}, \bibinfo {author}
  {\bibfnamefont{A.}~\bibnamefont{Millis}},\ and\ \bibinfo {author}
  {\bibfnamefont{C.}~\bibnamefont{Homes}},\ }%
  \bibfield{journal}{%
  \bibinfo {journal} {Europhysics Letters}\ }%
  \textbf{\bibinfo {volume} {70}},\ \bibinfo {pages} {225 } (\bibinfo {year}
  {2005/04/15})%
  \bibAnnoteFile{NoStop}{Zimmers:2005}%
\bibitem{Moriya:2000}%
  \BibitemOpen
  \bibfield{author}{%
  \bibinfo {author} {\bibfnamefont{T.}~\bibnamefont{Moriya}}\ and\ \bibinfo
  {author} {\bibfnamefont{K.}~\bibnamefont{Ueda}},\ }%
  \bibfield{journal}{%
  \bibinfo {journal} {Advances in Physics}\ }%
  \textbf{\bibinfo {volume} {49}},\ \bibinfo {pages} {555 } (\bibinfo {year}
  {2000/07})%
  \bibAnnoteFile{NoStop}{Moriya:2000}%
\bibitem{Kyung:2003}%
  \BibitemOpen
  \bibfield{author}{%
  \bibinfo {author} {\bibfnamefont{B.}~\bibnamefont{Kyung}}, \bibinfo {author}
  {\bibfnamefont{J.-S.}\ \bibnamefont{Landry}},\ and\ \bibinfo {author}
  {\bibfnamefont{A.~M.~S.}\ \bibnamefont{Tremblay}},\ }%
  \bibfield{journal}{%
  \bibinfo {journal} {Phys. Rev. B}\ }%
  \textbf{\bibinfo {volume} {68}},\ \bibinfo {pages} {174502} (\bibinfo {year}
  {2003})%
  \bibAnnoteFile{NoStop}{Kyung:2003}%
\bibitem{Hankevych:2005}%
  \BibitemOpen
  \bibfield{author}{%
  \bibinfo {author} {\bibfnamefont{V.}~\bibnamefont{Hankevych}}, \bibinfo
  {author} {\bibfnamefont{B.}~\bibnamefont{Kyung}}, \bibinfo {author}
  {\bibfnamefont{A.-M.}\ \bibnamefont{Dar{\'e}}}, \bibinfo {author}
  {\bibfnamefont{D.}~\bibnamefont{S{\'e}n{\'e}chal}},\ and\ \bibinfo {author}
  {\bibfnamefont{A.-M.}\ \bibnamefont{Tremblay}},\ }%
  in\ \emph{\bibinfo {booktitle} {Proceedings of SNS2004}}\ (\bibinfo {year}
  {2005})\ \Eprint{http://arxiv.org/abs/cond-mat/0407085}{cond-mat/0407085}%
  \bibAnnoteFile{NoStop}{Hankevych:2005}%
\bibitem{BergeronPhD:2011}%
  \BibitemOpen
  \bibfield{author}{%
  \bibinfo {author} {\bibfnamefont{D.}~\bibnamefont{Bergeron}},\ }%
  \emph{\bibinfo {title} {Conductivit\'e dans le mod\`ele de Hubbard
  bidimensionnel a faible couplage}},\ Ph.D. thesis,\ \bibinfo {school}
  {Universit\'e de Sherbrooke} (\bibinfo {year} {2011})%
  \bibAnnoteFile{NoStop}{BergeronPhD:2011}%
\bibitem{Kaminski:2003}%
  \BibitemOpen
  \bibfield{author}{%
  \bibinfo {author} {\bibfnamefont{A.}~\bibnamefont{Kaminski}}, \bibinfo
  {author} {\bibfnamefont{S.}~\bibnamefont{Rosenkranz}}, \bibinfo {author}
  {\bibfnamefont{H.~M.}\ \bibnamefont{Fretwell}}, \bibinfo {author}
  {\bibfnamefont{Z.~Z.}\ \bibnamefont{Li}}, \bibinfo {author}
  {\bibfnamefont{H.}~\bibnamefont{Raffy}}, \bibinfo {author}
  {\bibfnamefont{M.}~\bibnamefont{Randeria}}, \bibinfo {author}
  {\bibfnamefont{M.~R.}\ \bibnamefont{Norman}},\ and\ \bibinfo {author}
  {\bibfnamefont{J.~C.}\ \bibnamefont{Campuzano}},\ }%
  \bibfield{journal}{%
  \Doi{10.1103/PhysRevLett.90.207003}{\bibinfo {journal} {Phys. Rev. Lett.}}\
  }%
  \textbf{\bibinfo {volume} {90}},\ \bibinfo {pages} {207003} (\bibinfo {month}
  {May}\ \bibinfo {year} {2003})%
  \bibAnnoteFile{NoStop}{Kaminski:2003}%
\bibitem{Hussey:2008}%
  \BibitemOpen
  \bibfield{author}{%
  \bibinfo {author} {\bibfnamefont{N.~E.}\ \bibnamefont{Hussey}},\ }%
  \bibfield{journal}{%
  \bibinfo {journal} {Journal of Physics: Condensed Matter}\ }%
  \textbf{\bibinfo {volume} {20}},\ \bibinfo {pages} {123201} (\bibinfo {year}
  {2008})%
  \bibAnnoteFile{NoStop}{Hussey:2008}%
\bibitem{Cooper:2009}%
  \BibitemOpen
  \bibfield{author}{%
  \bibinfo {author} {\bibfnamefont{R.~A.}\ \bibnamefont{Cooper}}, \bibinfo
  {author} {\bibfnamefont{Y.}~\bibnamefont{Wang}}, \bibinfo {author}
  {\bibfnamefont{B.}~\bibnamefont{Vignolle}}, \bibinfo {author}
  {\bibfnamefont{O.~J.}\ \bibnamefont{Lipscombe}}, \bibinfo {author}
  {\bibfnamefont{S.~M.}\ \bibnamefont{Hayden}}, \bibinfo {author}
  {\bibfnamefont{Y.}~\bibnamefont{Tanabe}}, \bibinfo {author}
  {\bibfnamefont{T.}~\bibnamefont{Adachi}}, \bibinfo {author}
  {\bibfnamefont{Y.}~\bibnamefont{Koike}}, \bibinfo {author}
  {\bibfnamefont{M.}~\bibnamefont{Nohara}}, \bibinfo {author}
  {\bibfnamefont{H.}~\bibnamefont{Takagi}}, \bibinfo {author}
  {\bibfnamefont{C.}~\bibnamefont{Proust}},\ and\ \bibinfo {author}
  {\bibfnamefont{N.~E.}\ \bibnamefont{Hussey}},\ }%
  \bibfield{journal}{%
  \Doi{10.1126/science.1165015}{\bibinfo {journal} {Science}}\ }%
  \textbf{\bibinfo {volume} {323}},\ \bibinfo {pages} {603} (\bibinfo {year}
  {2009}),\
  \Eprint{http://arxiv.org/abs/http://www.sciencemag.org/content/323/5914/603.%
full.pdf}{http://www.sciencemag.org/content/323/5914/603.full.pdf}%
  \bibAnnoteFile{NoStop}{Cooper:2009}%
\bibitem{Doiron-Leyraud:2009_1}%
  \BibitemOpen
  \bibfield{author}{%
  \bibinfo {author} {\bibfnamefont{N.}~\bibnamefont{{Doiron-Leyraud}}},
  \bibinfo {author} {\bibfnamefont{P.}~\bibnamefont{{Auban-Senzier}}}, \bibinfo
  {author} {\bibfnamefont{S.~R.}\ \bibnamefont{{de Cotret}}}, \bibinfo {author}
  {\bibfnamefont{A.}~\bibnamefont{{Sedeki}}}, \bibinfo {author}
  {\bibfnamefont{C.}~\bibnamefont{{Bourbonnais}}}, \bibinfo {author}
  {\bibfnamefont{D.}~\bibnamefont{{Jerome}}}, \bibinfo {author}
  {\bibfnamefont{K.}~\bibnamefont{{Bechgaard}}},\ and\ \bibinfo {author}
  {\bibfnamefont{L.}~\bibnamefont{{Taillefer}}},\ }%
  \bibfield{journal}{%
  \bibinfo {journal} {ArXiv e-prints}}%
   (\bibinfo {month} {May}\ \bibinfo {year} {2009}),\
  \Eprint{http://arxiv.org/abs/0905.0964}{arXiv:0905.0964 [cond-mat.supr-con]}%
  \bibAnnoteFile{NoStop}{Doiron-Leyraud:2009_1}%
\bibitem{Doiron-Leyraud:2009}%
  \BibitemOpen
  \bibfield{author}{%
  \bibinfo {author} {\bibfnamefont{N.}~\bibnamefont{Doiron-Leyraud}}, \bibinfo
  {author} {\bibfnamefont{P.}~\bibnamefont{Auban-Senzier}}, \bibinfo {author}
  {\bibfnamefont{S.}~\bibnamefont{Ren\'e~de Cotret}}, \bibinfo {author}
  {\bibfnamefont{C.}~\bibnamefont{Bourbonnais}}, \bibinfo {author}
  {\bibfnamefont{D.}~\bibnamefont{J\'erome}}, \bibinfo {author}
  {\bibfnamefont{K.}~\bibnamefont{Bechgaard}},\ and\ \bibinfo {author}
  {\bibfnamefont{L.}~\bibnamefont{Taillefer}},\ }%
  \bibfield{journal}{%
  \Doi{10.1103/PhysRevB.80.214531}{\bibinfo {journal} {Phys. Rev. B}}\ }%
  \textbf{\bibinfo {volume} {80}},\ \bibinfo {pages} {214531} (\bibinfo {month}
  {Dec}\ \bibinfo {year} {2009})%
  \bibAnnoteFile{NoStop}{Doiron-Leyraud:2009}%
\bibitem{Abdel-Jawad_9957290}%
  \BibitemOpen
  \bibfield{author}{%
  \bibinfo {author} {\bibfnamefont{M.}~\bibnamefont{Abdel-Jawad}}, \bibinfo
  {author} {\bibfnamefont{J.~G.}\ \bibnamefont{Analytis}}, \bibinfo {author}
  {\bibfnamefont{L.}~\bibnamefont{Balicas}}, \bibinfo {author}
  {\bibfnamefont{A.}~\bibnamefont{Carrington}}, \bibinfo {author}
  {\bibfnamefont{J.~P.~H.}\ \bibnamefont{Charmant}}, \bibinfo {author}
  {\bibfnamefont{M.~M.~J.}\ \bibnamefont{French}},\ and\ \bibinfo {author}
  {\bibfnamefont{N.~E.}\ \bibnamefont{Hussey}},\ }%
  \bibfield{journal}{%
  \bibinfo {journal} {Physical Review Letters}\ }%
  \textbf{\bibinfo {volume} {99}},\ \bibinfo {pages} {107002 } (\bibinfo {year}
  {7 Sept. 2007})%
  \bibAnnoteFile{NoStop}{Abdel-Jawad_9957290}%
\bibitem{Ando:2004}%
  \BibitemOpen
  \bibfield{author}{%
  \bibinfo {author} {\bibfnamefont{Y.}~\bibnamefont{Ando}}, \bibinfo {author}
  {\bibfnamefont{S.}~\bibnamefont{Komiya}}, \bibinfo {author}
  {\bibfnamefont{K.}~\bibnamefont{Segawa}}, \bibinfo {author}
  {\bibfnamefont{S.}~\bibnamefont{Ono}},\ and\ \bibinfo {author}
  {\bibfnamefont{Y.}~\bibnamefont{Kurita}},\ }%
  \bibfield{journal}{%
  \bibinfo {journal} {Physical Review Letters}\ }%
  \textbf{\bibinfo {volume} {93}},\ \bibinfo {pages} {267001 } (\bibinfo {year}
  {2004/12/31})%
  \bibAnnoteFile{NoStop}{Ando:2004}%
\bibitem{Palsson:2001}%
  \BibitemOpen
  \bibfield{author}{%
  \bibinfo {author} {\bibfnamefont{G.}~\bibnamefont{P\'alsson}},\ }%
  \emph{\bibinfo {title} {Computational studies of thermoelectricity in
  strongly correlated electron systems}},\ Ph.D. thesis,\ \bibinfo {school}
  {Rutgers University} (\bibinfo {year} {2001})%
  \bibAnnoteFile{NoStop}{Palsson:2001}%
\bibitem{Terletska:2010}%
  \BibitemOpen
  \bibfield{author}{%
  \bibinfo {author} {\bibfnamefont{H.}~\bibnamefont{{Terletska}}}, \bibinfo
  {author} {\bibfnamefont{J.}~\bibnamefont{{Vu\v{c}i\v{c}evi{\'c}}}}, \bibinfo
  {author} {\bibfnamefont{D.}~\bibnamefont{{Tanaskovi{\'c}}}},\ and\ \bibinfo
  {author} {\bibfnamefont{V.}~\bibnamefont{{Dobrosavljevi{\'c}}}},\ }%
  \bibfield{journal}{%
  \bibinfo {journal} {ArXiv e-prints}}%
   (\bibinfo {month} {Dec.}\ \bibinfo {year} {2010}),\
  \Eprint{http://arxiv.org/abs/1012.5833}{arXiv:1012.5833 [cond-mat.str-el]}%
  \bibAnnoteFile{NoStop}{Terletska:2010}%
\bibitem{Dare:1994}%
  \BibitemOpen
  \bibfield{author}{%
  \bibinfo {author} {\bibfnamefont{A.-M.}\ \bibnamefont{Dar\'e}}, \bibinfo
  {author} {\bibfnamefont{L.}~\bibnamefont{Chen}},\ and\ \bibinfo {author}
  {\bibfnamefont{A.-M.~S.}\ \bibnamefont{Tremblay}},\ }%
  \bibfield{journal}{%
  \Doi{10.1103/PhysRevB.49.4106}{\bibinfo {journal} {Phys. Rev. B}}\ }%
  \textbf{\bibinfo {volume} {49}},\ \bibinfo {pages} {4106} (\bibinfo {month}
  {Feb}\ \bibinfo {year} {1994})%
  \bibAnnoteFile{NoStop}{Dare:1994}%
\bibitem{Mahan:3rd}%
  \BibitemOpen
  \bibfield{author}{%
  \bibinfo {author} {\bibfnamefont{G.}~\bibnamefont{Mahan}},\ }%
  \emph{\bibinfo {title} {Many-Particle Physics}}\ (\bibinfo {publisher}
  {Kluwer Academic},\ \bibinfo {year} {2000})%
  \bibAnnoteFile{NoStop}{Mahan:3rd}%
\end{thebibliography}

\end{document}